\documentclass[12pt]{iopart}
\usepackage{amsgen}
\usepackage{amsbsy}
\usepackage{amsfonts}
\usepackage{amssymb}
\usepackage{iopams}
\usepackage{color,colortbl}
\usepackage{multirow}
\usepackage{bigstrut}
\usepackage[pdftex]{graphicx}

\definecolor{Gr}{rgb}{0.20,0,0.6}
\definecolor{Dblue}{rgb}{0.20,0,0.6}

\newcommand{\cH}{\mathcal{H}}
\newcommand{\cE}{\mathcal{E}}
\newcommand{\cP}{\mathcal{P}}
\newcommand{\quotes}[1]{\textquotedblleft #1\textquotedblright}

\definecolor{Gray}{gray}{0.9}
\begin{document}
\topical[Nonlinear lattice waves in heterogeneous media]{Nonlinear lattice waves in heterogeneous media}
\author{T V Laptyeva$^1$, M V Ivanchenko$^{2}$, and S Flach$^{3}$}
\address{$^1$ Theory of Control Department, Lobachevsky State University of Nizhny Novgorod, Nizhny Novgorod, Russia}
\address{$^1$ Department of Bioinformatics,  Lobachevsky State University of Nizhny Novgorod, Nizhny Novgorod, Russia}
\address{$^3$ New Zealand Institute for Advanced Study, Massey University, Auckland, New Zealand}
\eads{\mailto{tetyana.laptyeva@gmail.com},\mailto{ivanchenko.mv@gmail.com},\mailto{s.flach@massey.ac.nz}}
\begin{abstract}
We review recent progress in the dynamics of nonlinear lattice waves in heterogeneous media,
which enforce complete wave localization in the linear wave equation limit, especially
Anderson localization for random
potentials, and Aubry-Andre localization for quasiperiodic potentials.
 Additional nonlinear terms in the wave equations
can either preserve the phase-coherent localization of waves, or destroy it through nonintegrability and
deterministic chaos. Spreading wave packets are observed to show universal features in their dynamics which are related to properties of nonlinear diffusion equations. 
\end{abstract}
\submitto{\JPA}
\maketitle
\section{Linear waves}
\label{LRM}
Propagating lattice waves often encounter heterogeneities e.g. due to impurities or defects, spatial gradients caused by external fields, or due
the destruction of perfect spatial periodicity with a secondary incommensurate lattice. 
Heterogeneity leads to strong modifications of the transport properties of waves, starting
from expected renormalizations of transport coefficients and ending with completely new
states of the system. 
\subsection{Anderson localization}
\label{AL}      
The propagation of a free electron in a perfectly periodic crystal is described in terms of extended Bloch waves, and results in ballistic transport \cite{Kitt2004}. Since artificial as well as conventional materials are not 
ideally periodic, a natural question arises: what happens in a realistic situation when the periodicity of the system is broken by impurities, defects, or other imperfections? 

Classical transport theory viewed disordered materials as multiple-scattering media, where scattered Bloch waves loose phase coherence on the length scale of the mean-free-path $l_m$ -- an average distance traveled by the electron between collisions. The transport of electrons in such media can be thus described as a diffusion process. One then arrives at the Ohmic dependence of conductance on the length of a sample  with the Drude conductivity 
\cite{Zim72}.  

A breakthrough was achieved in the seminal paper by Anderson \cite{And58}, who proposed that at a certain amount of disorder the diffusive motion of the electron will come to a complete halt. Anderson considered the tight-binding approximation for an electron on a lattice, with on-site energies being randomly distributed and only nearest-neighbor tunneling. In brief, the electron inability to diffuse away from an initial position was related to the convergence of a renormalized perturbation series, up to all multiple scattering electronic paths incorporated. In terms of the electron wave function this means, that its amplitude falls off exponentially with growing distance from its original location. Most importantly the electronic state keeps phase coherence, reflecting the fact that wave localization
is a phase-coherent phenomenon, and loss of phase coherence means wave delocalization \cite{Rayanov2013}.

Being essentially a product of wave interference, Anderson localization (AL) is not restricted to electrons, but manifests itself for many types of wave propagation in inhomogeneous media. Indeed, AL has been experimentally observed for electromagnetic, acoustic and matter waves propagating in random media. S. John  \textit{et al.} predicted phonon localization in disordered elastic media \cite{Joh83a} and, similarly, the critical behavior of electromagnetic waves in disordered dielectric media \cite{Joh87} (for a comprehensive review see \cite{Tigg98}).

\begin{figure}[ht]
\begin{center}
\includegraphics[width=0.7\columnwidth,keepaspectratio,clip]{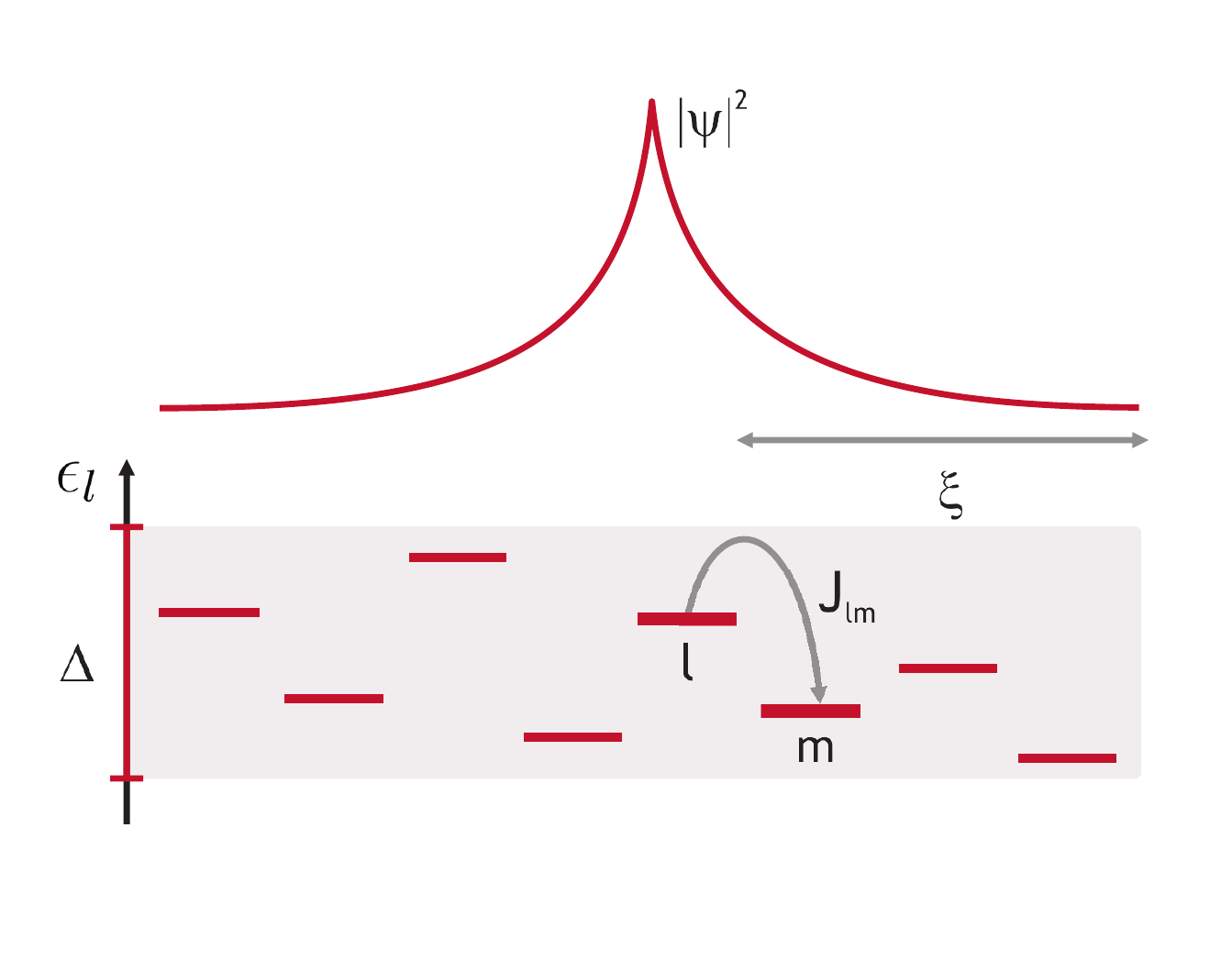}
\caption{\textit{Bottom:} schematic representation of the Anderson tight-binding model for an electron on disordered lattice. 
\textit{Top:} due to disorder, the electronic wave-function $\psi$ becomes spatially localized with the exponentially decaying tails. 
} 
\label{fig:1p2}
\end{center}
\end{figure}

Anderson model is governed by the Hamiltonian \cite{And58}
\begin{equation}
\hat{\cH}_{\rm A} = \sum_l \epsilon_l \left| l \right\rangle \left\langle l \right| + \sum_{l,m} J_{lm} \left| l \right\rangle \left\langle m \right|,
\label{eq:1p3}	
\end{equation}
where $\left| l \right\rangle$ is a Wannier state localized about the $l$-th potential well, \textit{i.e.} the $l$-th lattice site. The non-diagonal elements $J_{lm}$ describe the electron hopping from the site $l$ to the site $m$. In the original Anderson model (see the sketch in \fref{fig:1p2}), the electron is allowed to tunnel between nearest-neighbor sites with a constant rate $J_{lm}=1$ and the disorder is introduced by taking the on-site energies $\epsilon_l$ from the box probability distribution $\mathcal{P}(\epsilon_l) = 1/W$ for $\left|\epsilon_l\right| < W/2$ and zero otherwise. Here $W$ parametrizes the disorder strength. 
A more detailed account on different models of disorder, such as structural, topological, orientational can be found in \cite{Kra93}.

The wave-function amplitudes on lattice sites $ \Psi = \sum_l \varphi_l \left| l \right\rangle$
then evolve according to the discrete Schr\"odinger equation (DLS):
\begin{equation}
i \dot{\varphi}_l = \epsilon_l \varphi_l + \varphi_{l+1} + \varphi_{l-1}.
\label{eq:1p4}	
\end{equation}

The stationary states $\varphi_l(t)=\psi_l \e^{-i E t}$ satisfy the eigenvalue problem
\begin{equation}
E \psi_l = \epsilon_l \psi_l + \psi_{l+1} + \psi_{l-1},
\label{eq:1p4p}	
\end{equation}
which eigenstates are exponentially localized $\left|\psi_l\right|^2 \propto \e^{-\left|l-l_0\right|/\xi_E}$ with the eigenvalue-dependent localization length $\xi_E$. 
Alternative methods for quantifying localized states are e.g. the inverse participation number, intensity-intensity correlation functions, Lyapunov exponents \cite{Kra93, Cris93}.

Approaches for evaluating $\xi_E$ include the transfer matrix method, random matrix theory, perturbative techniques (see \cite{Kra93, Cris93, Ish73} and references therein). For a one-dimensional (1D) system with weak uncorrelated disorder $W \ll 1$ and eigenvalues close to  the band center 
\begin{equation}
\xi_E \approx 24 \frac{(4-E^2)}{W^2}.
\label{eq:1p6ppppp}	
\end{equation}
The largest localization length of the mode in the band center ($E=0$) reads $\xi_0 \approx 96/W^2$. In case of strong disorder $W \gg 1$, perturbation theory yields $\xi_0 \approx \left[\ln(W/2)\right]^{-1}$.

\begin{figure}[ht]
\begin{center}
\includegraphics[width=0.65\columnwidth,keepaspectratio,clip]{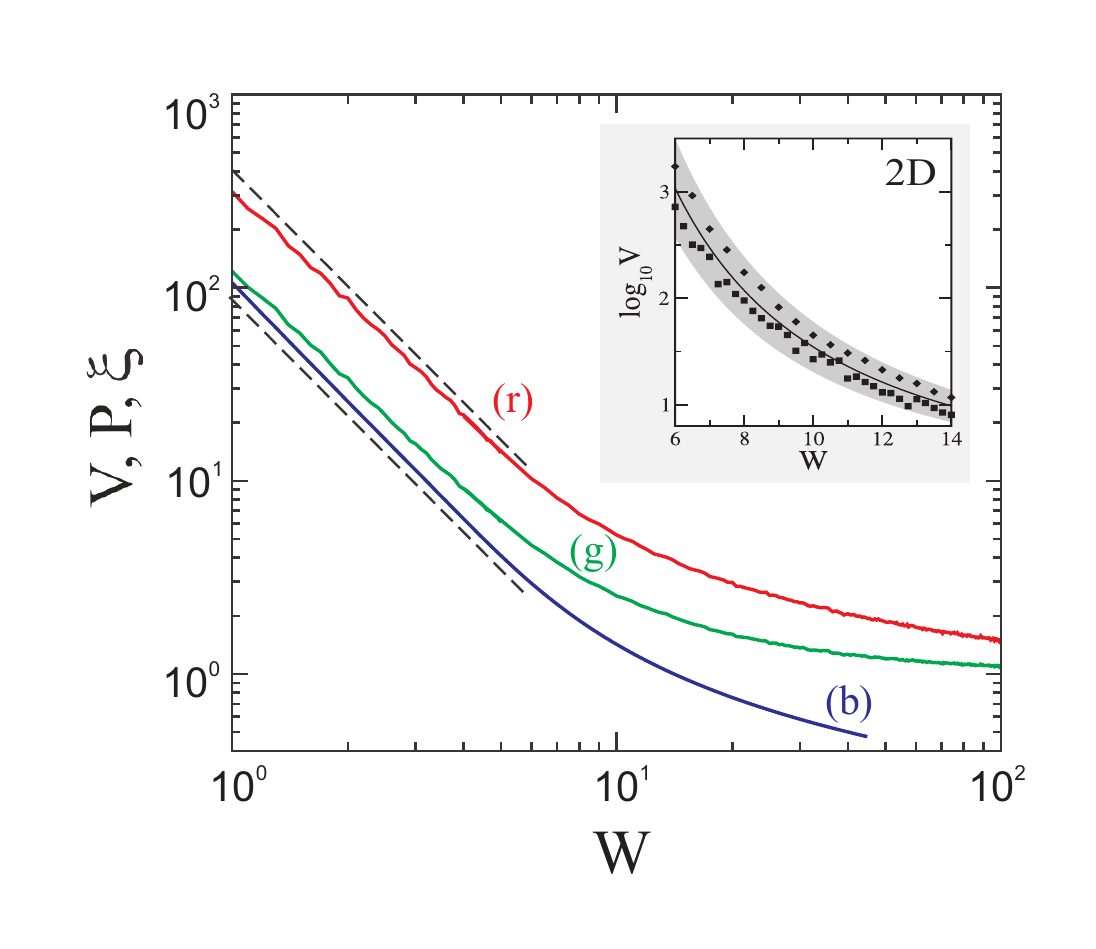}
\caption{Numerical evaluation of various length scales of (\ref{eq:1p4p}). Average localization volume $V$ (red (r)), participation number $P$ (green (g)), and localization length $\xi$ (blue (b)) \textit{vs.} disorder strength $W$ are shown for the most extended eigenstates (with eigenfrequencies located about the band center). Dashed lines correspond to $330/W^2$ and $100/W^2$ for $V$ and $\xi$ in the small disorder limit, respectively. Reprinted from \cite{Krim10}. \textit{Inset:} numerically calculated dependence of $\log_{10}V$ on $W$ for the 2D DLS (squares) and KG (diamonds) lattices, see (\ref{eq:2DdDLS}) and (\ref{eq:1DdKG})). The gray region denotes standard deviation. The solid
line guides the eye. Reprinted from \cite{Laptyeva2012}.} 
\label{fig:1p3}
\end{center}
\end{figure}

Alternative measures of the degree of localization  are the second moment $m_2$ and the participation number $P$.  $m_2^{(\nu)}$ quantifies characteristic width of the $\nu$-th eigenstate and can be computed as $m_2^{(\nu)} = \sum_l (m_1^{(\nu)}-l)^2\left|\psi_{\nu,l}\right|^2$  with $m_1^{(\nu)} = \sum_l l \left|\psi_{\nu,l}\right|^2$ being the spatial center. The participation number $P^{(\nu)}=1/\sum_l \left|\psi_{\nu,l}\right|^4$ roughly counts the number of strongly populated 
lattice sites. Both quantities averaged over modes and disorder realizations may approximate the spatial extent of eigenstates -- the localization volume $V$. In case of flat compact distributions in the 1D problem both $m_2=\bar{m}_2^{(\nu)}$ and $P=\bar{P}^{(\nu)}$ can define localization volume as $V=\sqrt{12m_2}+1$ or $V=P$. 
For weak disorder and fluctuating  distributions the participation number underestimates the localization volume.
Numerical results \cite{Krim10} show that the mean localization volume $V \approx 3.3 \xi_0$ in the limit of weak disorder and tends to unity as the disorder gets sufficiently large (see \fref{fig:1p3}). 

Generalization to higher-dimensional, for example 2D, lattices is straightforward:
\begin{equation}
i \dot{\psi}_{l,m} = \epsilon_{l,m} \psi_{l,m} - (\psi_{l+1,m} + \psi_{l-1,m}+\psi_{l,m+1} + \psi_{l,m-1}).
\label{eq:2DdDLS}
\end{equation}	   

The above equations of motion can be derived from the following Hamiltonians by 
$i\dot{\psi}_l = \partial \cH / \partial \psi^\ast_l$:
\begin{eqnarray}
\mbox{1D:}\;
\qquad 
\cH_{\rm D} = \sum_l \left[\epsilon_l \left| \psi_l \right|^2 - \left(\psi_{l+1}\psi_l^\ast + {\rm c.c.}\right)\right],
\label{eq:1DHdDLS}
\\
\mbox{2D:}\;
\cH_{\rm D} = \sum_{l,m} \left\{\epsilon_{l,m}\left|\psi_{l,m} \right|^2 - \left[\psi_{l,m}^\ast\left(\psi_{l+1,m}+\psi_{l,m+1}\right) + {\rm c.c.}\right]\right\}.
\label{eq:2DHdDLS}
\end{eqnarray}	   

\begin{figure}[ht]
\begin{center}
\includegraphics[width=0.45\columnwidth,keepaspectratio,clip]{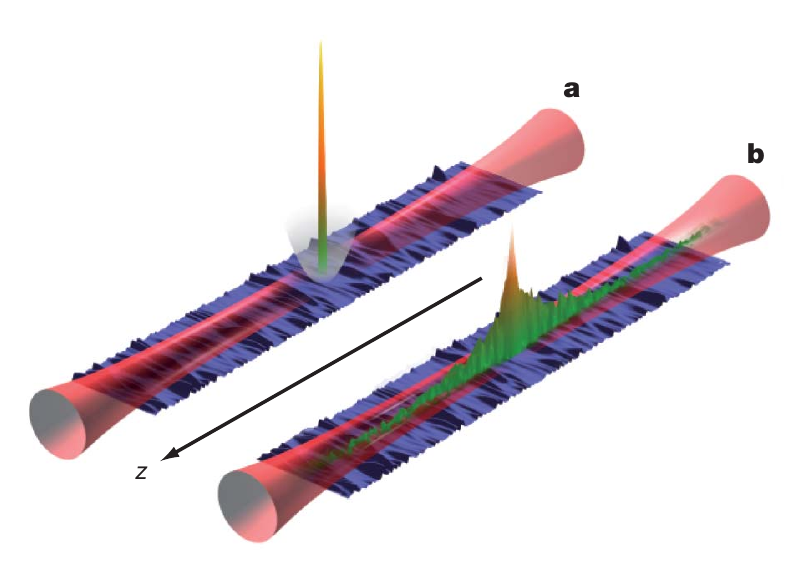}
\includegraphics[width=0.33\columnwidth,keepaspectratio,clip]{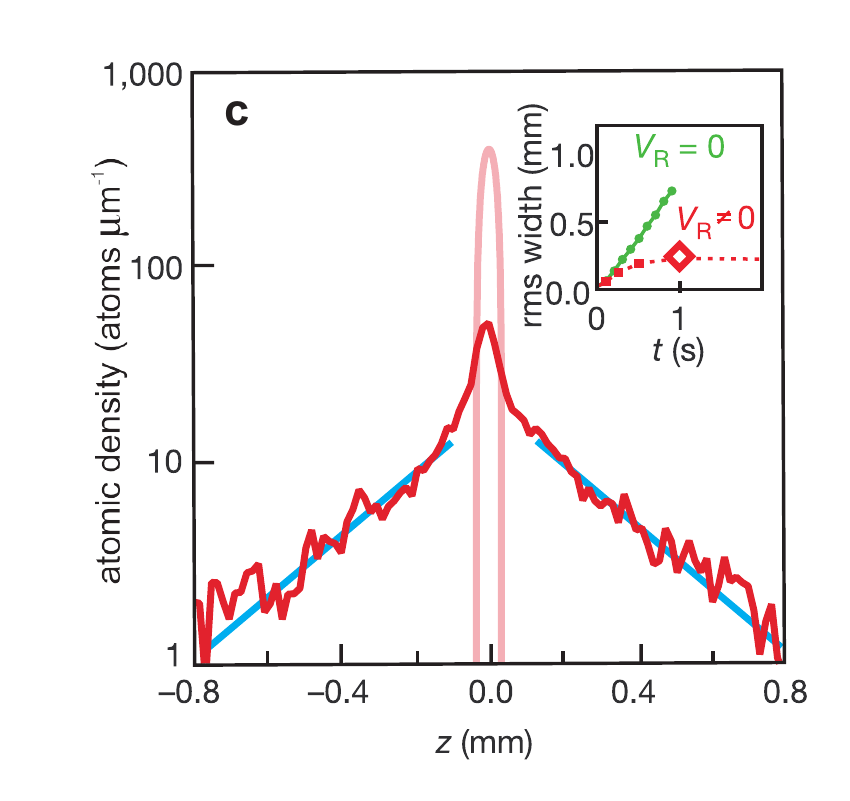}
\caption{AL of ultra-cold atoms in a random potential \cite{Billy08}.The  BEC is initially prepared in a magnetic trap and confined transversely to the $z$-axis in a 1D optical waveguide (a). When the trap is switched off, the condensate is allowed to expand along the waveguide superimposed with a speckle potential (b). The BEC stops it expansion after about 0.5 s. Its atomic density distribution shows exponentially decaying tails (c). The inset displays the time dependence of the root mean square width of the density profile with/without disordered potential $V_{\rm R}$. Reprinted from \cite{Billy08}.} 
\label{fig:BECs}
\end{center}
\end{figure}

Persistent efforts to directly observe AL have accumulated in a bevy of various experimental works. An exhaustive review on the past state of art can be found, \textit{e.g.} in \cite{Kra93}. Here we briefly mention more recent studies.
Bose-Einstein-Condensates (BECs) are an ideal tool to realize and experimentally study AL. Experiments are commonly based on dilute BEC of $^{39}$K or $^{87}$Rb prepared in a trap. The BEC is usually loaded onto an optical lattice and a magnetic trap confines its expansion \cite{Morsch2006,Dalfovo1999}. Then the trap is turned off letting the gas to diffuse across an optically generated disordered potential. It is created by a speckle field, the result of passing a laser beam through a diffusive medium \cite{Good07}. The influence of atom-atom interactions in the condensate is either reduced to a negligible level by Feshbach resonances or compensated by deep optical potentials. Under these conditions the almost non-interacting Bose gas can be characterized by a single wave function. The localization of the atomic wave function can be studied \textit{in situ}, employing absorption or fluorescence techniques to image the atomic density. This approach has led to groundbreaking results demonstrating AL of BEC in random potentials in one \cite{Billy08,Fort05,Clement05}, and three \cite{Kondov11,Jendr2011} dimensions. \Fref{fig:BECs} illustrates one of the pioneering results supporting AL of ultra-cold atoms in random potentials \cite{Billy08}. 
\begin{figure}
\begin{center}
\includegraphics[width=0.5\columnwidth,keepaspectratio,clip]{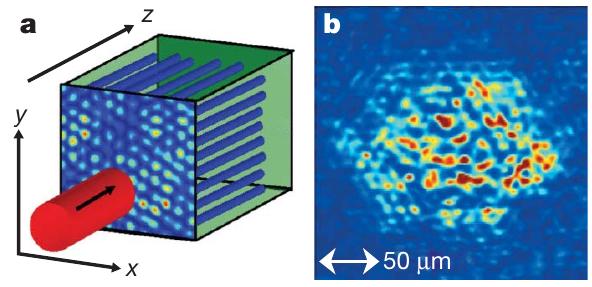}
\includegraphics[width=0.25\columnwidth,keepaspectratio,clip]{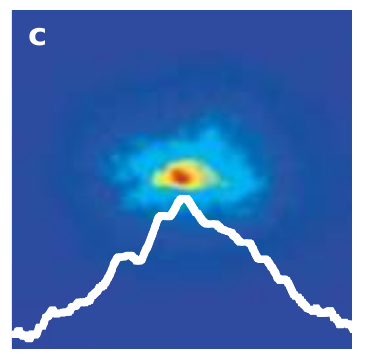}
\caption{AL of light in 2D disordered photonic lattices \cite{Schwartz07}. Panel (a) shows the scheme of experimental setup: a light beam enters the optically induced photonic lattice. Disorder is introduced by adding a speckle beam. 
In the absence of disorder the lattice is perfectly periodic in $x$- and  $y$-directions. The intensity distribution of light is measured at the lattice output. Panel (b) shows the output diffraction pattern of light in the perfectly periodic lattice. Panel (c) displays the average output intensity distribution of light in the presence of disorder (100 disorder realizations). The logarithm of the average intensity profile (white line in (c)) manifests linearly decaying tails, the hallmark of  AL. Reprinted from \cite{Schwartz07}.} 
\label{fig:phot2d}
\end{center}
\end{figure}
Localization of microwaves in a 2D random medium composed of dielectric cylinders and placed between two parallel aluminum plates was reported in \cite{Dalich91}. Sharp peaks observed in the transmission spectrum and the energy density of the microwave probe were attributed to the existence of localized modes. The same concept of wave localization by scattering can be applied to acoustic waves. Indeed, localization of ultrasound in 3D random elastic network of aluminum beads has been reported \cite{Hu08}. 

\begin{figure}[ht]
\begin{center}
\includegraphics[width=0.5\columnwidth,keepaspectratio,clip]{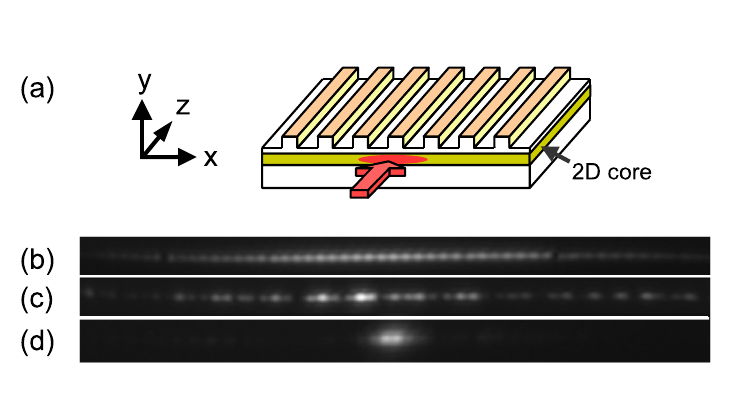}
\includegraphics[width=0.35\columnwidth,keepaspectratio,clip]{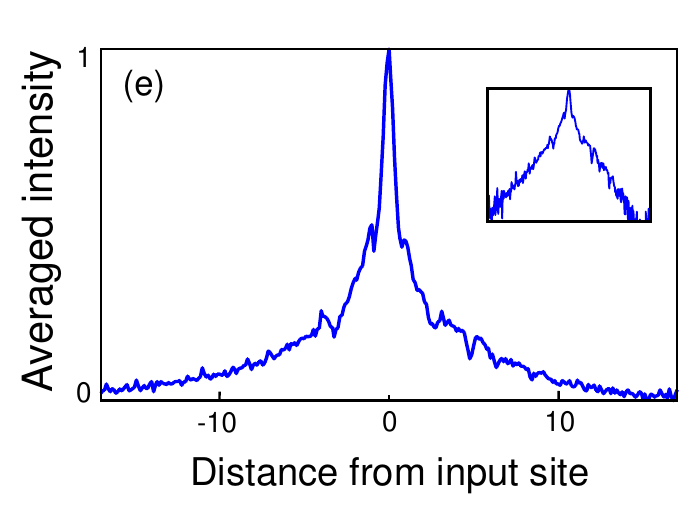}
\caption{AL of light in a 1D disordered waveguide lattice, obtained in \cite{Lahini08}. Panel (a) shows a sketch of the experiment and its geometry. The (red) arrow indicates the input beam, that generally covers a few lattice sites. Panels (b)-(d) display the output light distribution in the case of a periodic lattice (b), and in the case of a disordered lattice (c, d). In (e) the average output light distribution is shown. Initially, light was injected into a single lattice site and, at strong enough disorder, the outcome light distribution reveals localization. The inset in (e) shows the light distribution on a semi-log scale, clearly indicating the exponential tails. Adopted from \cite{Lahini08}.} 
\label{fig:phot1d}
\end{center}
\end{figure}

Direct observation of AL of light was performed in photonic lattices. The light intensity distribution can be  measured and visualized at the lattice output. Recent experiments report on the transverse localization of light due to presence of disorder potentials in 2D photonic lattices \cite{Schwartz07}. Disorder was introduced by adding a speckle beam so that the properties of the random potential could be varied easily and controlled quite precisely. The authors demonstrated that ballistic transport turns diffusive in the presence of weak disorder and a crossover to AL occurs at a higher level of disorder (\fref{fig:phot2d}). The transition from ballistic wave packet expansion to exponential localization has also been observed in 1D disordered lattices of coupled optical waveguides \cite{Lahini08}, where disorder was introduced through a randomly varying waveguide width (see \fref{fig:phot1d}). 

\subsection{Quasi-periodic lattices and Aubry-Andre localization}
\label{AAM}
\begin{figure}[h]
\begin{center}
\includegraphics[width=0.9\columnwidth,keepaspectratio,clip]{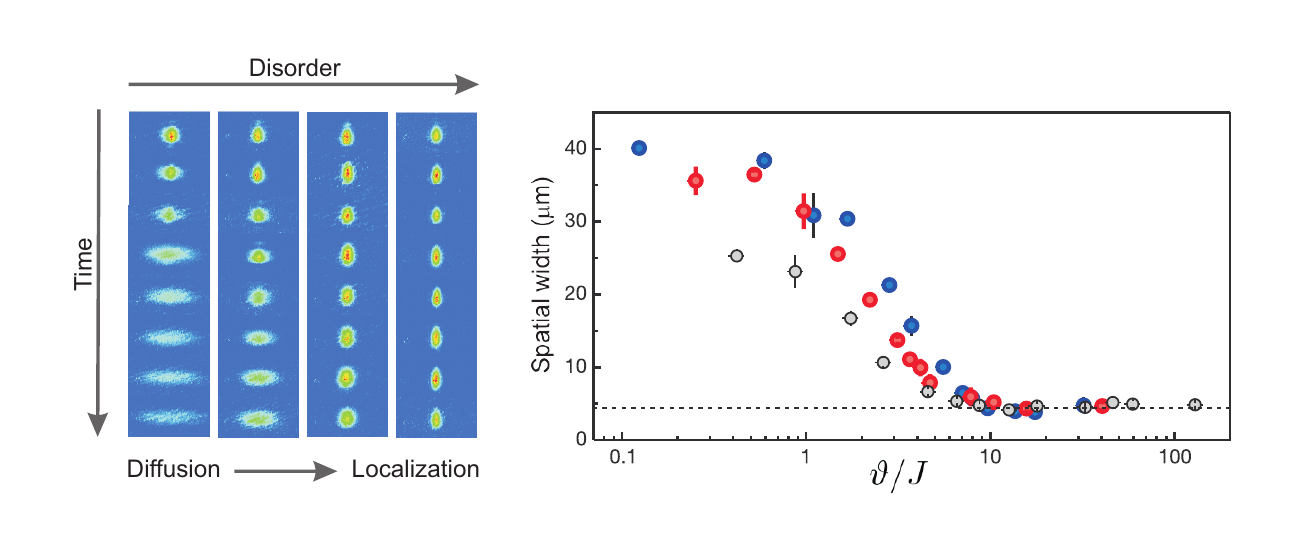}
\caption{The metal-insulator transition of a non-interacting BEC loaded into a quasi-periodic lattice. \textit{Left panel:} absorption images of the BEC for different values of modulation strength $\vartheta$  and a constant tunneling rate $J$. In a regular lattice (the leftmost part), the condensate expands ballistically. The speed of condensate expansion reduces with increasing modulation. In the limit of strong modulation, the condensate width remains constant, manifesting the onset of localization. \textit{Right panel:} The condensate width is measured at a fixed evolution time of 750 ms. The dashed line guides the eye to the initial condensate size. Reprinted from \cite{Roati08}.} 
\label{fig:quasi}
\end{center}
\end{figure}

The translational symmetry of a perfectly periodic system is also destroyed
by a secondary periodic potential with incommensurate frequency. Quasiperiodic potentials
gained much attention due to quasicrystals \cite{Levine84,Vekilov10} and different generation
schemes (Fibonacci, Thue-Morse, Cantor, \textit{etc.}) \cite{Albuq03}. 
The primary periodic lattice can be easily created via interference patterns of two counter-propagating beams and quasi-periodicity is introduced by superimposing a weak secondary lattice with incommensurate wavelength \cite{Guid97}. This bichromatic lattice configuration has recently been used in ultra-cold atomic physics for direct observation and exploration of Aubry-Andre localization \cite{Roati08,Edwards08} (see \fref{fig:quasi}).

The corresponding model 
is obtained by a slight change of the Anderson
tight-binding model (\ref{eq:1p3}):
\begin{equation}
\hat{\mathcal{H}}_{\rm AA}= \sum_{l} \vartheta \cos(2 \pi \alpha l) \left|l\right\rangle \left\langle l\right| + \sum_{l, m} J_{lm} \left|l\right\rangle \left\langle m \right|,
\label{eq:1p2p0}
\end{equation}	   
where the hopping rate $J_{lm}=\delta_{l,m \pm 1}$. The parameter $\vartheta$ controls the potential modulation depth, similar to the disorder strength $W$ for the Anderson model. The  incommensurability ratio $\alpha$ is 
usually chosen as the inverse golden mean $\alpha=\left(\sqrt{5}-1\right)/2$.
Such quasi-periodic lattices constitute a paradigmatic class of systems mediating between the ordered and disordered cases. While all eigenstates are localized in 1D disordered systems and extended in periodic ones, quasi-periodic potentials display a metal-insulator transition from extended to localized states at a critical value of modulation depth \cite{Aubry80}.

The quasi-periodic version of the 1D discrete Schr\"odinger equation reads
\begin{equation}
i \dot{\varphi}_l = \vartheta \cos(2 \pi \alpha l) \varphi_l+\varphi_{l+1}+\varphi_{l-1}
\label{eq:1p2p1}
\end{equation}	   
also known as the Aubry-Andr\'e model \cite{Aubry80} (for the critical case $\vartheta = 2$ 
it is identical to the Harper equation \cite{Harper55}). 
The ansatz $\varphi_l = A_l \e^{-i E t}$ yields the eigenvalue problem
\begin{equation}
E A_l = \vartheta \cos(2 \pi \alpha l) A_l + A_{l+1} + A_{l-1}.
\label{eq:1p2p2}
\end{equation}	   

\begin{figure}
\begin{center}
\includegraphics[width=0.8\columnwidth,keepaspectratio,clip]{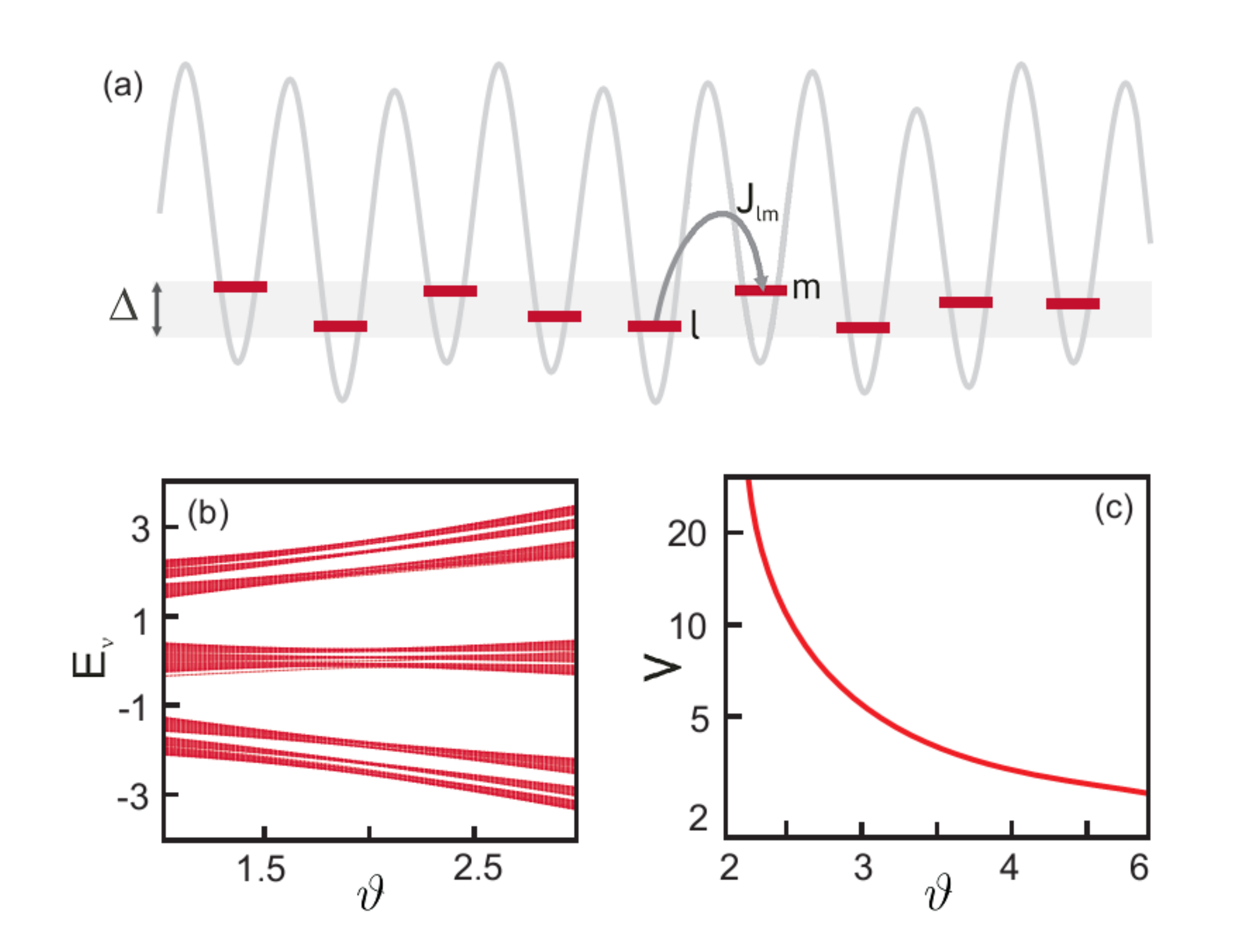}
\caption{(a) Schematic representation of the 1D Aubry-Andr\'e model. (b)
Eigenvalues $E_\nu$ of the system (\ref{eq:1p2p1}) obtained from numerical diagonalization of \eref{eq:1p2p2} as a function of the modulation strength $\vartheta$. (c) Average localization volume $V$ of eigenstates as a function of $\vartheta$, numerically calculated in \cite{Larcher2012}.} 
\label{fig:1p2p2}
\end{center}
\end{figure}

At variance to the Anderson model, here the spectrum of the 
eigenvalues $E_\nu$ has a complex fractal structure. For a given $\alpha$ it consists of a hierarchy 
of bands (\fref{fig:1p2p2}(b)), related to the Hofstadter butterfly \cite{Hofst76}.

The transform $\varphi_l=\sum_{k}\e^{2 \pi \alpha i k l}\phi_k$ to the quasi-momentum basis $\left\{\phi_k\right\}$ leads to 
\begin{equation}
i \dot{\phi}_k = 2 \cos(2 \pi \alpha k) \phi_k + \frac{\vartheta}{2}(\phi_{k+1}+\phi_{k-1}).
\label{eq:1p2p3}
\end{equation}	   

Equation (\ref{eq:1p2p1})
describes  dynamics in real space and $\vartheta$ controls the on-site energy terms, while \eref{eq:1p2p3}
describes the dynamics in the momentum representation, and $\vartheta$ controls the kinetic coupling of the momentum modes. Even though the two equations are
strictly different, through a swapping of prefactors they appear to be equivalent. This property known as self-duality, was first discovered by Aubry and
Andr\'e \cite{Aubry80}. 

Both equations become identical at the self-dual point $\vartheta=2$, where the system states undergoes a transition from extended to localized. For $\vartheta<2$ all states are extended in real space (and localized in momentum space). At the critical point $\vartheta=2$, the two representations have identical spectra and eigenstates. For $\vartheta>2$ the eigenstates
become exponentially localized in real space with localization length $\xi=\left[\ln(\vartheta/2)\right]^{-1}$, while they become delocalized in momentum space \cite{Aubry80,Sokol81}.

The spectrum of the Aubry-Andr\'e model at $\vartheta=2$ reveals criticality and fractal properties of eigenstates that has remained a subject of intensive studies. Approaches include analyzing the density of states \cite{Soukoul82}, density-density correlations \cite{Boers07}, Husimi distributions \cite{Aulbach04}, and spreading of density moments \cite{Hu00,Diener01}. 
The Aubry-Andre model shows an abrupt metal insulator transition simultaneously for all eigenstates.
Mobility edges in the spectrum can be obtained by e.g. the replacement $\cos\left(2 \pi \alpha l\right) \mapsto \cos\left(2 \pi \alpha l^\varsigma \right)$. Then one obtains a mobility edge already in dimension one if $0<\varsigma<1$ and $\vartheta<2$ \cite{Cris02,Varga92}. Another generalization to the Aubry Andre model, that is varying the radius of interactions, leads to a shift of the transition point and again to
the appearance of a mobility edge \cite{Biddle09}.


\section{Nonlinear waves}
\label{NRM}
Interactions (e.g. between atoms) and nonlinear response (e.g. of a medium to propagating light fields) lead to nonlinear terms in wave equations. Nonlinearity couples the localized eigenmodes of the linear wave equation, altering their characteristics. The problem of localization in nonlinear systems can be considered in different contexts: (i) the evolution of wave packets in the zero temperature limit or (ii) finite temperature conductivity. The major part of the section is devoted to the first problem.
A brief discussion of conductivity issues is also provided. We employ the basic models, widely utilized to study the joint impact of nonlinearity and spatial heterogeneity. Guided by the previous results, discrete Nonlinear Schr\"odinger and Klein-Gordon equations with cubic nonlinearity are chosen as specific examples. 

\subsection{Discrete nonlinear Schr\"odinger equations} 
A variety of nonlinear wave processes are described by the discrete nonlinear Schr\"odinger (DNLS) equation
\begin{equation}
i \dot{\psi}_l = \epsilon_l \psi_l - J(\psi_{l+1} + \psi_{l-1}) \pm \beta \left|\psi_l\right|^2\psi_l,
\label{eq:genericDNLS}
\end{equation}
which is used also to describe nonlinear localization phenomena in arrays of identical anharmonic oscillators $\epsilon_l=0$ \cite{Eilback1985}. For a review on the major results and historical aspects, we also refer the reader to \cite{Kevrekidis2001,Eilbeck2003}. Beside the total energy, DNLS arrays conserve the total norm $S=\sum_l \left|\psi_l\right|^2$. Note, that depending on the experimental context that DNLS would describe, the physical meaning of the norm density could be e.g. the density of an atomic condensate or the intensity of light. It is also easy to see that varying the nonlinearity strength $\beta$ is strictly equivalent to varying the norm. Due to this fixing $S=1$ and having $\beta$ as a control parameter is a convenient and frequent choice for finite norm studies.

\subsubsection{Nonlinear disordered lattices} 
Introducing purely random $\epsilon_l$, we obtain the disordered DNLS (dDNLS), which in one dimension (1D) reads
\begin{equation}
i \dot{\psi}_l = \epsilon_l \psi_l - \psi_{l+1} - \psi_{l-1} + \beta \left|\psi_l \right|^2\psi_l,
\label{eq:1DdDNLS}
\end{equation}
where $\psi_l$ is the complex valued function subjected to the $l$-th site of a lattice and the strength of nearest-neighbor interactions $J=1$. Similar to equation (\ref{eq:1p4}) random on-site energies $\epsilon_l$ are drawn from an uncorrelated uniform distribution on $\left[-W/2, W/2\right]$ parametrized by the disorder strength $W$. We restrict ourselves to considering only positive nonlinearity strength $\beta \geq 0$, which corresponds to defocusing nonlinearity or repulsive interactions. A sign-alternating staggering and complex conjugation transformation of the wave amplitudes 
$(-1)^{l} \psi^*_l$ together with a change of sign of $\beta$ leaves the equations (\ref{eq:1DdDNLS}) invariant.

 The 2D generalization is straightforward:
\begin{equation}
i \dot{\psi}_{l,m} = \epsilon_{l,m} \psi_{l,m} - (\psi_{l+1,m} + \psi_{l-1,m}+\psi_{l,m+1} + \psi_{l,m-1}) + \beta \left|\psi_{l,m}\right|^2\psi_{l,m}.
\label{eq:2DdDNLS}
\end{equation}

dDNLS equations are Hamiltonian systems with
\begin{eqnarray}
1D:\;  \cH_{\rm dD} = & \sum_l (\epsilon_l \left| \psi_l \right|^2 - \left(\psi_{l+1}\psi_l^\ast + {\rm c.c.}\right) + \frac{\beta}{2}\left|\psi_l\right|^{4}),
\label{eq:1DHdDNLS}
\\
2D:\;  \cH_{\rm dD} = & \sum_{l,m} (\epsilon_{l,m}\left|\psi_{l,m} \right|^2 - \left[\psi_{l,m}^\ast\left(\psi_{l+1,m}+\psi_{l,m+1}\right) + {\rm c.c.}\right]  
\nonumber
\\ 
&
+
\frac{\beta}{2}\left|\psi_{l,m}\right|^{4}).
\label{eq:2DHdDNLS}
\end{eqnarray}	   

The two conserved quantities (energy and norm) ensure integrability of dDNLS equations only in the case of two coupled sites, the well-studied dimer \cite{Eilbeck2003,Kenkre1986}. An integrable arbitrary size DNLS-type lattice was proposed by Ablowitz and Ladik \cite{Ablowitz1976,Ablowitz1991}, though its physical relevance is less evident.

The above dDNLS equations (\ref{eq:1DdDNLS}), (\ref{eq:2DdDNLS}) embody only the effect of cubic (third-order) nonlinearity, although at certain conditions higher-order terms may become relevant. In certain optical materials, semiconductors, doped glasses \cite{Christian2007},  and at the BEC-BCS crossover in ultra-cold Fermi gases \cite{Yan2011} one may also parametrize the index of nonlinearity. Thus, studying generalizations to arbitrary nonlinearity index is challenging. 

The Hamiltonians of such generalized DNLS (gDNLS) equations are
\begin{eqnarray}
\rm{1D:} \quad \cH_{\rm gD} = & \sum_l \left[\epsilon_l \left| \psi_l \right|^2 - \left(\psi_{l+1}\psi_l^\ast + {\rm c.c.}\right) 
+
\frac{2\beta}{\sigma+2}\left|\psi_l\right|^{\sigma+2}\right],
\label{eq:1DHgDNLS}
\\
\mbox{2D:} \quad \cH_{\rm gD} = & \sum_{l,m} \left[\epsilon_{l,m}\left|\psi_{l,m} \right|^2 - (\psi_{l,m}^\ast\left(\psi_{l+1,m}+\psi_{l,m+1}\right) + {\rm c.c.}\right] 
\nonumber
\\
&
 + \frac{2\beta}{\sigma+2}\left|\psi_{l,m}\right|^{\sigma+2} ),
\label{eq:2DHgDNLS}
\end{eqnarray}   
where $\sigma>0$. The respective equations of motion read 
\begin{eqnarray}
\mbox{1D:} \quad i \dot{\psi}_l = & \epsilon_l \psi_l - \psi_{l+1} - \psi_{l-1} + \beta \left|\psi_l \right|^\sigma \psi_l,
\label{eq:1DgDNLS}
\\
\mbox{2D:} \quad i \dot{\psi}_{l,m} = & \epsilon_{l,m} \psi_{l,m} - (\psi_{l+1,m} + \psi_{l-1,m}+\psi_{l,m+1} + \psi_{l,m-1}) 
\nonumber
\\
&
+ \beta \left|\psi_{l,m}\right|^\sigma \psi_{l,m}.
\label{eq:2DgDNLS}
\end{eqnarray}

The linear dDNLS and gDNLS equations ($\beta=0$) reduce to the Anderson model, and thus display exponential localization of all eigenstates (cf. section \ref{LRM}).

\subsubsection{Nonlinear quasi-periodic chains}
The second type of localizing inhomogeneity is the quasi-periodic potential $\epsilon_l=\vartheta \cos(2\pi\alpha l)$. Then the 1D DNLS equation (qDNLS) takes the form
\begin{equation}
i \dot{\psi}_l = \vartheta \cos(2\pi\alpha l) \psi_l - \left(\psi_{l+1} + \psi_{l-1}\right) + \beta|\psi_l|^2\psi_l. 
\label{eq:qDNLS}
\end{equation}
gDNLS equations approximate dynamics of interacting BECs in optic traps \cite{Lucioni2011,Deissler2010}, or propagation of high-intensity light pulses in Kerr photonics \cite{Lahini2009}, where quasi-periodic potentials are easily reproduced by bichromatic optical lattices. 

The equations of motion are associated to the Hamiltonian
\begin{equation}
\cH_{\rm qD} = \sum_l \left[\vartheta \cos(2\pi\alpha l)|\psi_l|^2  - \left(\psi_{l+1}\psi_j^\ast+ {\rm c.c.}\right)+\frac{\beta}{2}|\psi_l|^4 \right].
\label{eq:HqDNLS}
\end{equation}
Its key parameters are the strengths of nonlinearity $\beta$ and quasi-periodic potential $\vartheta$, and the irrational number $\alpha$, responsible for quasi-periodicity. As before, we use the inverse of the golden mean $\alpha=(\sqrt{5}-1)/2$ as a convenient choice (cf. \sref{AAM}).

In the linear case ($\beta=0$), equations (\ref{eq:qDNLS}), (\ref{eq:HqDNLS}) coincide with Aubry-Andr\'e model that was discussed in \sref{AAM}. Therefore, the eigenstates of the linear qDNLS model undergo a transition from extended to exponentially localized shape at $\lambda=2$. The qDNLS equation, thus, offers a convenient framework to study the impact of nonlinearity on a localization-delocalization transition.

\subsection{Klein-Gordon lattices} 

Alternative oscillatory lattice models -- arrays of coupled nonlinear oscillators -- have also been studied intensively in variety of physical applications, including material science, and biophysics (see, for instance, the reviews \cite{Ford1992,Braun2004,Peyrard2004}). The number of models and their applications are enormous: Fermi-Pasta-Ulam (FPU) chains \cite{Fermi1955}, Frenkel-Kontorova model \cite{Kontorova1983}, Peyrard-Bishop model \cite{Dauxois1993} are just a few examples. 

In the following we focus on the Klein-Gordon (KG) class of such lattices. It has a fourth-order anharmonicity in the on-site potential and harmonic nearest-neighbor interactions. The KG chain is a suitable model of atomic arrays subject to external fields, e.g. anharmonic lattice vibrations in crystals \cite{Ovchinnikov2001}. 

In analogy to the DNLS equations, we introduce the disorder in the linear on-site potential and the model (dKG) reads
\begin{equation}
\ddot{u}_l=-{\tilde \epsilon}_l u_l+\frac{1}{W}\left(u_{l+1}+u_{l-1}-2u_l\right)-\beta_{\rm K} u_l^3
\label{eq:1DdKG}
\end{equation}
with the corresponding Hamiltonian
\begin{equation}
\cH_{\rm dK}=\sum_l \left[\frac{p_l^2}{2}+\frac{{\tilde \epsilon}_l u_l^2}{2}+\frac{1}{2W}\left(u_{l+1}-u_l\right)^2+\frac{\beta_{\rm K}}{4}u_l^4\right].
\label{eq:H1DdKG}
\end{equation}
We remind that $u_l$ and $p_l$ are the generalized coordinates and momenta of the $l$-th oscillator in the chain. The coefficients ${\tilde \epsilon}_l$ take uncorrelated random values drawn uniformly from the interval $[1/2,3/2]$. The set of dynamical equations (\ref{eq:1DdKG}) conserves only the total energy $\mathcal{H}_{\rm dK} = \sum_l \cE_l$, where 
\begin{equation}
\cE_l=\frac{p_l^2}{2}+\frac{{\tilde \epsilon}_l u_l^2}{2}+\frac{\beta_{\rm K}}{4}u_l^4++\frac{1}{4W}\left(u_{l+1}-u_l\right)^2
+ +\frac{1}{4W}\left(u_{l-1}-u_l\right)^2  
\label{energypersite}
\end{equation} 
is the energy associated with the $l$-th lattice site. Note, that rescaling $\beta_{\rm K}$ is equivalent to rescaling the total energy. In the following we set $\beta_{\rm K}=1$ and use the total energy as a control parameter.

It is also straightforward to extend the KG model to two dimensions
\begin{eqnarray}
\ddot{u}_{l,m}= & -{\tilde\epsilon}_{l,m}u_{l,m}+\frac{1}{W}\left(u_{l+1,m}+u_{l-1,m}+u_{l,m+1}+u_{l,m-1}-4u_l\right)
\nonumber
\\
&
-u_{l,m}^3
\label{eq:2DdKG}
\end{eqnarray}
with the corresponding Hamiltonian
\begin{eqnarray}
\cH_{\rm dK}= & \sum_{l,m}  ( \frac{p_{l,m}^2}{2}+\frac{{\tilde \epsilon}_{l,m} u_{l,m}^2}{2}+ \frac{1}{2W}\left[(u_{l+1,m}-u_{l,m})^2+(u_{l,m+1}-u_{l,m})^2\right]
\nonumber
\\
&
+\frac{u_{l,m}^4}{4} ) .
\label{eq:H2DdKG}
\end{eqnarray}

Similarly to DNLS, the linear parts of equations (\ref{eq:1DdKG}) and (\ref{eq:2DdKG}) are reducible to the Anderson model. Therefore, all linear eigenstates of KG model are exponentially localized. Exact mapping between DNLS and KG in their linear limits is summarized in table \ref{tbl:linmap}. 

\begin{table}[ht]
\caption{Characteristic quantities of the DLS eigenvalue problem and their mapping to the KG case.}
\begin{center}
\begin{tabular}{|c|c|c|c|}
\hline
\rowcolor{Gray} ${\rm D}$ & DLS & mapping & KG \bigstrut \\
\hline 
${\rm D} = 1$ & $\epsilon_l \in \left[-\frac{W}{2}, \frac{W}{2}\right]$ & $1+\epsilon_l/W \mapsto {\tilde \epsilon}_l$ &
${\tilde \epsilon}_l\in\left[\frac{1}{2}, \frac{3}{2}\right]$ \bigstrut \\
\hline 
${\rm D} = 2$ & $\epsilon_{l,m} \in \left[-\frac{W}{2}, \frac{W}{2}\right]$ & $1+ \epsilon_{l,m}/W \mapsto {\tilde \epsilon}_{l,m}$ &
${\tilde \epsilon}_{l,m}\in\left[\frac{1}{2}, \frac{3}{2}\right]$ \bigstrut \\
\hline 
\multirow{2}{*}{${\rm D} = 1$} & $E_\nu$ & 
$(E_\nu + 2)/W + 1 \mapsto \omega_\nu^2$ & $\omega_\nu^2$ \bigstrut \\ \cline{2-4}
& $\Delta = W + 4$	& $\Delta \mapsto \Delta/W$ & $\Delta = 1 + 4/W$ \bigstrut \\ 
\hline
\multirow{2}{*}{${\rm D} = 2$} & $E_\nu$ & 
$(E_\nu + 4)/W + 1 \mapsto \omega_\nu^2$ & $\omega_\nu^2$ \bigstrut \\ \cline{2-4}
& $\Delta = W + 8$	& $\Delta \mapsto \Delta/W$ & $\Delta = 1 + 8/W$ \bigstrut \\ 
\hline
\end{tabular}
\end{center}
\label{tbl:linmap}
\end{table}

Several publications suggest a strong similarity between their nonlinear versions as well \cite{Kivshar1992, Johansson2006, Johansson2004}.  In the small energy limit, applying slow modulation/rotating wave approximations, one recovers dDNLS equations (\ref{eq:1DdDNLS}), (\ref{eq:2DdDNLS}) from KG versions under an approximate mapping $\beta S \approx 3W\mathcal{H}_{\rm dK}$. It connects the dKG initial parameters $\mathcal{H}_{\rm dK}$ and $W$ to the total initial norm $S$ and nonlinear parameter $\beta$ of the corresponding dDNLS model.

Generalizations to different powers of nonlinearity yield the generalized Klein-Gordon lattice (gKG)
\begin{eqnarray}
\mbox{1D:} \quad \ddot{u}_l= & -{\tilde \epsilon}_l u_l+\frac{1}{W}\left(u_{l+1}+u_{l-1,m}-2u_l\right)-\left|u_l\right|^\sigma u_l,
\label{eq:1DgHKG}
\\
\mbox{2D:} \quad \ddot{u}_{l,m}= & -{\tilde\epsilon}_{l,m}u_{l,m}+\frac{1}{W}\left(u_{l+1,m}+u_{l-1,m}+u_{l,m+1}+u_{l,m-1}-4u_l\right)
\nonumber
\\
&
-\left|u_{l,m}\right|^\sigma u_{l,m}.
\label{eq:2DgHKG}
\end{eqnarray}
with corresponding Hamiltonians
\begin{eqnarray}
\mbox{1D:} \quad \cH_{\rm gK}= & \sum_l \left[ \frac{p_l^2}{2}+\frac{\tilde{\epsilon}_l u_l^2}{2}+\frac{1}{2W}\left(u_{l+1}-u_l\right)^2+\frac{\left|u_l\right|^{\sigma+2}}{\sigma+2}\right],
\label{eq:H1DgHKG}
\\
\mbox{2D:} \quad \cH_{\rm gK}= &\sum_{l,m} ( \frac{p_{l,m}^2}{2}+\frac{{\tilde \epsilon}_{l,m} u_{l,m}^2}{2}+ \frac{1}{2W}\left[(u_{l+1,m}-u_{l,m})^2+(u_{l,m+1}-u_{l,m})^2\right]
\nonumber
\\
&
+\frac{\left|u_{l,m}\right|^{\sigma+2}}{\sigma+2} ) .
\label{eq:H2DgHKG}
\end{eqnarray}

The approximate mapping from the KG to DNLS model can be generalized to arbitrary order of nonlinearity $\sigma$,
and reads in the 2D case
\begin{equation}
\beta \sum_{l,m} \left|\psi_{l,m}\right|^\sigma \approx a_\sigma W \sum_{l,m} \cE_{l,m}^{\sigma/2}, \quad a_\sigma \equiv \frac{8(\sigma+1)\Gamma(\sigma)}{\sigma(\sigma+2)\Gamma^2(\sigma/2)}.
\label{eq:Gmap}
\end{equation}

Finally we introduce a quasi-periodic version of a KG (qKG) chain governed by the Hamiltonian 
\begin{equation}
\cH_{\rm qK}=\sum_l \left[\frac{p_l^2}{2}+\left(2+\cos(2\pi\alpha l)\right)\frac{u_l^2}{4}+\frac{1}{2\vartheta}\left(u_{l+1}-u_l\right)^2+\frac{u_l^{4}}{4}\right],
\label{eq:HqHKG}
\end{equation}
Parameters $\alpha$ and $\vartheta$ has the same meaning as for qDNLS \eref{eq:qDNLS}. The equations of motion read
\begin{equation}
\ddot{u_l}=-\left(2+\cos(2\pi\alpha l)\right)\frac{u_l}{2}+\frac{1}{2\vartheta}(u_{l+1}+u_{l-1}-2 u_l)- u_l^3 \, .
\label{eq:qKG}
\end{equation}

Similar to qDNLS, the linear counterpart of \eref{eq:qKG} can be reduced to the Aubry-Andr\'e eigenvalue
problem (\sref{AAM}). 

The small amplitude mapping from nonlinear KG to DNLS models
suggests that both models will behave similarly, and it is enough to study one of them, whichever is more suitable for practical or technical reasons. That view might fail, since the KG model class conserves only one integral of motion - the energy, while the DNLS class conserves in addition also the norm. That has consequences as e..g the appearance of a strict selftrapping regime and a non-Gibbs statistical state in the DNLS case, which is related to the well-known phenomenon of blowup and finite time singularities in corresponding space-continuous nonlinear Schr\"odinger equations. 
However, the results for spreading wave packets discussed below has turned out to be surprisingly independent of the chosen model class.
Therefore we will discuss results using the framework of the DNLS models, yet represent numerical results for both DNLS and KG systems. In some cases only KG results have been published. This is due to the fact that  long time numerical simulations are needed, and computational restrictions arise (e.g. in two dimensions).
The study of the KG model remains appealing for at least two reasons. First, it allows for testing the generality of phenomena in the lattice with just one conserved quantity. Secondly, the KG model is advantageous from the numerical point of view, allowing for up to two orders of magnitude faster integration speed within the same integration error (see, e.g. \cite{Skokos2009}). Lastly, existing simulations of both models \cite{Flach2009,Skokos2009,Laptyeva2010,Bodyfelt2011} show similar qualitative results in a wide range of energy and disorder.

\subsection{Key measurables}
\label{KeyM}
In \sref{LRM} the participation number $P$ (measures the number of effectively excited sites) and the second moment $m_2$ (quantifies the squared width of the packet) were used to quantify the localization of eigenstates in heterogeneous systems. Clearly, they can also be utilized to characterize the spatio-temporal evolution of wave packets. 

Consider first 1D arrays. In the case of DNLS equations, we follow the time-dependence of normalized norm density distribution $z_l\equiv\left|\psi_l\right|^2 /\sum_{l^\prime}\left|\psi_{l^\prime}\right|^2$, for the KG -- of the normalized energy density distribution $z_l \equiv \cE_l/\sum_{l^\prime}\cE_{l^\prime}$. We compute the participation number $P=1/\sum_l z_l^2$ and the second moment $m_2=\sum_l (l-\bar{l})^2 z_l$, where $\bar{l}=\sum_l l z_l$ is the distribution center. Additionally, the wave packet sparseness can be quantified by the compactness index $\zeta=P^2/m_2$ \cite{Skokos2009}. Flat compact distributions have $\zeta=12$, `` well-thermalized'' distributions have $\zeta \approx 3$, while $\zeta \ll 3$ indicates very inhomogeneous sparse wave packets (see \cite{Skokos2009} for more details).

Similarly, for 2D DNLS the normalized norm density distribution is $z_{l,m}\equiv \left|\psi_{l,m}\right|^2/ \sum_{l^\prime,m^\prime}\left|\psi_{l^\prime,m^\prime}\right|^2$ and $z_{l,m}\equiv\cE_{l,m}/ \sum_{l^\prime,m^\prime} \cE_{l^\prime,m^\prime}$ is the normalized energy density distribution for 2D KG lattice. The measurables are calculated as  $P=1/\sum_{l,m} z_{l,m}^2$, $m_2=\sum_{l,m}\left[(l-\bar{l})^2+(m-\bar{m})^2\right]z_{l,m}$ (here the density center lattice coordinates are $\bar{l}=\sum_{l,m} l z_{l,m}$, $\bar{m}=\sum_{l,m} m z_{l,m}$), and $\zeta=P/m_2$.

In the same way, we can calculate $P$, $m_2$ and $\zeta$ in the NM space. In addition, for some numerical simulations we measure the fraction of wave packet norm $S_{V}$ for DNLS and the fraction of wave packet energy $\mathcal{H}_{V}$ for KG within the localization volume $V$ around the initially excited state in real space. For a localized state this fraction asymptotically tends to a constant nonzero value, while in case of spreading wave packets it goes to zero.

\subsection{First numerical experiments}
\label{FNE}
Already the first numerical experiments on the evolution of an initial single site excitation in nonlinear disordered chains (\ref{eq:1DdDNLS}), (\ref{eq:1DdKG}) showed the destructive effect of nonlinearity on Anderson localization \cite{Molina1998,Pikovsky2008,GarciaMata2009,Skokos2009, Flach2009}. The above studies reported spreading of a wave packet beyond the limits set by the linear theory 
The common fundamental observation was subdiffusive wave packet  spreading of initial single site excitations
with the second moment diverging as $m_2\propto t^{\alpha_m}$ with $\alpha_m <1$. 

The first prediction for $\alpha_m$ was made in \cite{Pikovsky2008}, where $\alpha_m=2/5$ was suggested, relying on the analogy with the kicked rotator model \cite{Shepelyansky1993a}. 
The basic assumption was dynamical chaos, and  fast decoherence of phases of all NMs participating in the wave packet dynamics. This regime, which {\sl is accessible} using spreading wave packets, was however {\sl not} reachable using the initial conditions in \cite{Pikovsky2008} (see below). Moreover, if reached, this regime will yield $\alpha_m=1/2$. Therefore the result $2/5$ is first of all  a consequence of incorrect treating the underlying diffusion equations, which are obtained by the above assumption. Second, the initial assumption is not sufficient, since when correctly treated, it yields $1/2$ and is  clearly much larger than the numerical observations, namely  $\alpha \approx 0.31 ... 0.34$ \cite{Pikovsky2008} and $\alpha \approx 0.33 $ \cite{GarciaMata2009}. Note that Ref. \cite{Molina1998} suggested even smaller $\alpha_m<0.3$, but that is clearly
due to very short integration times $t=10^4$. 

Ref. \cite{Flach2009} initially also assumed dynamical chaos and phase dceoherence of NMs. The diffusion equation approach
resulted in $\alpha_m=1/2$ and was numerically confirmed by enforcing NM dephasing. The same theoretical approach
was phenomenologically modified by using notions of resonance probabilities. The main conclusion was that only a part of the
NMs, which are participating in the wave packet dynamics, are also resonant, chaotic and quickly dephaising. Consequently, the authors of
\cite{Flach2009, Skokos2009} obtained the exponent $\alpha = 1/3$. Accurate numerical long time
simulations have confirmed that exponent impressively for both DNLS and KG models \cite{Flach2009, Skokos2009}.

\begin{figure}[ht]
\begin{center}
\includegraphics[width=0.65\columnwidth,keepaspectratio,clip]{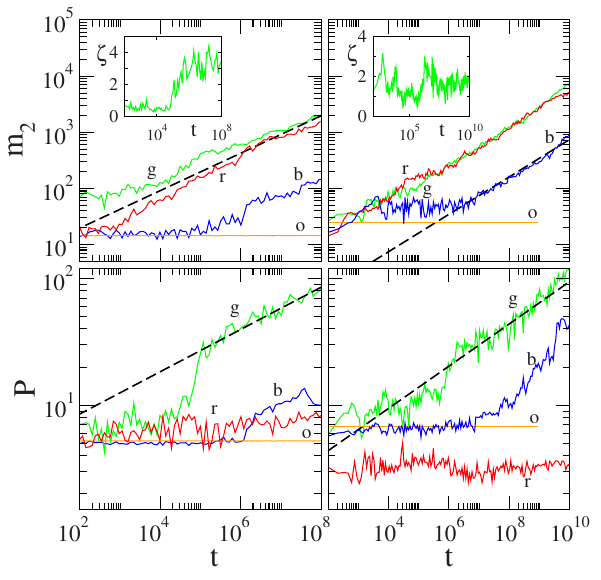}
\caption{Spatio-temporal evolution of single-site excitations in disordered nonlinear chains: the measures of $m_2$ and $P$ are shown versus $t$ in log-log plots. \textit{Left panel}: dDNLS (\ref{eq:1DdDNLS}) with parameters $W=4$ and $\beta=0, 0.1, 1.0, 4.5$  ((o) orange, (b) blue, (g) green, (r) red). \textit{Right plots}: dKG (\ref{eq:1DdKG}) with $W=4$ and initial energies $E=0.05, 0.4, 1.5$ ((b) blue, (g) green, (r) red). The orange (o) curves correspond to the solution of the linear equations ($\beta_{\rm K}=0$). Dashed straight lines guide the eye for exponents $1/3$ for $m_2(t)$ and  $1/6$ for $P(t)$, respectively. \textit{Insets}: $\zeta(t)$ shown in linear-log plots for $\beta=1$ (dDNLS) and $E=0.4$ (dKG). Adapted from \cite{Skokos2009}.} 
\label{fig:SKKF_PRE}  
\end{center}
\end{figure}

Further tests revealed a regime of strong nonlinearity and self trapping. For the DNLS model with $\beta=1.0$ and KG model with $E=1.5$ the participation number $P$ does not grow significantly, while the second moment diverges again as $m_2 \propto t^{\alpha}$ with $\alpha=1/3$ (\fref{fig:SKKF_PRE}, (r) curves). 
For dDNLS arrays the energy and norm conservation arguments allow for a rigorous proof that at least part of initial wave packet with high enough energies must remain self trapped \cite{Kopidakis2008,Skokos2009}. In dKG arrays, where only energy is conserved, the theorem is not applicable though numerics does display self-trapping, at least on the computationally attainable time scales. 

For values of $\beta=1.0$ (dDNLS) and $E=0.4$ (dKG) self-trapping is avoided and subdiffusive spreading is observed with $m_2\propto t^{1/3}$ and $P\propto t^{1/6}$ (\fref{fig:SKKF_PRE}, (g) curves). For smaller values of $\beta$ and $E$ one finds no visible spreading up to some time $\tau_0$, which increases with decreasing nonlinearity. Nevertheless for $t>\tau_0$, both $m_2(t)$ and $P(t)$ behaviors do not change and reveal power-laws $m_2\propto t^{1/3}$ and $P\propto t^{1/6}$. Finally, the simulation of the equations of motion in the absence of nonlinear terms (\fref{fig:SKKF_PRE}, (o) curves) shows AL. 

Remarkably, in all simulations for single-site initial wave packets the spreading exponent did not show any substantial dependence on $\beta$ (dDNLS), $E$ (dKG) or $W$.Their variation affects only the transient time $\tau_{0}$ and prefactors. A consistent protocol has been elaborated to ensure accuracy, consistency and reproducibility of the results. Fittings were performed by analyzing 20 runs with different realizations of disorder \cite{Skokos2009}. For each realization $\alpha_m$ was fitted and then averaged over computational measurements: $\alpha_m=0.33 \pm 0.02$ and $\alpha_m=0.33 \pm 0.05$ respectively were found for DNLS and KG \cite{Flach2009, Skokos2009}. Therefore, the theoretically predicted exponent $\alpha=1/3$ appears to explain the data \cite{Skokos2009}.

The authors of \cite{Flach2009, Skokos2009} also attempted to mimic strongly incoherent wave packet spreading to probe limitations of the theory. They took the same disorder realizations and single-site initial conditions an were artificially dephasing the normal modes in a random way every one hundred time units \cite{Skokos2009}. In that case the subdiffusion speeded up to $m_2\propto t^{1/2}$. An extremely intriguing result that inspired further progress in the subject concerned single mode excitations (occupying more than one site in direct space). In this case, authors detected a growth of $m_2$, which was also subdiffusive but faster than $t^{1/3}$, at least on some intermediate time scales. Although the theory presented in \cite{Flach2009, Skokos2009} gave a convincing description of the single-site excitation spreading, it clearly missed answers for multiple-site initial wave packets and further analytical and computational studies were needed.

Numerical integration of qDNLS (\ref{eq:qDNLS}) \cite{Larcher2009} and wsDNLS (\ref{eq:qDNLS}) \cite{Krimer09,Datta1998,Kolovsky2010,Kolovsky2010a} equations also confirmed that nonlinearity destroys initial localization and wave packet spreads subdiffusively (unless suppressed by self-trapping). This result is consistent with the previously discussed finding for purely disordered nonlinear chains. However, the spreading exponents $\alpha_m$ have shown some dependency on the nonlinearity strength. 

\subsection{Theories}
\label{TP}  

A predictive theory of the spreading explaining its subdiffusive nature has been first proposed in Ref. \cite{Flach2009,Skokos2009} and further shaped in Ref. \cite{Laptyeva2010,Flach2010,Bodyfelt2011}. Here we present its key ideas using 1D dDNLS \eref{eq:1DdDNLS}, with subsequent extensions to other models. Corresponding numerical experiments will be discussed in \sref{NR}.
\subsubsection{NM representation} 
\label{EMR}
Nonlinearity induces interaction between normal modes.  Due to the localized character of NMs no spreading is possible in the absence of their interaction, i.e. of nonlinearity.
The observed wave packet spreading phenomena occur well below the self trapping regime. The latter can be viewed as a nonperturbative strong nonlinearity effect. Its natural space is the direct space, in which the pure nonlinear terms decouple. The wave packet spreading
however is a perturbative effect of nonlinearity on the linear wave equation, whose natural space is the NM one.
 In order to rewrite the equations of motion, we apply the transformation $\psi_l = \sum_\nu A_{\nu,l}\phi_\nu$, where $\phi_\nu$ is the time-dependent amplitude of the $\nu$-th NM $A_\nu$. Taking into account the orthogonality of NMs the equations of motion (\ref{eq:1DdDNLS}) become
\begin{equation}
i \dot{\phi}_\nu = E_\nu \phi_\nu + \beta \sum \limits_{\nu_1, \nu_2, \nu_3}^{}I_{\nu, \nu_1, \nu_2, \nu_3}\phi_{\nu_1}^\star\phi_{\nu_2}\phi_{\nu_3}
\label{eq:NMRep}
\end{equation}	   
with the overlap integrals  
\begin{equation}
I_{\nu, \nu_1, \nu_2, \nu_3} = \sum \limits_{l}^{}A_{\nu,l}A_{\nu_1,l}A_{\nu_2,l}A_{\nu_3,l}.
\label{eq:overlap}
\end{equation}	   
It follows that interactions between normal mode oscillators are formally infinite-range. However, each NM is effectively coupled only to a finite number of neighboring modes due to exponential localization. Therefore  the effective interaction range is finite.

{\it Average eigenvalue spacing inside a localization volume -- }
Consider an eigenstate $A_{\nu,l}$ for a given disorder realization. How many of the neighboring eigenstates will have non-exponentially small amplitudes inside its localization volume $V_{\nu}$?
Note that there is a one-to-one correspondence between the number of lattice sites, and the number of eigenstates. Therefore, on average the number of neighboring eigenstates
will be simply $V_{\nu}$. Let us consider sets of neighboring eigenstates. Their eigenvalues will be in general different, but confined to the interval $\Delta$ of the spectrum.
Therefore the average spacing $d$ of eigenvalues of neighboring NMs
within the range of a localization volume is of the order of $d \approx \Delta / V$,
which becomes $d \approx \Delta W^2 /300 $ for weak disorder.
The two scales $ d \leq \Delta $ are expected to determine the
packet evolution details in the presence of nonlinearity.

{\it The secular normal form -- }
Let us perform a further transformation $\phi_{\nu} = {\rm e}^{-i \lambda_{\nu} t} \chi_{\nu}$ and insert
it into Eq. (\ref{eq:NMRep}):
\begin{equation}
i \dot{\chi}_{\nu} = \beta \sum_{\nu_1,\nu_2,\nu_3}
I_{\nu,\nu_1,\nu_2,\nu_3} \chi^*_{\nu_1} \chi_{\nu_2} \chi_{\nu_3} {\rm e}^{i(\lambda_{\nu} + 
\lambda_{\nu_1}-\lambda_{\nu_2}-\lambda_{\nu_3})t}
\;.
\label{NMeqchi}
\end{equation}
The right hand side contains oscillating functions with frequencies
\begin{equation}
\lambda_{\nu,\vec{n}} \equiv \lambda_{\nu} + 
\lambda_{\nu_1}-\lambda_{\nu_2}-\lambda_{\nu_3}\;,\;\vec{n} \equiv (\nu_1,\nu_2,\nu_3)\;.
\label{dlambda}
\end{equation}
For certain values of $\nu,\vec{n}$ the value $\lambda_{\nu,\vec{n}}$ becomes exactly zero. These secular terms
define some slow evolution of (\ref{NMeqchi}). Let us perform an averaging over time of all terms
in the rhs of (\ref{NMeqchi}), leaving therefore only the secular terms. The resulting secular
normal form equations (SNFE) take the form
\begin{equation}
i \dot{\chi}_{\nu} = \beta \sum_{\nu_1}
I_{\nu,\nu_,\nu_1,\nu_1} |\chi_{\nu_1}|^2 \chi_{\nu}
\;.
\label{NMeqRNF}
\end{equation}
Note that possible missing factors due to index permutations can be absorbed into the overlap integrals,
and are not of importance for what is following.
The SNFE can be now solved for any initial condition $\chi_{\nu}(t=0)=\eta_{\nu}$ and yields
\begin{equation}
\chi_{\nu}(t) = \eta_{\nu} {\rm e}^{-i \Omega_{\nu} t}\;,\; \Omega_{\nu} = \beta \sum_{\nu_1}
I_{\nu,\nu_,\nu_1,\nu_1} |\eta_{\nu_1}|^2 
\;.
\label{SNFE}
\end{equation}
Since the norm of every NM is preserved in time for the SNFE, it follows that Anderson localization
is preserved within the SNFE. The only change one obtains is the renormalization of the eigenfrequencies
$\lambda_\nu$ into $\tilde{\lambda}_{\nu} = \lambda_{\nu}+\Omega_{\nu}$. Moreover, the phase coherence
of NMs is preserved as well. Any different outcome will be therefore due to the nonsecular terms,
neglected within the SNFE. We note that $I_{\nu,\nu_,\nu,\nu} \equiv p_{\nu}^{-1}$. Then the sum in  (\ref{eq:overlap}) contains only
nonnegative terms. By normalization $A_{\nu,l} \sim 1/\sqrt{V}$ inside its localization volume, and therefore $I_{\nu,\nu_,\nu,\nu} \sim 1/V$.
Similar argumentation leads to $I_{\nu,\nu_,\nu_1,\nu_1} \sim 1/V$.


\subsubsection{Mechanisms of spreading}
\label{SprMech}
As indicated in the above, an effective interaction in disordered systems is only possible between whose modes that lie in the same localization volume.  Such modes in the ``cold'' exterior on the wave packet boundary can be excited in two possible ways: incoherently ``heated'' up by the packet (non-resonant process) or directly excited by some packet mode from a boundary layer (resonant process).

{\it Incoherent heating beyond the secular normal form -- }
In general, nonlinearity leads to the loss of integrability, chaotic dynamics and ergodic properties. The chaotic dynamics inside a wave packet will
enforce a decoherence of the NM phases after suitable times (typically the inverse of some Lyapunov exponents). It is well known
that wave localization relies entierly on keeping the phase coherence of the participating waves \cite{Rayanov2013}.
Loss of phase coherence replaces the wave equation by a diffusion equation \cite{Rayanov2013}.
In this case expansion of wave packets has much in common with the process of heat transfer from a hot droplet (the wave packet) into the
cold exterior. Here the average energy density becomes an analog of temperature. 

The time-averaged secular norm form (\ref{NMeqRNF}) keeps the integrability of the nonlinear wave equation, and therefore also 
keeps Anderson localization. Any deviation from Anderson localization is therefore due to the omitted time-dependent oscillating terms
in (\ref{NMeqchi}). Let us isolate one of the many terms in the rhs sum in (\ref{NMeqchi})
\begin{equation}
 \dot{\chi}_{\nu} = \beta 
I_{\nu,\vec{n}} \chi^*_{\nu_1} \chi_{\nu_2} \chi_{\nu_3} {\rm e}^{i\lambda_{\nu,\vec{n}} t}
\;.
\label{oneterm}
\end{equation}
Assume a solution of the secular normal form equations (\ref{NMeqRNF}) in the limit of weak nonlinearity. 
Consider the solution of (\ref{oneterm}) as the first order correction. This correction has an amplitude
\begin{equation}
|\chi_{\nu}^{(1)}| = |\beta \eta_{\nu_1}\eta_{\nu_2}\eta_{\nu_3}|
R_{\nu,\vec{n}}^{-1}\;,\; R_{\nu,\vec{n}} \sim
\left|\frac{\lambda_{\nu,\vec{n}}}{I_{\nu,\vec{n}}}\right| \;,
\label{PERT1}
\end{equation}
The
perturbation approach breaks down, and resonances set in, when $|\eta_{\nu}|
< |\chi_{\nu}^{(1)}|$ for at least one triplet $\vec{n}$, and for at least one excited reference mode $\nu$:
\begin{equation}
|\eta_{\nu}| < |\eta_{\nu_1}\eta_{\nu_2}\eta_{\nu_3}|
\frac{\beta}{R_{\nu,\vec{n}}}\;.
\label{PERT2}
\end{equation}
Let us discuss this result. The eigenfrequencies contribute through the quadruplet  $\lambda_{\nu,\vec{n}}$ (\ref{dlambda}). Resonances will be triggered for small quadruplets. The above quadruplet can become small for eigenvalues which are separated way beyond $d$. An extreme example is an equidistant spectrum which allows for
exact zeros of quadruplets.  In the disordered case with $V \gg 1$, for one reference mode $\nu$ we consider $V$ states in its localization volume, which allow for about
$V^3$ quadruplet combinations. It is reasonable to assume that the set of $V$ eigenvalues will show correlations on energy separations of the order of $d$ (level spacing), but a decay
of these correlations at larger energy distances. Therefore, for most of the $V^3$ combinations, the participating eigenvalues can be considered to be uncorrelated.
With that assumption, the PDF $\mathcal{W}_{\lambda}(\lambda_{\nu,\vec{n}})$, which is a sum of four random numbers, 
can be expected to be close to a normal distribution due to the central limit theorem, i.e.
\begin{equation}
\mathcal{W}_{\lambda}(x) \approx \frac{1}{\sqrt{2\pi} \sigma } {\rm e}^{-\frac{x^2}{2\sigma^2}}\;,\;\sigma^2 = \frac{\Delta^2}{12}\;.
\label{gauss}
\end{equation}
In a recent study of a one-dimensional ladder geometry \cite{Yu2014} the closeness of $\mathcal{W}_{\lambda}$ to the normal distribution was numerically confirmed.
Since we are interested in small quadruplet values, we stress that the normal distribution has a finite value at zero argument, i.e. 
\begin{equation}
\mathcal{W}_{\lambda}(0) \approx \frac{\sqrt{3}}{\sqrt{2\pi} \Delta }\;.
\label{gausszero}
\end{equation}
Again the predicted value is only a factor of two off the actual numbers computed in \cite{Yu2014}.

The second important quantity which enters (\ref{PERT2}) through the definition of $R_{\nu,\vec{n}}$ in (\ref{PERT1}) are the overlap integrals $I_{\nu,\vec{n}}$. Much less is known about
these matrix elements (however see \cite{Krim10}). 
This is mainly due to the strong correlations between eigenvectors of states residing in the same localization volume but having sufficiently well separated eigenvalues. Let us ignore those difficulties for the moment, and assume that we can operate with one characteristic (average)
overlap integral $\langle I \rangle$. 
Then the PDF $\mathcal{W}_R$ of $R$ becomes 
\begin{equation}
\mathcal{W}_R (x) = \langle I \rangle \mathcal{W}_{\lambda}(\langle I \rangle x)\;,\; \mathcal{W}_R(0) = \frac{\sqrt{3} \langle I \rangle}{\sqrt{2\pi} \Delta }\;.
\label{pdfR}
\end{equation}
With the additional assumption that all amplitudes $\eta \sim \sqrt{n}$ (note that this excludes a systematic consideration of a single normal
mode excitation) we arrive at the resonance condition
\begin{equation}
\beta n < R_{\nu,\vec{n}}\;.
\label{resonance_R}
\end{equation}
For a given set $\{\nu,\vec{n}\}$ the probability of meeting such a resonance is given by
\begin{equation}
\mathcal{P}_{\nu,\vec{n}} = \int_0^{\beta n} \mathcal{W}_R (x) dx \;,\; \mathcal{P}_{\nu,\vec{n}}|_{\beta n \rightarrow 0} \rightarrow \frac{\sqrt{3} \langle I \rangle}{\sqrt{2\pi} \Delta } \beta n\;.
\label{indres}
\end{equation}
For a given reference mode $\nu$ there are $V^3$ combinations of quadruplets. The probability that at least one of these quadruplets satisfies the resonance condition
is equivalent to the probability that the given mode violates perturbation theory:
\begin{equation}
\mathcal{P}_{\nu} = 1 - \left(  1 -  \int_0^{\beta n} \mathcal{W}_R (x) dx           \right) ^{V^3}\;,\; \mathcal{P}_{\nu} |_{\beta n \rightarrow 0} 
\rightarrow \frac{\sqrt{3} V^3 \langle I \rangle}{\sqrt{2\pi} \Delta } \beta n\;.
\label{resonance}
\end{equation}
The main outcome is that the probability of resonance is proportional to $\beta n$ for weak nonlinearity. Moreover, within the disorder interval $1 \leq W \leq 6$
a numerical evaluation of the average overlap integral $\langle I \rangle \approx 0.6 \; V^{-1.7}$ \cite{Krim10}. This yields 
$\mathcal{P}_{\nu} |_{\beta n \rightarrow 0}  \approx 0.43\;V^{0.3} (\beta n/d)$. The uncertainty of the correct estimate of the overlap integral average, and the restricted
studied disorder range may well address the weak dependence $V^{0.3}$. What remains however is evidence that the resonance probability for weak nonlinearity
is proportional to the ratio $(\beta n ) / d$. Therefore a practical outcome is that the average spacing $d$ sets the energy scale - for $\beta n \ll d$ the resonance probability
$\mathcal{P} \sim (\beta n) /d$, while for $\beta n \gg d$ the resonance probability $P \approx 1$. 

A straightforward numerical computation of the above probability can be performed avoiding a number of the above assumptions.
For a
given NM $\nu$ we define $ R_{\nu,\vec{n}_0} = \min_{\vec{n} }
R_{\nu,\vec{n}}$.  Collecting $R_{\nu,\vec{n}_0}$ for many $\nu$ and many
disorder realizations, we can obtain the probability density distribution
$\mathcal{W}(R_{\nu,\vec{n}_0})$. 
\begin{figure}[ht]
\begin{center}
\includegraphics[width=0.7\columnwidth,keepaspectratio,clip]{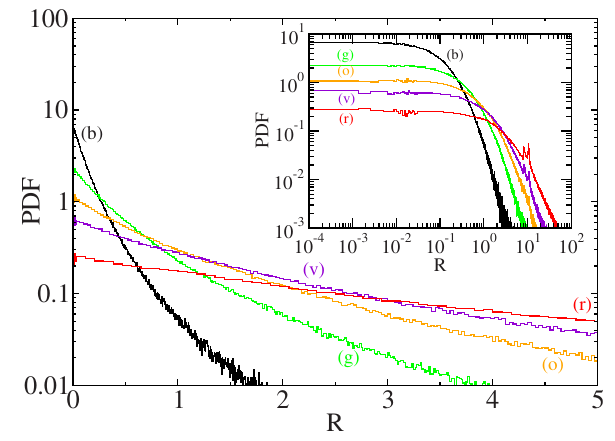}
\caption{The probability densities $\mathcal{W}(R_{\nu,\vec{\mu_0}})$ for NMs to be in resonance presented in linear-log (main figure) and log-log (inset) scales. Numerics for dDNLS \eref{eq:1DdDNLS} adapted from \cite{Krim10}.
Disorder strengths are $W=4$ ((b) black curve), $W=6$ ((g) green curve), $W=8$ ((o) orange curve), $W=10$ ((v) violet curve), $W=15$ ((r) red curve). Reprinted from \cite{Krim10}.} 
\label{fig:dDNLSpdf}
\end{center}
\end{figure}
The probability $\mathcal{P}$ for a mode, which is excited to a norm $n$ (the
average norm density in a packet of modes), to be resonant with at least one triplet
of other modes at a given value of the interaction parameter $\beta$ is again given
by \cite{Krim10,Skokos2009}
\begin{equation}
\mathcal{P} = \int_0^{\beta n} \mathcal{W}(x) {\rm d}x\;.
\label{resprob}
\end{equation} 
Therefore again $\mathcal{W}(R_{\nu,\vec{n}_0} \rightarrow 0)
\rightarrow C(W) \neq 0$ \cite{Skokos2009}.  For the cases studied, the constant $C$ drops with
increasing disorder strength $W$, in agreement with (\ref{resonance}), which suggests
$C = \frac{\sqrt{3} V^3 \langle I \rangle}{\sqrt{2\pi} \Delta }$ (see Fig.\ref{fig:dDNLSpdf}).

The large power $V^3$ in (\ref{resonance}) allows to make a simple exponential approximation
\begin{equation}
\mathcal{W} (R) \approx C {\rm e}^{-CR}\;,\; C = \frac{\sqrt{3} V^3 \langle I \rangle}{\sqrt{2\pi} \Delta }\;.
\label{approxp}
\end{equation}
which in turn can be expected to hold also for the case of weak disorder.
It leads to the approximative result
\begin{equation}
\mathcal{P} = 1-{\rm e}^{-C\beta n}\;.
\label{approxpp}
\end{equation}
Therefore the probability for a mode in the packet to be resonant is
proportional to $C \beta n$ in the limit of small $n$ \cite{Flach2009,Skokos2009}.  

We stress again that the discussed uncertainty  in the definition of an average overlap integral, and the fact that the distribution of quadruplets is expected to be controlled
by the {\it stiffness} of the set of eigenvalues of the normal mode set $\{ \nu,\vec{n} \}$ rather than its spacing $d$, might be related. This does become evident if assuming an equidistant
set. But then again, for a disordered system discussed here, the only scale on which the quadruplets can fluctuate close to zero, is the spacing $d$. 

{\it Resonant spreading. -- }
Now, we consider the process of resonant excitation of an exterior mode by a mode from the packet as a possible spreading mechanism. The number of packet modes in the boundary layer of width $V$, which are resonant with the mode of a cold exterior is proportional to $\beta n$. After long enough spreading the norm density decays and $\beta n \ll d$. Therefore, as a rule, there will be no mode inside the packet that efficiently resonates with an exterior mode. Then the wave packet becomes trapped at its edges, while the phases in the interior of the wave packet get randomized by nonlinear interactions (the wave packet thermalizes). Thus, on large time scales the packet will be able to incoherently excite the exterior and to increase its size, while the resonant excitation becomes an unlikely event and can be excluded.

{\it Measuring chaos -- }
Michaeli and Fishman studied the evolution of single site excitations for the DNLS model \cite{Michaely2012}. They considered the rhs of Eq.(\ref{NMeqchi})
as a function of time $i\dot{\chi}_{\nu} =F_{\nu}(t)$ for a mode $\nu=0$ which was strongly excited at time $t=0$. The statistical analysis of the time dependence of 
$F_0(t)$ shows a quick decay of its temporal correlations for spreading wave packets. Therefore the force $F_0(t)$ can be considered as a random noise function
on time scales relevant for the spreading process. This is a clear signature of chaos inside the wave packet.

Vermersch and Garreau (VG) \cite{bvjcg13} followed a similar approach for the DNLS model. They measured the time dependence of the participation number $P(t)$ of a spreading wave packet. VG then extracted a spectral entropy, i.e. a measure of the number of participating frequencies
which characterize this time dependence. Spectral entropies are convenient measure to discriminate between regular and chaotic dynamics. VG concluded
that the dynamics of spreading wave packets {\it is} chaotic. They also measured short time Lyapunov exponents to support their conclusion.

The long-time dependence of the largest  Lyapunov exponent $\Lambda$ as chaos strength indicators inside spreading wave packets for KG models was recently tested in \cite{csigsf13}. The crucial point
is that during spreading the energy density {\it is} decreasing, and therefore a weakening of the momentary chaos indicator is expected. Therefore $\Lambda (t)$ will
be not constant in time, but decrease its value with increasing time. Moreover, the calculation of Lyapunov exponents for integrable systems will also yield nonzero numbers
when integrating the system over any finite time. This is due to the method used - one evolves the original trajectory in phase space, and in parallel runs the
linearized perturbation dynamics of small deviations from the original trajectory in tangent space. Since this deviation is nonzero, any computer code
will produce nonzero estimates for the Lyapunov exponent at short times. The crucial point is that for integrable systems the long-time dependence of $\Lambda$
follows $\Lambda \sim 1/t$. This is also the result found in \cite{csigsf13} for the {\it linear} wave equation which obeys Anderson localization.
However the nonlinear case of wave packet spreading yields a dependence 
\begin{equation}
\Lambda (t) \sim \frac{1}{t^{1/4}} \gg \frac{1}{t}\;.
\label{mle}
\end{equation}
The authors of \cite{csigsf13} further compare  the obtained chaoticity time scale $1/\Lambda$ with the time scales governing the
slow subdiffusive spreading and conclude, that the assumption about persistent and fast enough chaoticity needed for thermalization
and inside the wave packet is correct. The dynamics inside the spreading wave packet is chaotic, and remains chaotic up to the largest simulation times,
without any signature of a violation of this assumption for larger times (no visible slowing down).

A further very important result concerns the seeds of
deterministic chaos and their {\sl meandering} through the packet in
the course of evolution. 
The motion of these chaotic seeds was visualized by following the
spatial evolution of the deviation vector distribution (DVD) used for the computation of
the largest Lyapunov exponent \cite{csigsf13}. This vector tends to align with the most unstable direction
in the system's phase space. Thus, monitoring how its components on
the lattice sites evolve allow to identify the most chaotic spots. Large
DVD values tell at which sites the sensitivity on initial
conditions (which is a basic ingredient of chaos) is larger.
The authors of \cite{csigsf13} observe that  the DVD stays localized, but the peak positions clearly meander
in time, covering distances of the order of the wave packet width. 

{\it Effective noise theories -- }
Having established that the dynamics inside a spreading wave packet is chaotic, let us proceed to construct an effective noise theory for spreading.
For that we replace the time dependence on the rhs of Eq. (\ref{NMeqchi}) by a random function in time:
\begin{equation}
i\dot{\chi}_{\nu} = F(t)\;,\;\langle F \rangle = 0\;,\; \langle F^2(t) \rangle = f^2;.
\label{ent1}
\end{equation}
Assume that the norm density (norm per site/mode) inside the wave packet is $n$.
Consider a normal mode $\mu$ which is outside the wave packet, but in a boundary layer of one of its edges. The boundary layer thickness is of the order of $V$.  The equation of motion for this mode is given by (\ref{ent1}).
At some initial time $t_0$ assume that the norm of the considered mode is close to zero $|\chi_{\mu}(t_0)|^2 = n_{\mu}(t_0) \ll n$.
Then the solution of the stochastic differential equation (\ref{ent1}) is yielding a diffusion process in norm/energy space of the considered
NM:
\begin{equation}
n_{\mu} (t) \sim f^2 t\;.
\label{ent2}
\end{equation}
The considered mode will reach the packet norm $n$ after a time $T$ whose inverse will be proportional to the
momentary diffusion rate of the wave packet $D \sim 1/T$:
\begin{equation}
D \sim \frac{f ^2}{n} \;.
\label{ent3}
\end{equation}
Let us estimate the variance $f$ for the nonlinear wave equation. It follows from (\ref{NMeqchi}) that
$f \sim \beta n^{3/2} \langle I \rangle$. With that, we arrive at $D \sim (\beta n \langle I \rangle )^2$. 
The main point here is that the diffusion coefficient is proportional to $n^2$, therefore the more the packet spreads, the lower its density, and the smaller $D$.
We obtain a time-dependent diffusion coefficient, and a tendency to spread slower than within a normal diffusion process. 
The second moment $m_2$ of a wave packet is inverse proportional to its squared norm density $m^2 \sim 1/n^2$. At the same time it should obey
$m_2 \sim D t$. Since $D \sim 1/m_2$ it follows $m_2 \sim t^{1/2}$. 

The second way is to write down a nonlinear diffusion equation \cite{zeldovich1950,Barenblatt1952} for the norm density distribution
(replacing the lattice by a continuum for simplicity, see also \cite{Kolovsky2010}):
\begin{equation}
\partial_t n = \partial_{\nu}(D \partial_{\nu} n) \;,\; D \sim n^{\kappa}\;.
\label{ent4}
\end{equation}
The solution $n(\nu,t)$ obeys the scaling $n(\nu,t/a) = bn(c\nu,t)$  with $b=c=a^{1/(\kappa+2)}$ if $n(\nu \pm \infty,t) \rightarrow 0$. Therefore the second moment
\begin{equation}
m_2 \sim t^{\alpha}\;,\;\alpha = \frac{2}{\kappa+2}\;.
\label{ent4b}
\end{equation}

With $\kappa=2$ we obtain the subdiffusive law $m_2 \sim t^{1/2}$ again. 
Note that the above nonlinear diffusion equation can be derived through a master equation and a Fokker-Planck equation for both norm and energy densities \cite{Basko2014}, or Boltzmann equations \cite{gsamf13}.

Until now we have obtained the subdiffusion exponent $1/2$, but the numerically observed $1/3$ remains unexplained. However, recalling that enforced randomization of NM phases during the spreading {\sl does yield the exponent} $1/2$ one finds himself on the right track. That is artificially randomized NM phases fulfill an assumption of the effective noise theory. What is then the reason for the even slower subdiffusion
with $\alpha=1/3$? We recall that perturbation theory leads to a probability $\mathcal{P}$ of a given NM being resonant which is 
small for small densities (\ref{resonance}): 
$\mathcal{P}_{\nu} |_{\beta n \rightarrow 0} 
\rightarrow \frac{\sqrt{3} V^3 \langle I \rangle}{\sqrt{2\pi} \Delta } \beta n$. In case when this probability is equal to one, the above diffusion constant assumption 
would make sense. In the case when the resonance probability is zero, perturbation theory should be applicable, the secular normal form 
yields Anderson localization, and spreading should stop. In that case $f=0$ and then $D=0$. Therefore another factor is missing in the
expression of $f$. This factor was assumed be a function of $\mathcal{P}$ such that the factor becomes one when $\mathcal{P}=1$ and zero when
$\mathcal{P}=0$. The simplest prefactor is $\mathcal{P}$ itself \cite{Flach2009,Skokos2009} and yields
\begin{equation}
f \sim \mathcal{P} \beta n^{3/2} \langle I \rangle\;,\; D \sim (\mathcal{P} \beta n \langle I \rangle )^2\;,\; \mathcal{P} = 1-{\rm e}^{-C\beta n}\;,\;
C=\frac{\sqrt{3} V^2 \langle I \rangle}{\sqrt{2\pi} d }\;.
\label{ent5}
\end{equation} 
Then the solution of the nonlinear diffusion equation (\ref{ent4}) reads
\begin{eqnarray}
m_2 &\sim& (\beta \langle I \rangle V)^{4/3} d^{-2/3} t^{1/3} \;,\; C\beta n \ll  1\; : \; {\rm weak \; chaos}\;,
\label{ent6a}
\\
m_2 &\sim& \beta \langle I \rangle t^{1/2}\;\;\;\;\;\;\;\;\;\;\;\;\;\;\;\;\;\;\;\;\;\;\;,\; C\beta n \gg  1\; : \; {\rm strong \; chaos}\;.
\label{ent6b}
\end{eqnarray}
We arrived at a construction which results in the correct weak chaos exponent $\alpha = 1/3$ \cite{Flach2009}. We also predict that there must be an intermediate regime
of strong chaos for which $\alpha = 1/2$ - {\sl without any enforcing of the randomization of NM phases} \cite{Flach2010}.  It has to be intermediate, since with an assumed further
spreading of the wave packet, the density $n$ will decrease, and at some point satisfy the weak chaos condition (\ref{ent6a}) instead of the strong chaos condition (\ref{ent6b}).
Therefore, a potentially long lasting regime of strong chaos has to cross over into the asymptotic regime of weak chaos \cite{Flach2010}. That crossover is not a sharp one in
the time evolution of the wave packet. It might take several orders of magnitude in time to observe the crossover. Therefore, instead of fitting the numerically
obtained time dependence $m_2(t)$ with power laws, it is much more conclusive to compute derivatives $d\langle \log_{10} m_2 \rangle / d \log_{10} t$ in order
to identify a potentially long lasting regime of strong chaos, crossovers, or the asymptotic regime of weak chaos.

\subsubsection{Dynamical regimes of spreading}
\label{SR}

Let us discuss evolution from several different initial states.
(a) If only one normal mode is initially excited to norm $n$, then it follows from (\ref{SNFE}) that 
its frequency renormalization $\Omega_{\nu} = \beta n p_{\nu}^{-1} \sim \beta n /V$
where $V$ is a typical localization volume of a normal mode. Comparing this value to the spacing $d \sim \Delta /V$ we conclude
that weak chaos holds if $\beta n \sim \Delta$.
(b) If however a large group of normal modes is excited inside a wave packet such that all normal modes have norm $n$, then
the sum in (\ref{SNFE}) will change the frequency renormalization to $\Omega_{\nu} \sim \beta n$ for each of the participating modes.
Comparing that to the spacing $d$ we now find that weak chaos breaks down at sufficiently weaker nonlinearities
$\beta n \sim \Delta / V$. (c) Finally assume that only one lattice site is initially excited with norm $n$. That means that $V$ normal modes
are excited each with norm $n/V$. After some short 
transient time the wave packet will occupy a localization volume region, and stay in there for all times for the linear wave equation. Then the frequency normalization for each
participating mode becomes $\Omega_{\nu} \sim \beta n /V$ as in (a), and weak chaos holds up to 
$\beta n \sim \Delta$.

All of the considered lattices allow for selftrapped states in the regime of strong nonlinearity. The natural basis for selftrapped states is the original lattice itself, rather than the normal modes of the linear wave
equation. This becomes evident when considering a lattice without any disorder, for which the normal modes of the linear wave equation
are extended states, yet selftrapping is present as well within the nonlinear wave equation.
Selftrapping is an example of a {\it nonperturbative} physics of strong nonlinearity.  
For the above cases of initial conditions, selftrapping can be effectively predicted whenever a single oscillator on one
site renormalizes its frequency $\epsilon_l+\beta |\psi_l|^2$ such that it exits the linear wave spectrum.
For the above initial state case (a) this happens when $\beta n  \sim V (\Delta/2-\lambda_{\nu})$, about $V$ times larger than the
weak chaos threshold. For case (b) the norm $n$ per normal mode is also the norm $n$ per lattice site. Therefore selftrapping 
is expected at $\beta n \sim \Delta /2 $, again about $V$ times larger than the corresponding weak chaos threshold.
However, case (c) is different. Here we place a norm $n$ {\it initially} on one site. If the selftrapping condition for that site holds, the dynamics
will stay from scratch in the nonperturbative discrete breather regime, without any chance to spread into a localization volume region
set by the linear wave equation. Therefore the selftrapping threshold reads $\beta n \sim \Delta /2 - \epsilon_l$ and becomes of the same
order as the weak chaos threshold. Single site excitations will be thus launched either in a weak chaos regime, or in a self trapped one.
The other initial states allow for a third regime - outside the weak chaos regime, but well below the selftrapping one.
For reasons to come, we coin this additional regime {\it strong chaos regime}.
We recapitulate again, that single site excitations are expected to be either in the regime of weak chaos, or selftrapping. Other initial states
allow for another intermediate regime of strong chaos.

Consider a wave packet at $t=0$ which has norm density $n$ and size $L$.
Let us wrap the above discussion into expected regimes of spreading \cite{Flach2010}. Note that due to the above ambiguities, 
the following estimates are at the best semi-quantitative.

{\sc Single site excitations} with norm $n$ and $\epsilon_l=0$ at the excitation site:
\begin{eqnarray}
\beta n  < \Delta / 2 \;:\; {\rm weak\; chaos}
\nonumber
\\
{\rm strong \; chaos\; not \; present}
\label{sser}
\\
\beta n > \Delta / 2 \;:\; {\rm selftrapping}
\nonumber
\end{eqnarray}
{\sc Single mode excitations} with norm $n$ and $\lambda_{\nu}=0$ for the excited mode:
\begin{eqnarray}
\beta n < \Delta \;:\; {\rm weak\; chaos}
\nonumber
\\
\Delta < \beta n  < V \Delta /2 \;:\; {\rm strong\; chaos}
\label{smer}
\\
V \Delta / 2 < \beta n \;:\; {\rm selftrapping}
\nonumber
\end{eqnarray}
{\sc Multi site/mode wave packet} with norm density $n$ per site/mode and size $V$:
 \begin{eqnarray}
\beta n < \Delta /V \;:\; {\rm weak\; chaos}
\nonumber
\\
\Delta / V < \beta n  < \Delta /2 \;:\; {\rm strong\; chaos}
\label{wper}
\\
\Delta / 2 < \beta n \;:\; {\rm selftrapping}
\nonumber
\end{eqnarray}

\begin{figure}[ht]
\begin{center}
\includegraphics[width=0.7\columnwidth,keepaspectratio,clip]{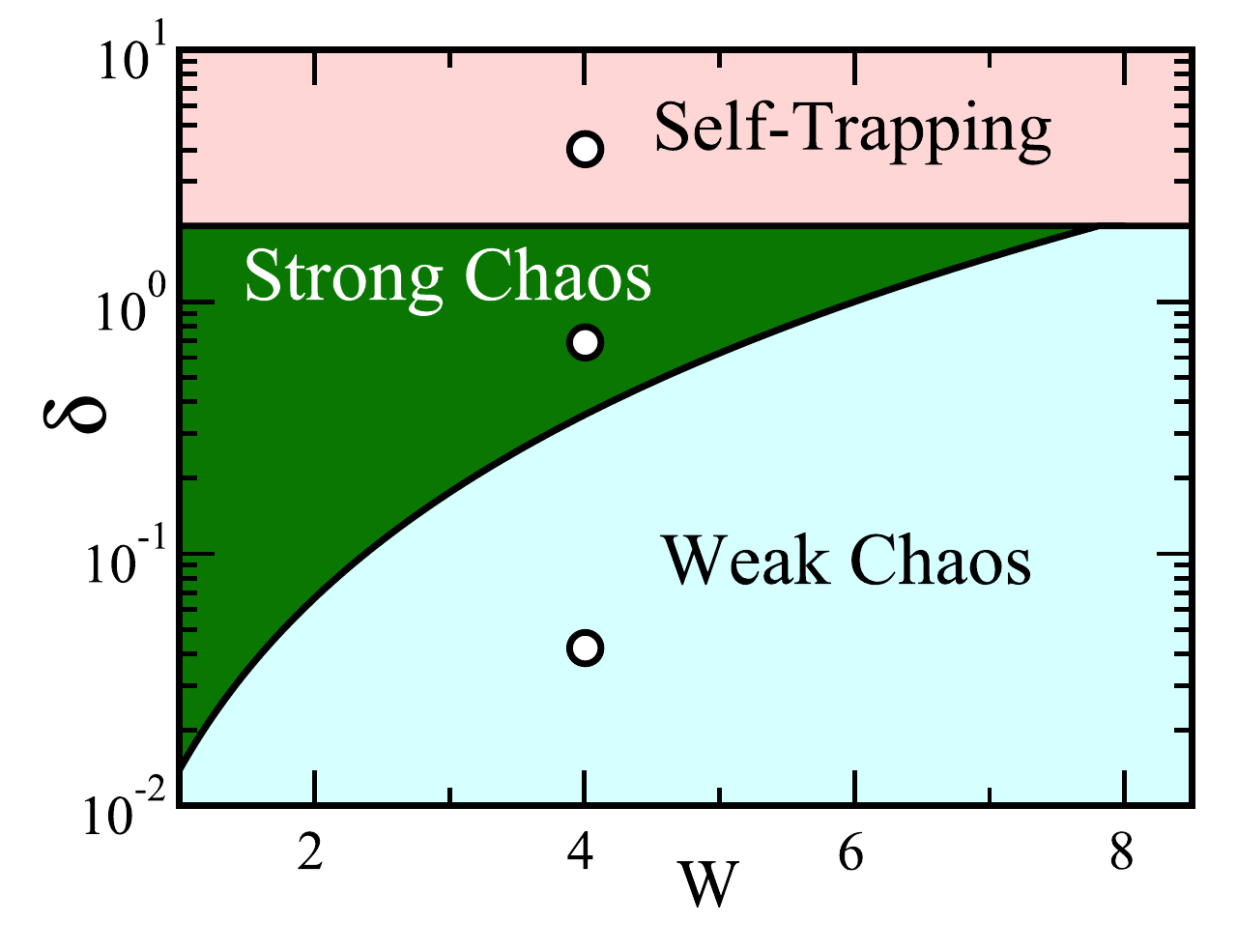}
\caption{Parametric space of disorder strength $W$ versus the nonlinear frequency shift $\delta \propto \beta n$ for the dDNLS \eref{eq:1DdDNLS}. Three spreading regimes are shown for dynamics dictated by: weak chaos (pale blue), strong chaos (green), and onset of self-trapping (pale red). The circles show the initial values used in numerics. Reprinted from \cite{Laptyeva2010}.} 
\label{fig:ParSpace}
\end{center}
\end{figure}

\Fref{fig:ParSpace} draws these regimes in the $\{\delta, W\}$ parameter space, where $\delta\equiv\beta n$, for $L=V$, the lines corresponding to the regime boundaries $\delta=d$ and $\delta=2$. The lower one is analytically found \textit{via} $d = \Delta/(3.3\xi_0)$ with the weak-disorder estimate $\xi_0 = 96W^{-2}$. More sophisticated numerical estimates of $d$ yield only slight corrections for $W > 6$ \cite{Krim10}. It is necessary to remark that the regime boundaries in \Fref{fig:ParSpace} are not sharp, rather there are transitional bands between the regimes.

Let us discuss \fref{fig:ParSpace}. Depending on disorder strength and initial norm density, the packet can be launched in one of the three regimes. A wave packet launched in the weak chaos regime stays in this regime for all times as the decrease of norm density does not lead to slowing down.  Conversely, launched in the strong-chaos regime the wave packet crosses over to the weak chaos  (which is asymptotic), when its norm density drops below the threshold of probability $1$ resonances. Note, that the time it takes may be quite long. A part of a packet launched in the self-trapped regime will remain localized, while the other part will be spreading and follow weak chaos subdiffusion asymptotically. The ratio between them and the dynamics of the spreading exponent remains poorly studied. 

\subsubsection{Generalization to higher dimensions and arbitrary nonlinearities}
\label{GTHDAN}

Let us consider $\bf D$-dimensional lattices with
nonlinearity order $\sigma > 0$:
\begin{equation}
i\dot{\psi_{\bf l}}= \epsilon_{\bf l} \psi_{\bf l}
-\beta |\psi_{\bf l}|^{\sigma}\psi_{\bf l}
-\sum\limits_{\vec{m}\in
{\bf D}({\bf l})}\psi_{\bf m}\;.
\label{RDNLS-EOMG}
\end{equation}
Here $\bf l$ denotes a $\bf D$-dimensional lattice vector with
integer components, and ${\bf m}\in
{\bf D}({\bf l})$ defines its set of nearest neighbor lattice sites.
We assume that (a) all NMs are spatially localized (which can be obtained on a lattice for strong
enough disorder $W$), (b) the property $\mathcal{W}(x \rightarrow 0) \rightarrow const \neq 0$
holds, and (c) the probability of resonances on the edge surface of a wave packet is tending
to zero during the spreading process. A wavepacket with average norm $n$ per excited mode has a second moment
$m_2 \sim 1/n^{2/\bf D}$. The nonlinear frequency shift is proportional to $\beta n^{\sigma/2}$.
The typical localization volume of a NM is still denoted by $V$, and the average spacing by $d$.

Consider a wave packet with norm density $n$ and volume $L < V$. A straightforward generalization of
the expected regimes of spreading leads to the following:
\begin{eqnarray}
\beta n^{\sigma/2} \left( \frac{L}{V}\right)^{\sigma/2} V   < \Delta \;:\; {\rm weak\; chaos}\;,
\nonumber
\\
\beta n^{\sigma/2} \left( \frac{L}{V}\right)^{\sigma/2} V > \Delta \;:\; {\rm strong\; chaos}\;,
\nonumber
\\
\beta n^{\sigma/2} > \Delta \;:\; {\rm selftrapping}\;.
\nonumber
\end{eqnarray}
The regime of strong chaos, which is located between selftrapping and weak chaos,
can be observed only if 
\begin{equation}
L > L_c = V^{1-2/\sigma}\;,\; n > n_c = \frac{V}{L} \left( \frac{d}{\beta}\right)^{2/\sigma}\;.
\end{equation}
For $\sigma =2$ we need $L>1$, for $\sigma \rightarrow \infty$ we need $L > V$,
and for $\sigma < 2$ we need $L \geq 1$. Thus the regime of strong chaos can be observed
e.g. in a one-dimensional system with a single site excitation and $\sigma < 2$.

If the wave packet size $L > V$ then the conditions for observing different regimes simplify to
\begin{eqnarray}
\beta n^{\sigma/2} < d \;:\; {\rm weak\; chaos}\;,
\nonumber
\\
\beta n^{\sigma/2} > d \;:\; {\rm strong\; chaos}\;,
\nonumber
\\
\beta n^{\sigma/2} > \Delta \;:\; {\rm selftrapping}\;.
\nonumber
\end{eqnarray}
The regime of strong chaos can be observed if 
\begin{equation}
n > n_c = \left( \frac{d}{\beta}\right)^{2/\sigma}\;.
\end{equation}

Similar to the above we obtain a diffusion
coefficient
\begin{equation}
D \sim \beta^2 n^{\sigma} (\mathcal{P}(\beta n^{\sigma/2}))^2
\;.
\label{ggeneralizeddiffusion}
\end{equation}
In both regimes of strong and weak chaos the spreading is subdiffusive
\cite{Flach2009,Flach2010}:
\begin{eqnarray}
m_2 \sim (\beta^2 t)^{\frac{2}{2+\sigma {\bf D}}}\;,\;{\rm strong}\;{\rm chaos}\;,
\label{sigma_strong}
\\
m_2 \sim (\beta^4 t)^{\frac{1}{1+\sigma {\bf D}}}\;,\;{\rm weak}\;{\rm chaos}\;.
\label{sigma_weak}
\end{eqnarray}
Note that the strong chaos result was also obtained within a Boltzmann theory approach \cite{gsamf13,Basko2014}.



\subsection{Other heterogeneities}
\subsubsection{Nonlinear quasi-periodic chains}
\label{QUASIgenerals}
The first attempt to discern the different dynamical regimes of initial excitation spreading in nonlinear quasi-periodic chain was done in \cite{Larcher2009}. Larcher et al. additionally incorporated a lattice phase shift, such that for qDNLS chain (\ref{eq:qDNLS}) the on-site potential depth becomes $\vartheta \cos(2\pi\alpha l +\theta)$. 
Three spreading regimes were found in dependence on the nonlinearity strength $\beta$, lattice phase $\theta$, and strength of potential $\vartheta$
: (I) strong self-trapping; (II) ballistic spreading, but with discrete breather structures being seen; (III) mixed behavior, when the initial states are self-trapped for $\theta=0$ and subdiffusive for $\theta=\pi$. The latter effect was also suggested in \cite{Johansson1995}. 

This paper was followed by a deeper theoretical and computational study of both DNLS and KG chains \cite{Larcher2012}. 
Employing the ideas of the wave packet spreading theory the nonlinear frequency shift $\delta$ should be compared with relevant frequency scales set by the linear spectrum $\Delta$ and $d$. While the nonlinear frequency shift stays the same as for dDNLS $\delta \propto \beta n$, the linear scales $d$ and $\Delta$ differ (for more details, see \sref{AAM}). Since the central issue is the localization-nonlinearity interplay, we restrict ourselves to the case of $\vartheta>2$. In this case, the spectrum is fractal-like and characterized by an infinite number of gaps and bands (see \fref{fig:1p2p2}). It has major gaps dividing the spectrum in main parts; each of them is divided in turn in smaller parts, etc. We will refer to subparts of spectrum as ``mini-bands''. 

The spectrum structure suggests that now the meaningful average frequency spacing $d$ must be defined taking into account the number of mini-bands $M$. Numerically it can be done as follows. The value of $\Delta$ (whether the width of the whole spectrum or a mini-band) can be measured as the difference between the largest and the smallest eigenvalues. Then, consider a given mini-band and the eigenstates that lie within it. For each $\nu$-th eigenstate one can calculate its localization volume $V_\nu$. Thereafter, form the subset of those eigenstates $\left\{\mu\right\}$, which effectively interact with mode $\nu$, i.e. fulfill the condition $|X_\nu-X_\mu|<V_\nu/2$, where $X_\nu=\sum_l l \left|A_{\nu,l}\right|^2$ are the center-of-norm coordinates. The average number of states in the subset can be estimated as $\langle V\rangle /M$ and the spacings within this subset can be calculated. This procedure is repeated for each eigenstate in the band and the average gives the mean spacing $d$. Note that the number of mini-bands $M$ one has to take into account to get a good approximation of the spectrum varies with $\vartheta$:   the authors' choice was $M=9$ for $\vartheta\lesssim2.1$, $M=3$ for $2.2\lesssim\vartheta\lesssim2.75$ and $M=1$ for $\vartheta \gtrsim 2.75$.  

\begin{figure}[ht]
\begin{center}
\includegraphics[width=0.7\columnwidth,keepaspectratio,clip]{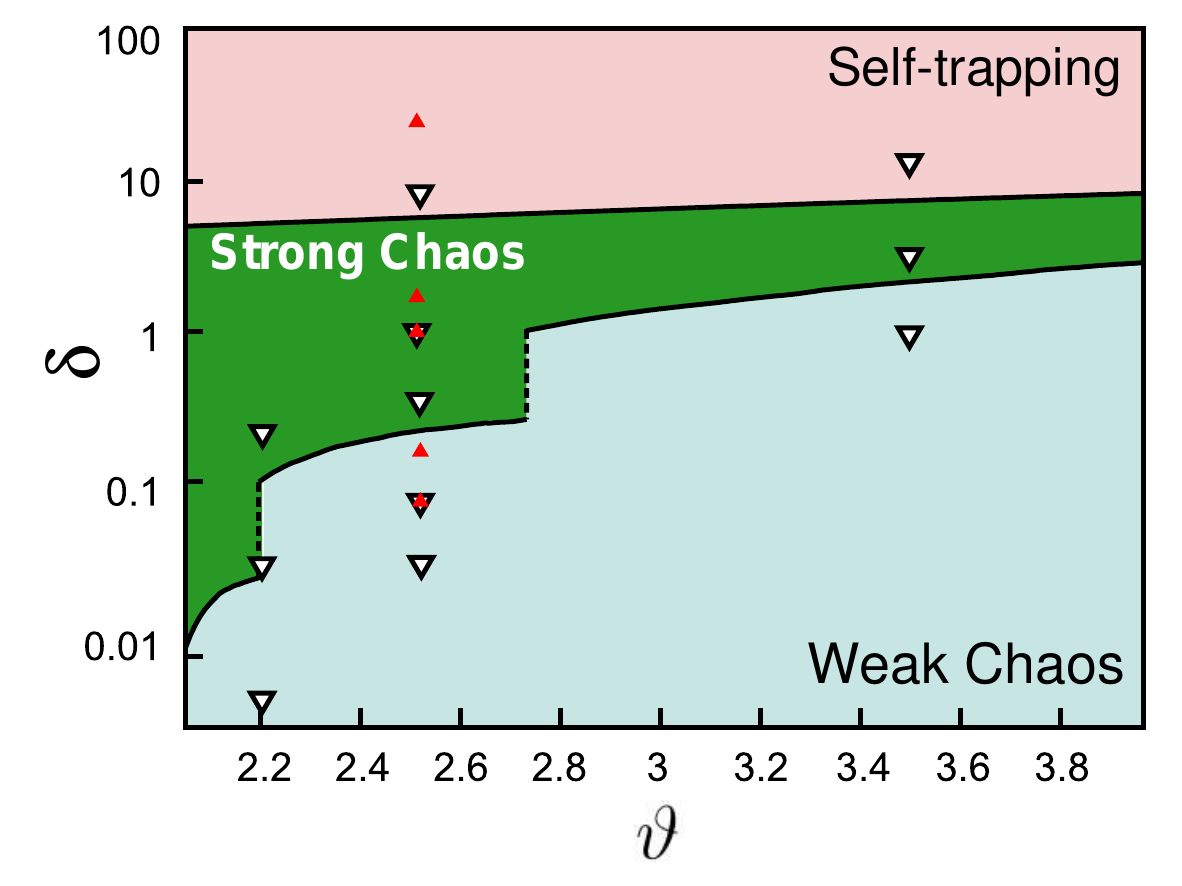}
\caption{Parametric space of quasi-periodic potential strength $\vartheta$ versus the frequency shift induced by nonlinearity $\delta \propto \beta n$. The ratio of scales $\delta$, $\Delta$, and $d$ conditions the three regimes of wave packet spreading for qDNLS model (\ref{eq:qDNLS}): weak chaos, strong chaos, and self-trapping. The separation between the regimes should be interpreted as a smooth crossover rather than as a sharp boundary. The downward (white) and upward (red) triangles show parameters used in computational experiments. Figure was adapted from \cite{Larcher2012}.} 
\label{fig:QDNSEregimes}
\end{center}
\end{figure}

Comparing the scales $\delta$, $d$, and $\Delta$, we expect qualitatively the same regimes of wave packet spreading, namely the weak chaos, strong chaos and self-trapping. It appears that at variance with disordered arrays partial self-trapping can occur even for arbitrary small $\beta n$ due to the presence of an infinite number of mini-bands and gaps in the linear spectrum. \Fref{fig:QDNSEregimes} sketches the suggested regimes in a parametric space, for the case of initial wave packet of norm $n$ and size $L \ge V$. 

\begin{figure}[ht]
\begin{center}
\includegraphics[width=0.7\columnwidth,keepaspectratio,clip]{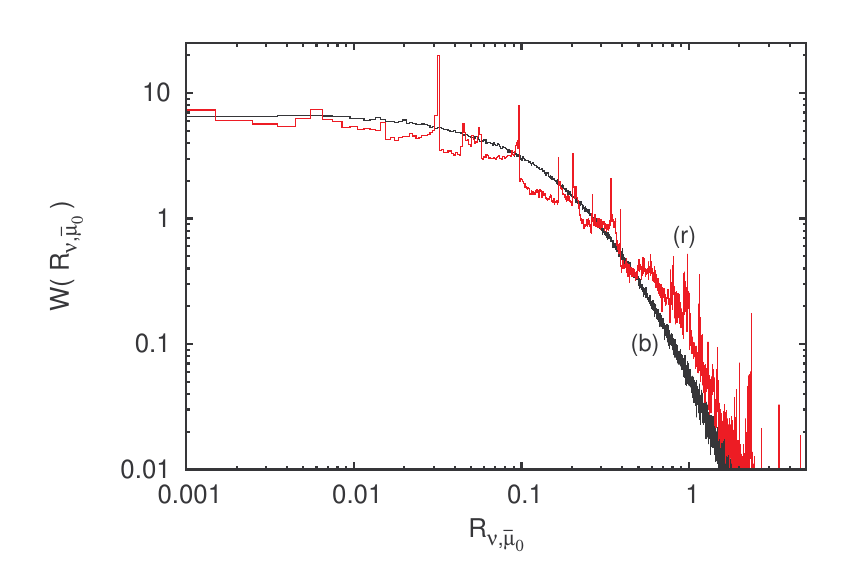}
\caption{Probability density function $\mathcal{W}(R_{\nu,\vec{\mu_0}})$ for the qDNLS model (red (r) curve) is shown in comparison with one for the dDNLS model (black (b) curve). For quasi-periodic case, the strength of potential is $\vartheta = 2.5$, while for the disordered case disorder strength is taken to give a similar localization length. Figure reprinted from \cite{Larcher2012}.} 
\label{fig:QDNSEres}
\end{center}
\end{figure}

The similarity of numerical observations between dDNLS and qDNLS suggests that the spreading theory may also be applicable here. Indeed, with ansatz $\psi_l=\sum_\nu A_{\nu,l}\phi_\nu$, the qDNLS equations of motion (\ref{eq:qDNLS}) can be rewritten in normal mode form identical to \eref{eq:NMRep}. Analogously, one considers the heating mechanism of spreading, for which the number of mode-mode resonances within the wave packet is a key parameter. The probability for the onset of a resonance can be calculated with the same numerical analysis \cite{Krim10,Skokos2009}. Let us reiterate that for a given normal mode $\nu$ on can calculate the quantity $R_{\nu,\vec{\mu_0}}= \min_{\vec{\mu}} \{R_{\nu,\vec{\mu}}\}$, which reflects the variations in amplitude of the $\nu$-th mode due to its interaction with a set $\vec{\mu}$ of three other modes. In case of the qDNLS model, $R_{\nu,\vec{\mu_0}}$ was calculated not only for many modes, but also for many values of the phase $\theta$. The obtained probability density distribution $\mathcal{W}(R_{\nu,\vec{\mu_0}})$ for $\vartheta = 2.5$ is shown in \fref{fig:QDNSEres}. 

The behavior of $\mathcal{W}(R_{\nu,\vec{\mu_0}})$ is qualitatively similar to the case of purely random systems apart from sharp peaks in the distribution (see \fref{fig:dDNLSpdf}). Still the most essential property $\mathcal{W}(R_{\nu,\vec{\mu_0}}\rightarrow 0)\rightarrow C\neq0$ is kept. As a consequence, the probability for a mode of norm $n$ to be resonant at a given nonlinearity strength $\beta$ is also approximated by $\mathcal{P} \approx 1-\e^{- C \beta n}$. Similarly to disordered systems, for $C\beta n \ll 1$ a finite fraction of packet modes interact resonantly with the probability $\mathcal{P} \propto C\beta n$ in the weak chaos regime. Conversely, for $C\beta n > 1$ almost all the modes interact resonantly and $\mathcal{P}\approx 1$ predicts the strong chaos regime. 

Following the analysis for the 1D dDNLS model, we expect to that the second moment of the wave packet grows as $m_2 \propto t^{1/3}$ in the weak chaos regime, and as $m_2 \propto t^{1/2}$ in the strong chaos regime. The strong chaos regime can only exist transiently for the same reason: as the wave packet spreads, its norm density $n$ decreases and eventually will satisfy the condition $\beta n < d$. Thus, a crossover from strong to weak chaos is expected to occur as time grows.

\subsubsection{Generalization to KG models}
\label{KGgenerals}
Due to the equivalence of KG and DNLS models (at least, within the small amplitude range) we expect the same regimes of the wave packet spreading. For the disordered KG chain (\ref{eq:1DdKG}) with cubic nonlinearity the nonlinear frequency shift for a single-site oscillator is proportional to its energy $\delta \propto \cE$. In order to compare the nonlinear frequency shift of the 1D dKG model with the scales of the 1D dDNLS \eref{eq:1DdDNLS}, we make use of the approximate mapping, which
in this case is $\beta S \approx 3W\cH_{\rm dK}$. Therefore, the dynamical regimes of wave packet spreading for 1D dDNLS model can be straightforwardly adapted for the dKG chain (in this case, for the spreading of the wave packet with initial energy density $\cE$):
\begin{eqnarray}
\cE \tilde{L} < d/3 &\qquad \mbox{weak chaos}, \nonumber \\ 
\cE \tilde{L} > d/3 &\qquad \mbox{strong chaos}, \label{eq:KGRegimes} \\
\cE \; \; \: > \Delta/3  &\qquad \mbox{self-trapping}, \nonumber
\end{eqnarray}	   
where $\tilde{L}=L/V$ for $L<V$ and $\tilde{L}\approx 1$ for $L\geq V$. As well as for DNLS equations, the average spacing between the eigenvalues can be approximated by $d=\Delta/3.3\xi_0$, where $\xi_0\approx 96 W^{-2}$ is the maximal localization length in the weak-disorder approximation and $\Delta=1+4/W$ is the width of eigenvalues spectrum of dKG linear problem. We do not show the parametric space of the above regimes, since it can be obtained from DNLS analog (cf. \fref{fig:ParSpace}) by the small-amplitude mapping, i.e. simply by dividing the regimes boundaries by $3W$.

\begin{figure}[ht]
\begin{center}
\includegraphics[width=0.65\columnwidth,keepaspectratio,clip]{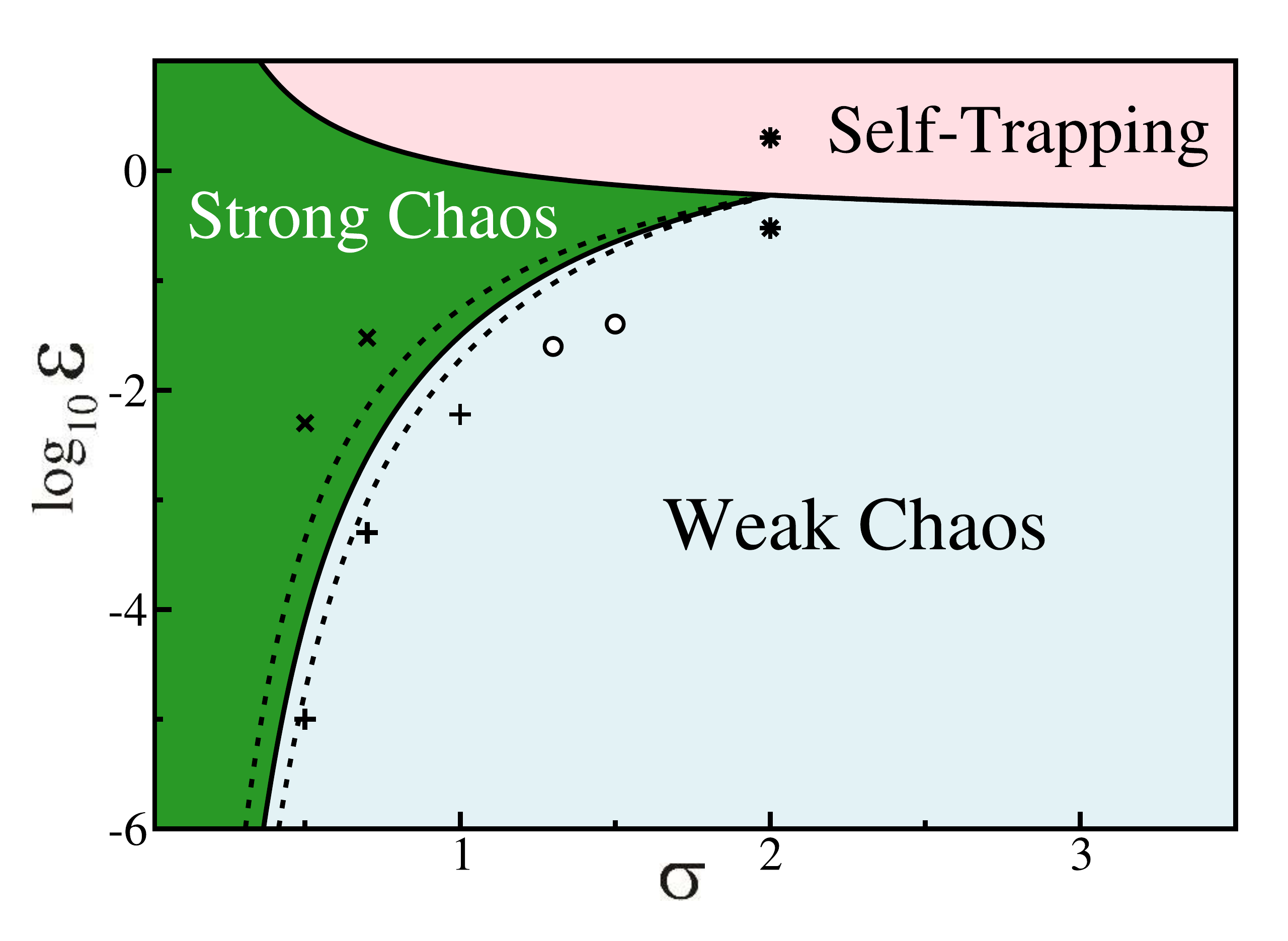}
\caption{Dynamical regimes for the 2D gKG model (\ref{eq:2DgHKG}) in the parameter space $\sigma$ and $\cE$ of a single site excitation ($L=1$, $W=10$). The different symbols correspond to the particular numerical simulations discussed in \sref{2dKG}. Dashed lines
show the variation of the the boundary obtained from variation of localization volume $V$.} 
\label{fig:gKGparspace}
\end{center}
\end{figure}

In a similar way, we can adapt the spreading regimes of the gDNLS model to the gKG model using the mapping formula (\ref{eq:Gmap}). Approximating the nonlinear frequency shift of a single-oscillator in gKG lattice as $\delta \propto \cE^{\sigma/2}$, we obtain
\begin{eqnarray}
(\cE\tilde{L})^{\sigma/2} < d/a_\sigma &\qquad \mbox{weak chaos}, \nonumber \\ 
(\cE\tilde{L})^{\sigma/2} > d/a_\sigma &\qquad \mbox{strong chaos}, \label{eq:gKGRegimes} \\
\cE^{\sigma/2} \ \ \ \ > \Delta/a_\sigma &\qquad \mbox{self-trapping}. \nonumber
\end{eqnarray}
Here, for the 1D gKG lattice the values of $d$ and $\Delta$ are the same as mentioned above, while for the 2D case $\Delta=1+8/W$. For a single-site excitation ($L = 1$) of energy $\cE$ expanding in the 2D gKG lattice, the regime boundaries (\ref{eq:gKGRegimes}) are shown in \fref{fig:gKGparspace}.

The quasi-periodic DNLS \eref{eq:qDNLS} also has its qKG counterpart (\ref{eq:qKG}).
In order to compare the nonlinear frequency shift $\delta \propto \cE$ of the qKG model with the energy scales of the qDNLS model, we use the approximate mapping $\beta S \approx 6 \vartheta \cH_{\rm qK}$. The quantity that is plotted in \fref{fig:QDNSEregimes} for the qKG model (upward red triangles) is $6\vartheta\cE$.

\subsection{Testing the theories}
\label{NR}
\subsubsection{Subdiffusion of nonlinear waves in 1D disordered lattices}
\label{S1DDL}

The numerical verification of spreading theory requires high precision computational methods for integration
and analysis \cite{Laptyeva2010,Bodyfelt2011}. For both models (\ref{eq:1DdDNLS}), (\ref{eq:1DdKG}) we considered initial distributions of width $L$ with: constant internal norm density $n=1$ (or energy density $\cE=\cH_{\rm dK}/L$ for the gKG) and zero outside this interval; a random phase at each site.  In the KG case one sets initial momenta equal to $p_l=\pm\sqrt{2\cE}$ with randomly assigned signs.

Equations (\ref{eq:1DdDNLS}), (\ref{eq:1DdKG}) were integrated by SABA-class symplectic schemes \cite{Laskar2001,Skokos2009}, with time-steps order $10^{-2}-10^{-1}$ up to a maximum time $t=10^{7} -10^{9}$. The particular scheme used for dDNLS is described in \cite{Bodyfelt2011}. The number of lattice sites was varied between $N=1000$ to $N=2000$ in order to exclude finite size effects in the wave packet evolution. In all simulations, the relative tolerance of the norm and energy conservation stayed below $0.1 \%$. 

For each set of parameters, we followed the dynamics of the three quantities $m_2(t)$, $P(t)$, and $\zeta(t)$, averaging the results over $10^3$ disorder realizations (unless otherwise stated). 
The averaged data $\langle\log_{10}m_2\rangle$ and $\langle\log_{10}P\rangle$ were additionally smoothed with a locally weighted regression algorithm \cite{Cleveland1988}, which allows to calculate local derivatives on log-log scales with central finite differences 
\begin{equation}
\alpha_m=\frac{d \langle \log_{10} m_2 \rangle}{d \log_{10} t}, \,\,\, \alpha_P=\frac{d \langle \log_{10} P \rangle}{d \log_{10} t}.
\label{eq:deriv}
\end{equation}

\begin{figure*}
\begin{center}
\includegraphics[width=0.9\columnwidth,keepaspectratio,clip]{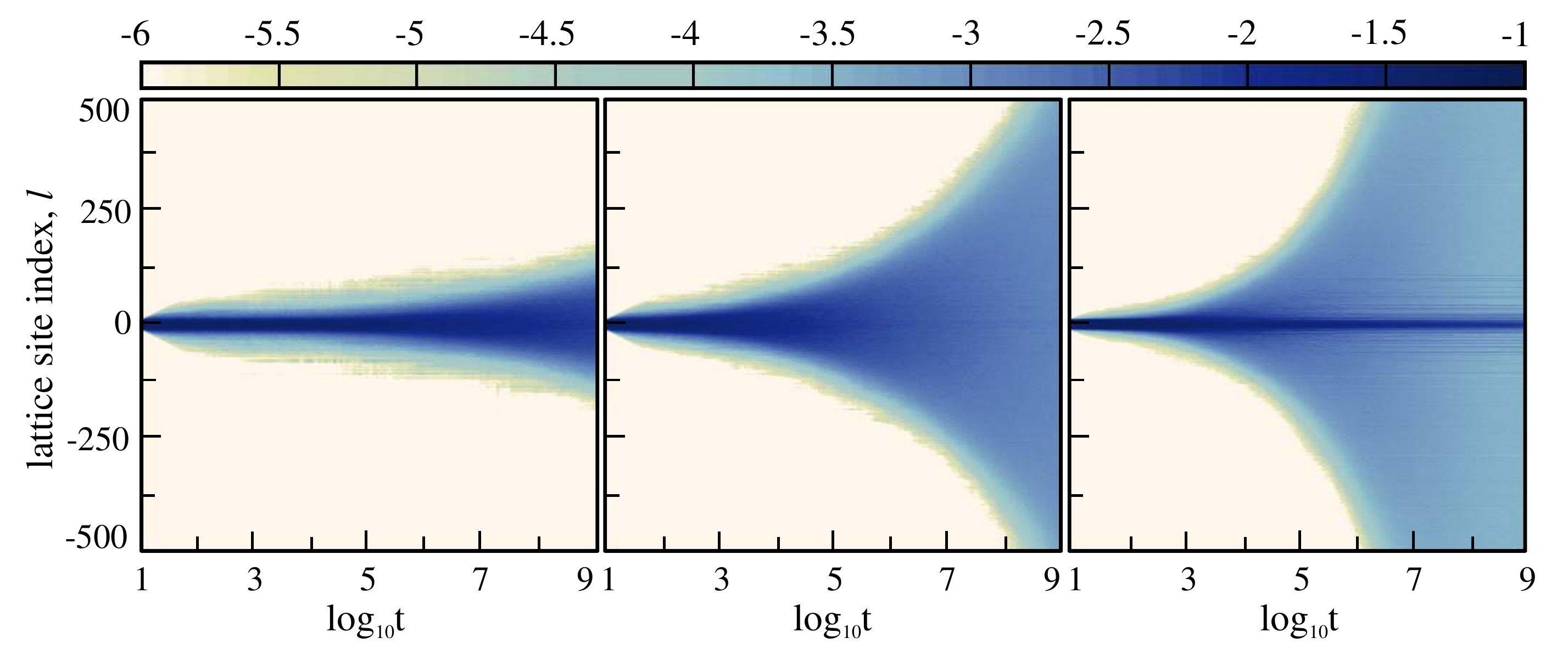}
\caption{Wave packet spreading in the dKG \eref{eq:1DdKG} with $W=4$ and $\cE=0.01$ (left), $\cE=0.2$ (middle), $\cE=3.0$ (right). The color code shows the real space averaged energy density distributions $\langle z_l \rangle $ on  logarithmic time and linear space scales.}
\label{fig:fig_DNLS_KG_W4_profiles}
\end{center}
\end{figure*}

\Fref{fig:fig_DNLS_KG_W4_profiles}(a)-(c) illustrates the dynamics of the averaged energy density distributions $\langle z_l \rangle $ in real space for three typical cases: $\cE=0.01$ (weak chaos), $\cE=0.2$ (strong chaos) and $\cE=3.0$ (self-trapping). The evolution starts from the same initial profile with size $L=V$. In the weak chaos regime (\fref{fig:fig_DNLS_KG_W4_profiles}(left)), the wave packets remain close to their initial configuration for some time, followed by delocalization. At $t=10^7$ the wave packet has spread to about $250$ sites, which is an order of magnitude larger than the localization volume set by the linear theory of Anderson localization. In the case of strong chaos (\fref{fig:fig_DNLS_KG_W4_profiles}(middle)), the spreading is considerably faster and leads to final time profiles spanning more than $1000$ sites. In the self-trapping regime (\fref{fig:fig_DNLS_KG_W4_profiles}(right)), the spreading part of the wave packet covers $1000$ sites about $100$ times faster, with another clearly visible part staying self-trapped at the initial excitation region. 

\begin{figure}
\begin{center}
\includegraphics[width=0.8\columnwidth,keepaspectratio,clip]{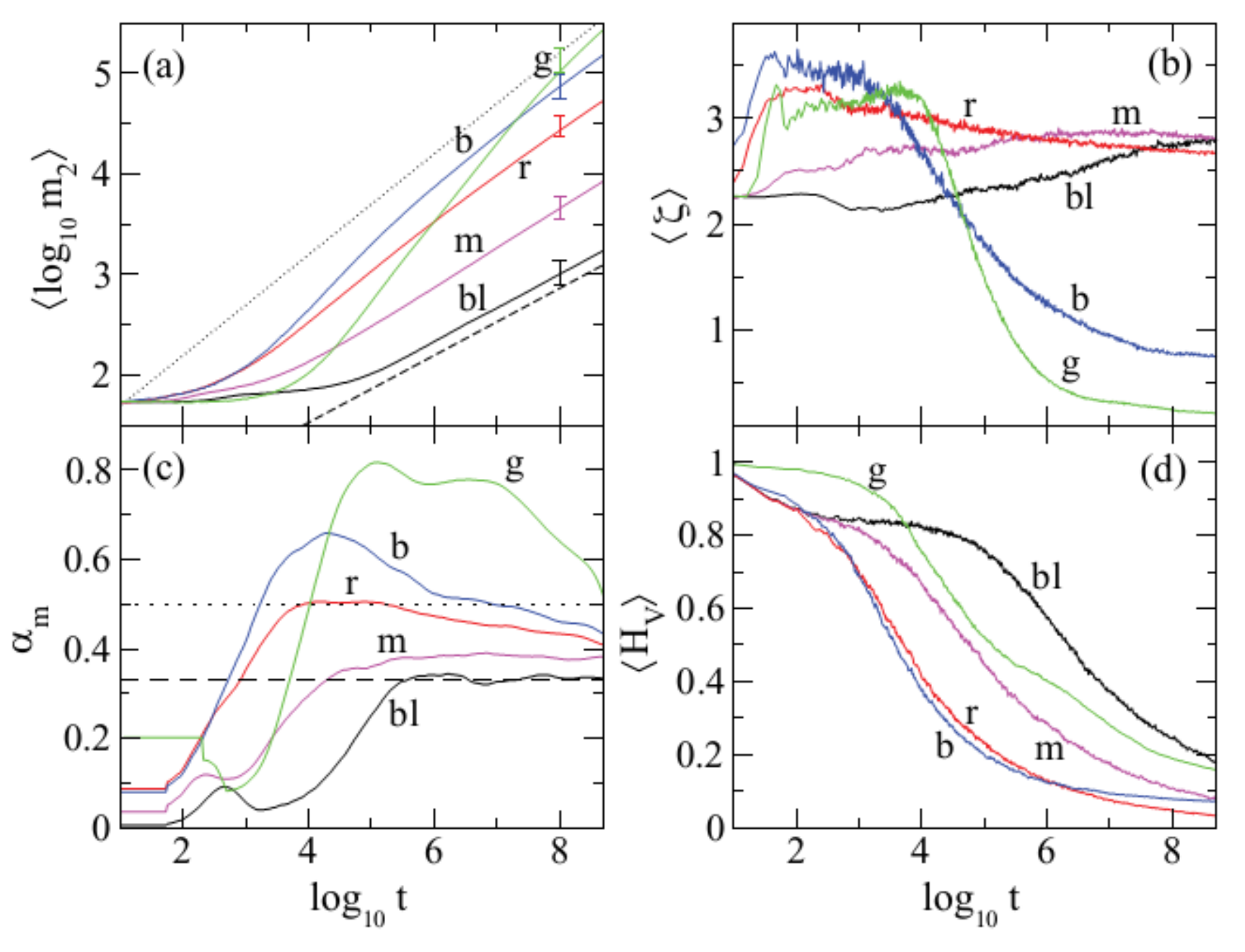} 
\caption{Numerical results for the dKG chain (\ref{eq:1DdKG}) with initial parameters $W=4$, $L=21$ and $\cE = 0.01, 0.04, 0.2, 0.75, 3.0$ (black (bl), magenta (m), red (r), blue (b), green (g)). Evolution of (a) $\langle \log_{10}m_2 \rangle$, (b) $\langle \zeta \rangle$, (c) $\alpha_m$, and (d) $\langle \cH_V \rangle$ are shown \textit{vs.} $\log_{10}t$. The straight lines in panels (a), and (c) correspond to the theoretically predicted power laws $m_2 \propto t^{\alpha_m}$ with $\alpha_m=1/3$ (dashed lines) and $\alpha_m=1/2$ (dotted lines). Error bars in panel (a) denote standard deviation errors. Reprinted from \cite{Bodyfelt2011}.}
\label{fig:fig_KG_4}
\end{center}
\end{figure}

The results of the quantitative analysis are presented in \fref{fig:fig_KG_4}.
For small initial energy density $\cE=0.01$, the characteristics of the weak chaos regime are observed. After a transient time $\tau_0 \approx 10^5$, we find a subdiffusive growth of $m_2$ according to $m_2 \propto t^{\alpha_m}$ with $\alpha_m \approx 1/3$.  The wave packets stay thermalized as they spread since $\langle \zeta \rangle \approx 3$ (\fref{fig:fig_KG_4}(b)), and the fraction $\langle \cH_V \rangle$ of the energy remaining in the initially excited region decreases (\fref{fig:fig_KG_4}(d)). For $\cE=0.04$ we enter the crossover region between weak and strong chaos: the spreading exponents $\alpha_m \approx 0.38$ lie between $1/3$ and $1/2$ and show slow
time dependence with a tendency to decrease towards the weak chaos value as time progresses.

The typical strong chaos scenario was observed for $\cE=0.2$: subdiffusive growth $m_a \propto t^{\alpha_m}$ with $\alpha_m \approx 1/2$ observed for about two decades $3.5 \lesssim \log_{10}t \lesssim 5.5$, and followed by a crossover to the weak chaos dynamics with $\alpha_m$ decreasing. 

\begin{figure}
\begin{center}
\includegraphics[width=0.6\columnwidth,keepaspectratio,clip]{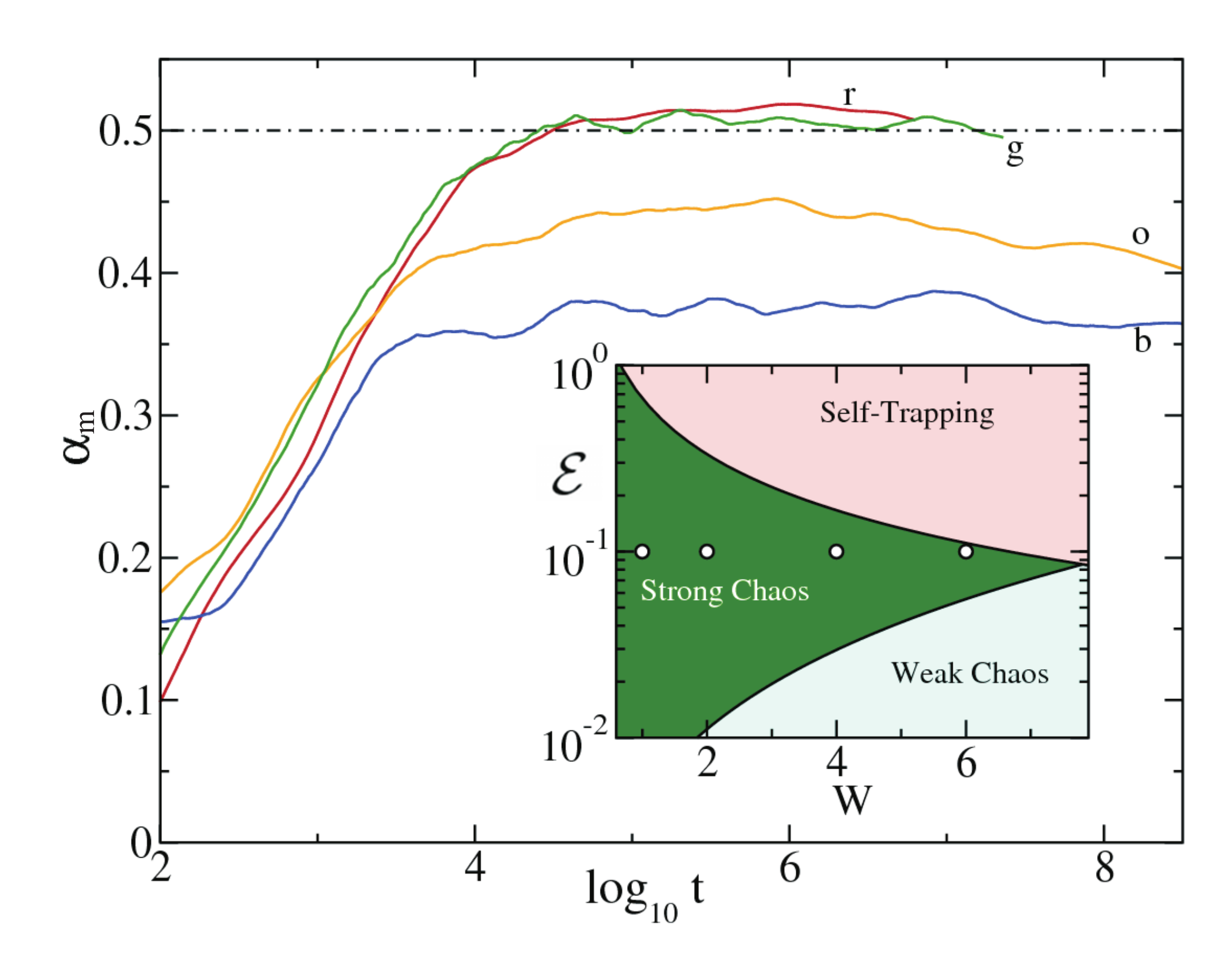}
\caption{Spreading behavior in the strong chaos regime for the dKG model (\ref{eq:1DdKG}) with initial energy density $\cE =0.1$: time behavior of $\alpha_m$ is shown \textit{vs.} $\log_{10}t$. The four curves are for the disorder strengths $W=1$, $W=2$, $W=4$, and $W=6$ (red (r), green (g), orange (o), blue (b)). \textit{Inset}: the dKG analog of the dDNLS parametric space (see \fref{fig:ParSpace}). The four points show the parameters used in the simulations presented in the main figure. Reprinted from \cite{Laptyeva2010}.}
\label{fig:SC_long}
\end{center}
\end{figure}

The duration of the strong chaos regime $\alpha_m = 1/2$ can be increased by simply reducing the strength of disorder $W$ in order to increase the distance to the weak chaos and selftrapping crossovers. This is illustrated in \fref{fig:SC_long} for the dKG chain. For $W=1, 2$ a long-lasting spreading in the strong chaos regime is clearly observed. For $W=4, 6$ the strong chaos window in the energy density is small (see inset in \fref{fig:SC_long}) and, even if initial conditions belong to it, the energy density decrease in the course of spreading will quickly transfer the dynamics into the weak chaos regime, $\alpha_m < 1/2$.

Getting closer to the self-trapping regime for $\cE=0.75$ or being deep inside it for $\cE=3.0$ one observes its typical features: the compactness index $\langle \zeta \rangle $ decreases (\fref{fig:fig_KG_4}(b)), and $\langle \cH_V \rangle$ tends to stabilize at small but finite non-zero values (\fref{fig:fig_KG_4}(d)). Besides, $m_2$ (\fref{fig:fig_KG_4}(a)) manifests initially the fast growth with $\alpha_m > 1/2$ (\fref{fig:fig_KG_4}(c)) followed by a decrease in $\alpha_m$ value.

The presented results convincingly support the predictions of weak and strong chaos, as well as the dynamical crossover between them. Self-trapping is observed as well, with a less understood dynamics of the trapped and spreading portion of norm/energy. We stress that none of the simulations exhibits a pronounced deviation or significant slowing down to values $\alpha_m < 1/3$. Qualitatively same results were obtained for the dDNLS model, including the case of strong disorder $W\in [15,40]$ \cite{Bodyfelt2011}.

As an additional test of the theory two cases with spatially inhomogeneous nonlinearity in dKG chain (\ref{eq:1DdKG}) have been studied \cite{Bodyfelt2011}. The first one had a non-zero nonlinearity in the central region of length $L=V$ only. Following theoretical predictions the wave packets evolved chaotically and spread subdiffusively only within the nonlinear part. 
Since the NMs with mass-centers far in the linear parts interact with the chaotic NMs in the inner nonlinear part exponentially weak, their excitation takes times that increase exponentially with growing distance, causing a slowing down of the wave packet spreading. 
In the second case the inner part of the chain is linear and the outer part is nonlinear. As long as most of the wave packet stays in the linear region, the dynamics remains close to regular and no spreading beyond the localization volume occurs for substantial times. However, after these large transient times parts of the packet leak into the nonlinear regions and the weak chaos spreading sets in again.

\subsubsection{Subdiffusion in 1D generalized DNLS and KG models}
\label{S1DGM}
\begin{figure}[ht]
\begin{center}
\includegraphics[width=0.7\columnwidth,keepaspectratio,clip]{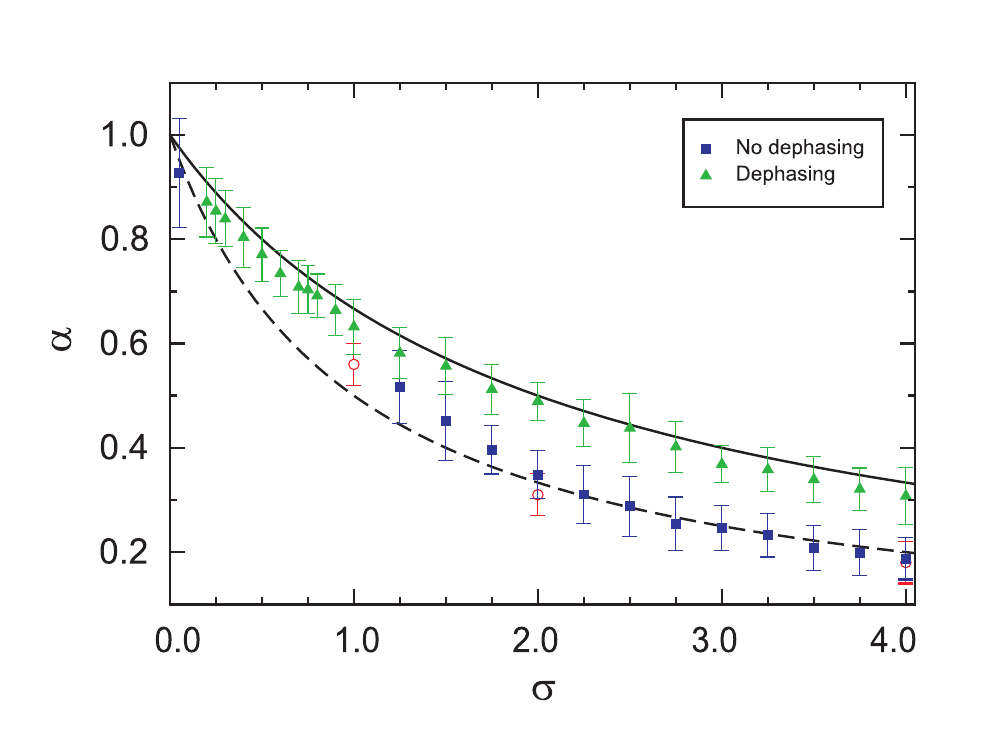}
\caption{Numerical results for 1D gKG model (\ref{eq:1DgHKG}): spreading exponent $\alpha_m$ \textit{\textit{vs.}} the nonlinearity index $\sigma$ for integration without dephasing (filled squares) and with dephasing of NMs (filled triangles). The theoretically predicted functions $\alpha_m = 1/(1+\sigma)$ for weak chaos and $\alpha_m = 2/(2+\sigma)$ for strong chaos are plotted by dashed and solid lines, respectively. Reprinted from \cite{Skokos2010}.}
\label{fig:SkokosTN}
\end{center}
\end{figure}

The numerical evidence for the validity of predictions (\ref{eq:gKGRegimes}) was presented in \cite{Skokos2010}, where
the evolution of single-site excitations ($L=1$) in gKG chain (\ref{eq:1DgHKG}) was addressed. The simulations were done for different integer and non-integer values of the nonlinearity index $\sigma$ and energies away from the self-trapping
regime. We show the main outcome of these results in \fref{fig:SkokosTN}.  The second moment of the wave packet is expected to increase in time according to the power law $m_2 \propto t^{\alpha_m}$. In these cases, the participation number follows the law $P \propto t^{\alpha_P}$ with $\alpha_P=\alpha_m/2$. For the case of the gKG chain, the dependence of the exponent $\alpha_m$ on the nonlinearity index $\sigma$ (see the general predictions (\ref{sigma_strong}),(\ref{sigma_weak})) was suggested to be \cite{Flach2010}:
\begin{eqnarray}
\alpha_ m= 
           1/(1+\sigma) \;,\; & \mbox{weak chaos}, \nonumber \\
~ \label{eq:1DgKGpow} \\
\alpha_ m=	         2/(2+\sigma) \;,\; & \mbox{strong chaos}. 
\nonumber
\end{eqnarray}

According to the estimates (\ref{eq:gKGRegimes}), a single-site excitation, if it is not self-trapped, belongs to the weak chaos regime for $\sigma \geq 2$. Additionally, the strong chaos regime can be entered for $\sigma < 2$. Indeed for $\sigma \geq 2$, the computed exponents are in good agreement with the theoretical prediction $\alpha=1/(1+\sigma)$ for weak chaos regime of subdiffusion (see dashed line in \fref{fig:SkokosTN}). 
The open red diamond data were obtained by Mylansky for the DNLS chain \cite{Mulansky2009}. It is instructive
that the results coincide witht the KG data, in particular e.g. for the value $\sigma=4$. 

Note that the exponents were obtained from power law fits in \cite{Skokos2010} and not using local derivatives. For $\sigma < 2$ 
the initial state happens to be deeper and deeper in the strong chaos regime. Therefore, the fits result in
numbers which tend towards the strong chaos values as $\sigma$ decreases. 
We stress that many other attempts to fit $m_2(t)$ with power laws suffer from the same deficiency - if the initial
state is launched in the strong chaos regime, the fits will produce exponent values located between the strong
and weak chaos values, which is due to the actual slow crossover dynamics. Instead the above described computations
of local derivatives do not suffer from the same deficiencies, and yield a much clearer picture of the studied
dynamics.

For $\sigma < 0.1$ the evolution of the wave packet in \cite{Skokos2010} shows even a destruction
of Anderson localization in the tails of the spreading packet, which is a phenomenon yet to be explored
analytically.

Note that for $\sigma \rightarrow 0$ both regimes yield normal diffusion. At the same time, the system dynamics should approach the behavior of the linear system ($\sigma=0$) which is characterized by Anderson localization and therefore absence of diffusion. This is possible, since the prefactor in $m_2\propto t^{\alpha_m}$ also depends on $\sigma$. As it was shown in \cite{Skokos2010}, the prefactors indeed tend to zero, leading to a vanishing of the corresponding diffusion constant.

\subsubsection{Subdiffusion in 2D disordered latices with tunable nonlinearity}
\label{2dKG}
The interplay between disorder and nonlinearity in 2D arrays  was first addressed numerically in \cite{GarciaMata2009}. However, the computational challenges precluded detailed studies and only more
recently systematic results have been obtained \cite{Laptyeva2012}.

For 2D gKG lattice (\ref{eq:2DgHKG}) the predicted regimes of subdiffusion from single-site excitations (\ref{eq:gKGRegimes}) are
\begin{eqnarray}
(\cE/V)^{\sigma/2} < d/a_\sigma &\qquad \mbox{weak chaos}, \nonumber \\ 
(\cE/V)^{\sigma/2} > d/a_\sigma &\qquad \mbox{strong chaos}, \label{eq:2DgKGRegimes} \\
\cE^{\sigma/2} \ \ \ \ \ \ > \Delta/a_\sigma &\qquad \mbox{self-trapping}. \nonumber
\end{eqnarray}
The second moment of the wave packet is expected to grow subdiffusively in time as $m_2 \propto t^{\alpha_m}$ with $\alpha_m<1$ and participation number as $P\propto t^{\alpha_P}$, $\alpha_P=\alpha_m$. The theoretically predicted dependence of the spreading exponent $\alpha_m$ on the nonlinearity index $\sigma$ (\ref{sigma_strong}),(\ref{sigma_weak}) is 
\begin{eqnarray}
\alpha_m = 
1/(1+2\sigma) \;,\;
& 
\; \mbox{weak chaos}, \nonumber \\
~ \label{eq:GenPowers_ss} \\
\alpha_m =
2/(2+ 2\sigma) \;,\; &  \mbox{strong chaos}. 
\nonumber
\end{eqnarray}

To test the predictions (\ref{eq:GenPowers_ss}),  the disorder strength $W=10$ was chosen, for which the 2D localization volume $V \approx 34$. Then varying the values of the energy density $\cE$ and nonlinearity power $\sigma$ one can probe the expected spreading regimes (\ref{eq:2DgKGRegimes}) as shown in \fref{fig:gKGparspace}. This also includes non-integer values of $\sigma$ and those $\sigma$ values at which the theory suggested an anomaly in the number of wave packet surface resonances (cf. \sref{GTHDAN} for  details), which would grow in the course of spreading. Initial states were single site excitations.


Numerics largely confirms the theoretical predictions, e.g. for $\sigma=2$ the numerical weak chaos exponent $\alpha_m \approx 0.21$ compares well with its theoretical prediction $\alpha_m=1/5$ \cite{Laptyeva2012}. Large integration times (up to $t=10^8$) are required to capture the asymptotic power-law, while insufficient times result in an overestimate \cite{GarciaMata2009}. Lowering the nonlinearity index  one also finds agreement for the weak chaos exponents for $1<\sigma<2$  (markers \quotes{$\circ$} in \fref{fig:gKGparspace}). Strong chaos spreading from single site excitations was also observed for $\sigma<2$ (markers \quotes{$\times$} in \fref{fig:gKGparspace}) with the expected spreading exponents.

\begin{figure}[ht]
\begin{center}
\includegraphics[width=0.7\columnwidth,keepaspectratio,clip]{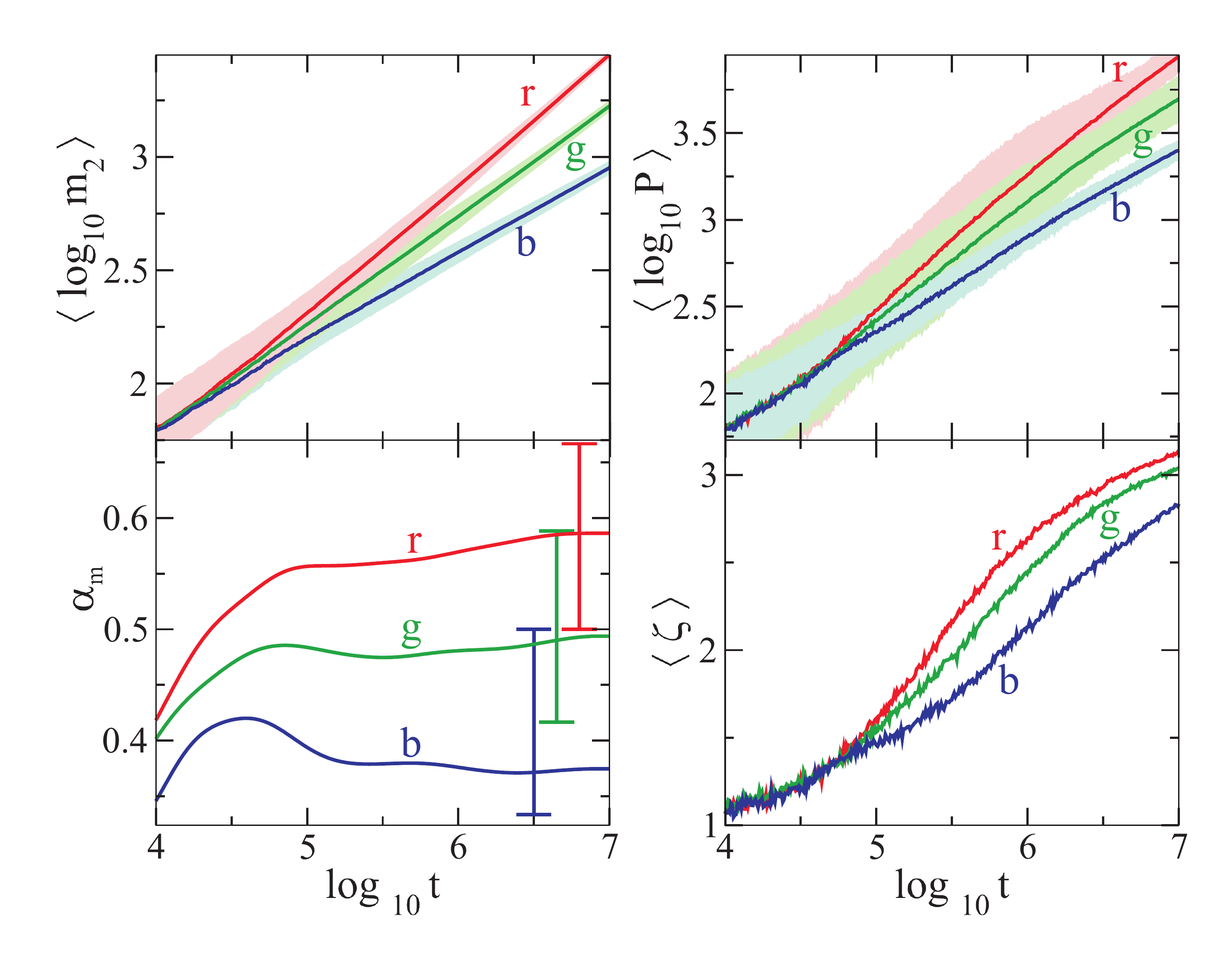}
\caption{Numerics for the parameter set marked by \quotes{$+$} in \fref{fig:gKGparspace}. The curves for parameters $(\sigma,\cE)=(0.5, 0.00001),(0.7, 0.0005),(1.0, 0.006)$ are colored respectively in red (r), green (g), and blue (b). \textit{Left column}: $\left\langle \log_{10}m_2\right\rangle$ (top) and its power-law exponent $\alpha_m$ (bottom) \textit{vs.} $\log_{10}t$. Similarly, \quotes{I}-bar bounds denote the theoretical expectations for weak chaos (lower bound) and strong chaos (upper bound). \textit{Right column}: $\left\langle \log_{10}P \right\rangle$ (top) and $\left\langle \zeta \right\rangle$ (bottom) \textit{vs.} $\log_{10}t$. In both columns of the upper row, the lighter clouds correspond to a standard deviation.}
\label{fig:sig_le1}
\end{center}
\end{figure}

Wave packet snapshots taken at the largest attainable times in different regimes of spreading indicate only a quantitative difference in their volumes, while the rough structure of their surfaces appears qualitatively the same (cf. \fref{fig:sig_le1}). 

\begin{figure}[ht]
\begin{center}
\includegraphics[width=0.9\columnwidth,keepaspectratio,clip]{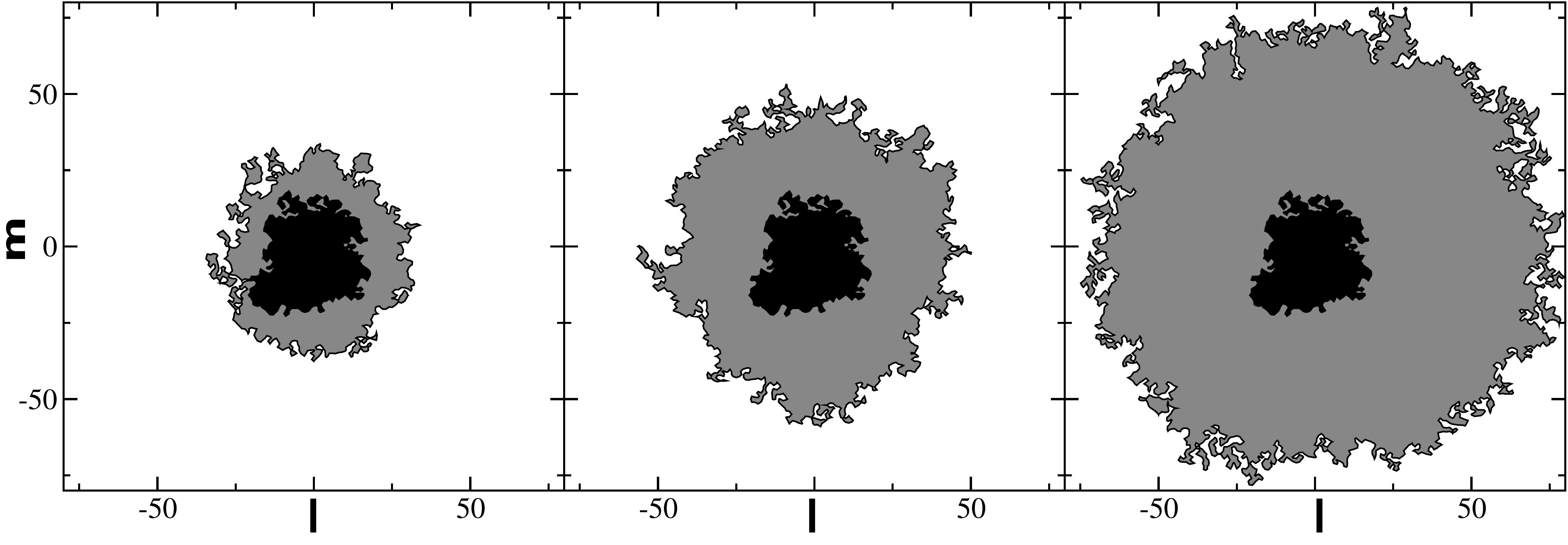}
\caption{Wave packet contours at $t=10^6$, the inner part defined as the set of sites with the local energy exceeding $10^{-5}$ . The black area in the center is the contour for the spreading in linear equations, while the gray region corresponds to subdiffusion in the following regimes (from left to right): weak chaos for $(\sigma,\cE)=(1.5,0.04)$, intermediate spreading for $(\sigma,\cE)=(0.5,0.00001)$, and strong chaos for $(\sigma,\cE)=(0.5,0.005)$.} 
\label{fig:waves}
\end{center}
\end{figure}

\subsubsection{Subdiffusion of nonlinear waves in quasi-periodic potentials}
\label{NAAnum}
Numerical \cite{Larcher2009,Larcher2012} and experimental \cite{Lucioni2011} studies show that a compact initial excitation in a 1D lattice with quasi-periodic potential spreads subdiffusively with its second moment growing as $m_2 \propto t^{\alpha_m}$ with $\alpha_m < 1$. 

According to the most recent and detailed results \cite{Larcher2012}, three dynamical spreading regimes are observed, analogously to purely disordered systems \cite{Skokos2009,Flach2009,Laptyeva2010,Bodyfelt2011,Flach2010}. If the nonlinear frequency shift exceeds the width of the linear spectrum, at least a major part of the wave packet is self-trapped. Otherwise, two outcomes are possible. If the nonlinear frequency shift is less than the average frequency spacing, the wave packet spreads in the weak chaos regime with the asymptotic divergence of the second moment $m_2 \propto t^{\alpha_m}$ with $\alpha_m=1/3$. For nonlinear frequency shifts larger than the average spacing, the wave packet temporarily evolves in the regime of strong chaos characterized by $m_2 \propto t^{\alpha_m}$ with $\alpha_m=1/2$, with the inevitable crossover to the asymptotic weak chaos later, when its density drops down. Let us reiterate that in the quasi-periodic case partial self-trapping is possible even for arbitrary weak nonlinearities due to the complicated multi-band structure of its linear spectrum (cf. \sref{QUASIgenerals}). Similar to random systems, this may give rise to larger transient exponents. This effect is an inherent property of quasi-periodic systems which inevitably manifests itself in all spreading regimes, while in the disordered case it was shown to occur only in the self-trapping regime \cite{Laptyeva2010,Bodyfelt2011}.

Due to the existence of an approximate mapping between qKG and qDNLS models, we expect to 
observe much the same spreading characteristics in both models. This has been already confirmed for purely random systems, where both dDNLS and dKG models reveal qualitatively similar results \cite{Skokos2009, Flach2009, Laptyeva2010, Bodyfelt2011,Flach2010}. The following numerical data are therefore shown for qKG chain (\ref{eq:qKG}) only.

\begin{figure}
\begin{center}
\includegraphics[width=0.75\columnwidth,keepaspectratio,clip]{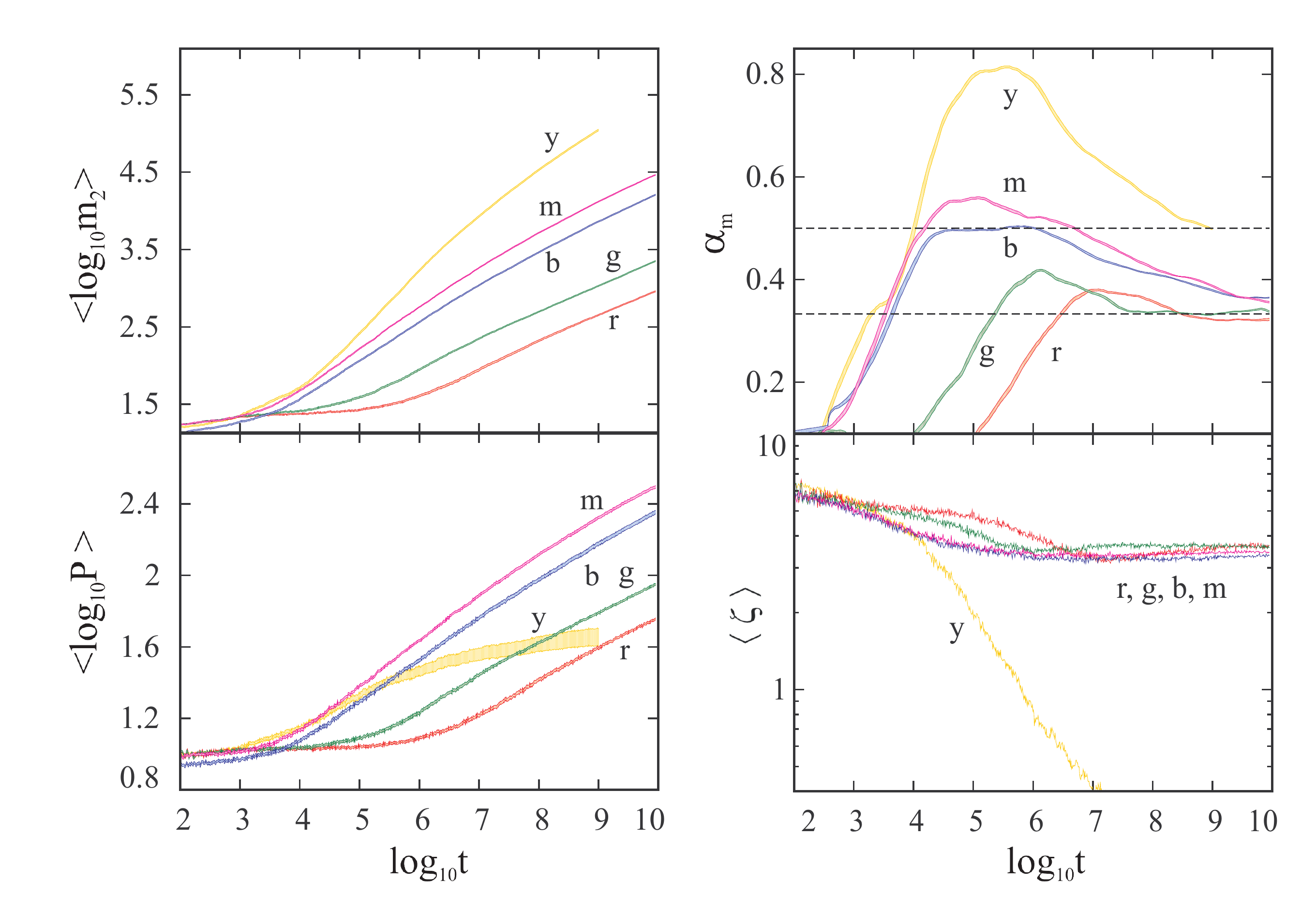}
\caption{Numerical integration of the qKG equations of motion (\ref{eq:qKG}): evolution of $\left\langle \log_{10} m_2\right\rangle$ (left panel, top), $\alpha_m$ (right panel, top), $\left\langle \log_{10} P \right\rangle$ (left panel, bottom), and $\left\langle \zeta \right\rangle$ (right panel, bottom) is shown \textit{vs.} $\log_{10}t$ for the spreading of wave packets of widths $L=13$ (red (r), green (g), magenta (m), and yellow (y)) and $L=11$ (blue (b)). The curves for parameters $(\vartheta, \cE)=(2.5, 0.005), (2.5, 0.01), (2.5, 0.055), (2.5, 0.075), (2.5, 1.0)$ are colored respectively in red (r), green (g), blue (b), magenta (m), and yellow (y). In the top right panel the two dashed lines correspond to the theoretically predicted $\alpha_m=1/3$ and $\alpha_m=1/2$.}
\label{fig:num_KG}
\end{center}
\end{figure}

The results of the simulations are shown in \fref{fig:num_KG} (see also \fref{fig:QDNSEregimes} for theoretically predicted diagram). After initial transients, which become longer with decreasing nonlinearity, all simulations reveal subdiffusive growth of the second moment $m_2$ according to power law $m_2 \propto t^{\alpha_m}$ with $\alpha_m<1$. If self-trapping is avoided, all simulations show a similar subdiffusive behavior for the participation numbers. Moreover, the wave packets remain compact as they spread since compactness 
indexes at the maximal computational times saturate around a constant value $\left\langle \zeta \right\rangle \approx 3.5 \pm 0.25$. 

For the two smallest values of initial energy density $\cE=0.05$ and $\cE=0.01$, the characteristics of the weak chaos regime are observed, namely, the exponent $\alpha_m$ saturates around  $1/3$ (red and green curves in \fref{fig:num_KG}) after a transient time. We stress, that the only difference from the purely random systems is the overshooting phenomenon at transient times, which is believed to be due to partial selftrapping in minigaps of
the spectrum, which is destroyed at later times.

For the two energy densities $\cE=0.055, 0.075$ the theory predicts the strong chaos behavior. The simulation with $\cE=0.055$ (blue curve in \fref{fig:num_KG}) exhibits the behavior typical of the strong chaos scenario: the characteristic exponent $\alpha_m$ increases up to $\alpha_m=1/2$ and keeps its value for about two time decades, followed by a crossover to the weak chaos dynamics with decreasing $\alpha_m$. However, there is also another possibility for larger $\cE=0.075$, when intermediate strong chaos is masked due to partial self-trapping. Here $\alpha_m$ attains values larger then $1/2$ but still with subsequent decay to slower subdiffusion. We would like to underline that none of the simulations exhibits a pronounced deviations from strong or weak chaos regimes of spreading, i.e. long-lasting  overshooting with $\alpha_m > 1/2$, or, significant slowing down to values $\alpha_m < 1/3$.

Finally, for $\cE=1.0$ the dynamics enters the self-trapping regime as the theory predicts. There a major part of an initial excitation stays localized, while the remainder spreads (yellow curves in \fref{fig:num_KG}). The participation numbers, therefore, do not grow significantly and $\left\langle\log_{10}P (t)\right\rangle$ levels off at large time (\fref{fig:num_KG}, left panel, bottom). In contrast, the non-selftrapped spreading portion contributes
to a continuous increase of the second moment $m_2$ (\fref{fig:num_KG}, left panel, top), which initially is characterized by large overshooting values of $\alpha_m > 1/2$. Due to selftrapping the compactness index $\left\langle\zeta\right\rangle$ (\fref{fig:num_KG}, right panel, bottom) drops down to small values, similar to purely random systems \cite{Laptyeva2010,Bodyfelt2011}. 

Numerical integration of qDNLS equations (\ref{eq:qDNLS})  shows similar results (cf. \cite{Larcher2012} for more details). For $\vartheta=2.5$ and small values of nonlinearity $\beta=0.5, 1.0$, the characteristics of the weak chaos regime $\alpha_m \approx 1/3$ were observed. Simulations with $\vartheta=2.5$ and $\beta=5,10$ reveals typical behavior of the strong chaos scenario: at intermediate times, spreading is characterized by a saturated $\alpha_m \approx 1/2$ for about two decades, followed by a crossover to the weak chaos dynamics with $\alpha_m$ decreasing. Finally, for $\vartheta=2.5$ and large nonlinearity $\beta=100$ self-trapping is observed.

Importantly, these results may provide a consistent interpretation of the experimental data \cite{Lucioni2011}, where the subdiffusive expansion of BEC in bichromatic optical lattice was observed with different diffusion exponents larger than $1/3$ even for weak nonlinearities (cf. \sref{NDMexp}). Within the short time window of the experimental observations, such values  $\alpha_m$ can be reasonably explained in terms of transient overshooting caused by partial self-trapping in mini-bands. Further numerical simulations of gDNLS equations (\ref{eq:qDNLS}) presented in \cite{Larcher2012} have also verified that the main results do not depend on the details of the shape of the initial wave packet.

\subsection{Heat conductivity}

So far we discussed the nonequilibrium process of wave packet spreading into an empty lattice.
Assuming this lattice has infinite size, the corresponding average energy and norm densities are zero. 
Nevertheless, the effective noise theory predicts the dependence of diffusion rates on the finite densities
inside the wave packet. Since the system size does not enter this dependence, we can use the effective noise theory
to predict the dependence of the diffusion rate and of corresponding conductivities on the average densities in an equilibrium state with finite average
densities. 

In linear systems heterogeneity-induced localization of modes yields zero conductivity.
Nonlinearity couples the normal modes and the resulting energy transfer can lead to intricate physics, as it has been recently discussed for the anomalous conductivity when a part of the linear normal modes remains extended in space \cite{Lepri2003,Dhar2008,Ivanchenko2011cond}. The case when all linear modes are localized demonstrates diffusive transport of energy and finite conductivity $\mathcal{K}$ \cite{Dhar2008a}. Its temperature dependence is 
however much less clear. Some numerical studies indicated that for low temperatures $\mathcal{K} \propto T^{1/2}$ \cite{Dhar2008a}, which would correspond to a singularity at zero temperatures. Other expectations claim that heat conductivity vanishes strictly for weak enough (but still finite) anharmonicity \cite{Dhar2008}. 

Assuming the validity of  effective noise theory, we arrive at the prediction that the heat conductivity of a thermalized system at small
temperature (density) must be proportional to the diffusion coefficient (\ref{ent5}) where the density $n$ is replaced by the temperature $T$. While one has to be careful 
in the DNLS case, where two conserved quantities (energy, norm) enforce Gibbs, or non-Gibbs distributions \cite{Basko2014}, the KG case might be 
again a better testing ground, where one conserved quantity (energy) can be expected to enforce a Boltzmann distribution.
\begin{figure}
\includegraphics[width=0.6\columnwidth]{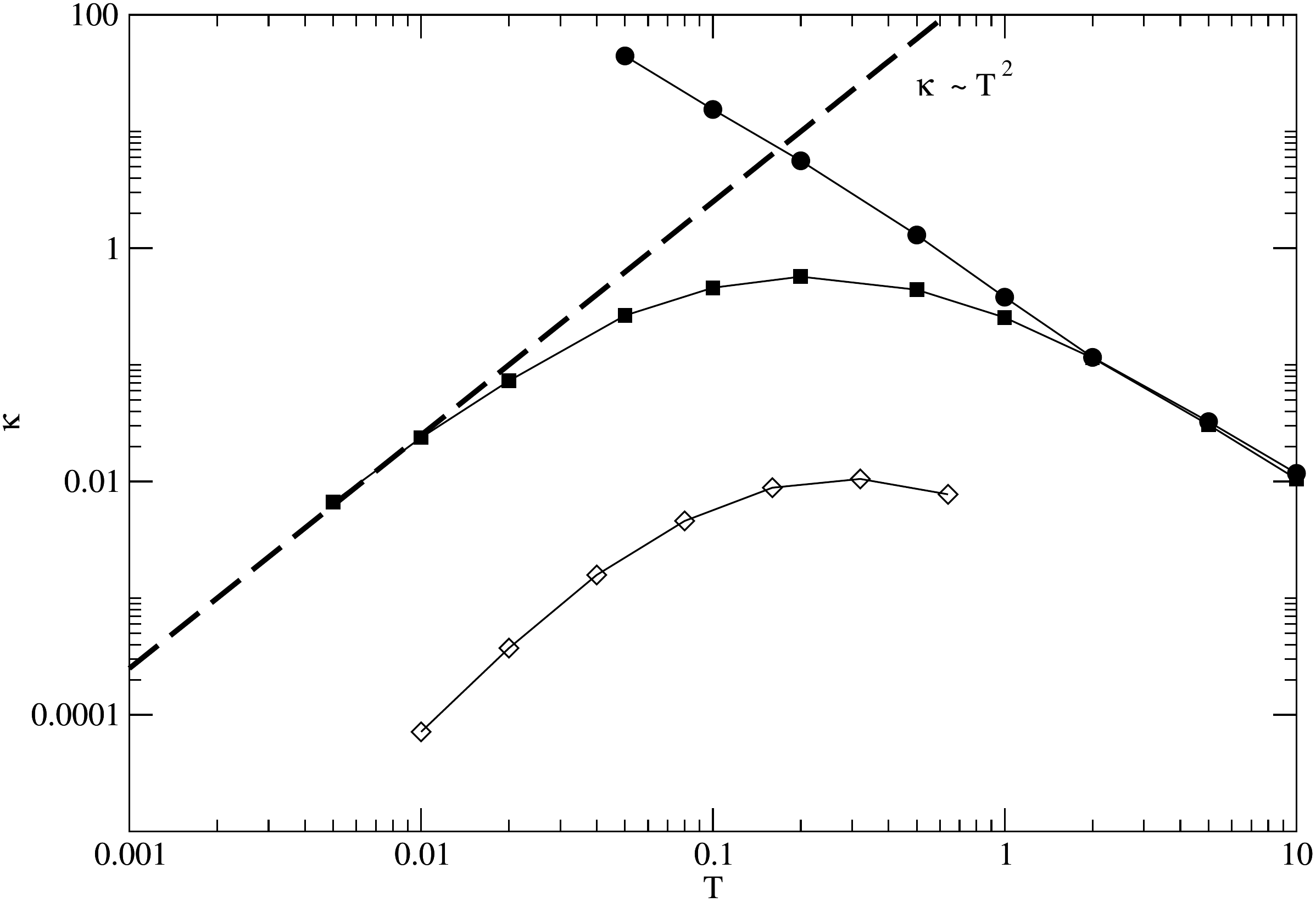}
\caption{KG chain: Heat conductivity $\kappa(T)$ for $W=2$ (filled squares). For comparison we also show the
data for the ordered case  $\tilde{\epsilon}_l\equiv 1$ (filled circles). 
Thin solid lines guide the eye.
The dashed line corresponds to the power law
$T^2$. 
The stronger disorder case $W=6$ corresponds to the open diamond data points.
Adapted from \cite{Flach2011}
}
\label{NEWfig13}
\end{figure}
The calculation of the heat conductivity for was performed in Ref.\cite{Flach2011}. Its dependence on the temperature is shown
in Fig.\ref{NEWfig13}. The strong chaos scaling $\kappa(T)\sim T^2$ is observed nicely. The expected weak chaos regime was not reachable by the
heavily extensive numerical efforts. Note that the decay of the heat conductivity for large temperatures is due to selftrapping, and observed even for
the ordered chain at $W=0$ (solid circles in Fig.\ref{NEWfig13}).


\subsection{First experiments on nonlinear heterogeneous media}
\label{NDMexp}
Significant theoretical and computational achievements in the understanding of the destruction of heterogeneity-induced localization by nonlinearity for lattice wave dynamics have been obtained.
In contrast, there are only few related experimental studies related to interacting BEC in quasi-periodic potential \cite{Deissler2010,Lucioni2011}, and the propagation of light in disordered photonic lattices \cite{Pertsch2004,Schwartz07,Lahini08,Lahini2009} with attempts to observe the subdiffusive processes.
The main reason is the very subdiffusion process - a very slow dynamics which needs many decades in time/space
in order to resolve universal exponents. The mentioned experimental efforts, when properly translated into
dimensionless units, achieve effective evolution times of the order of $10^4$. This is clearly not enough.
The subdiffusive spreading sets in after a quick ballistic explosion around $t\approx 10^2$. To resolve an exponent
$1/3$, with requested two decades of variation on both quantitites, up to six additional decades in time are 
needed, pushing the dynamics to $t\approx 10^8$. While this time can be reduced when using proper averaging techniques, and proper methods of analysis (local derivatives), experience tells that still one needs $t\approx 10^6$
at least. Therefore experiments still lack at least two decades. That is a challenging issue, since phase coherence of
the waves is absolutely needed to avoid slipping into a trivial regime of normal diffusion. 

\begin{figure}[h]
\begin{center}
\includegraphics[width=0.9\columnwidth,keepaspectratio,clip]{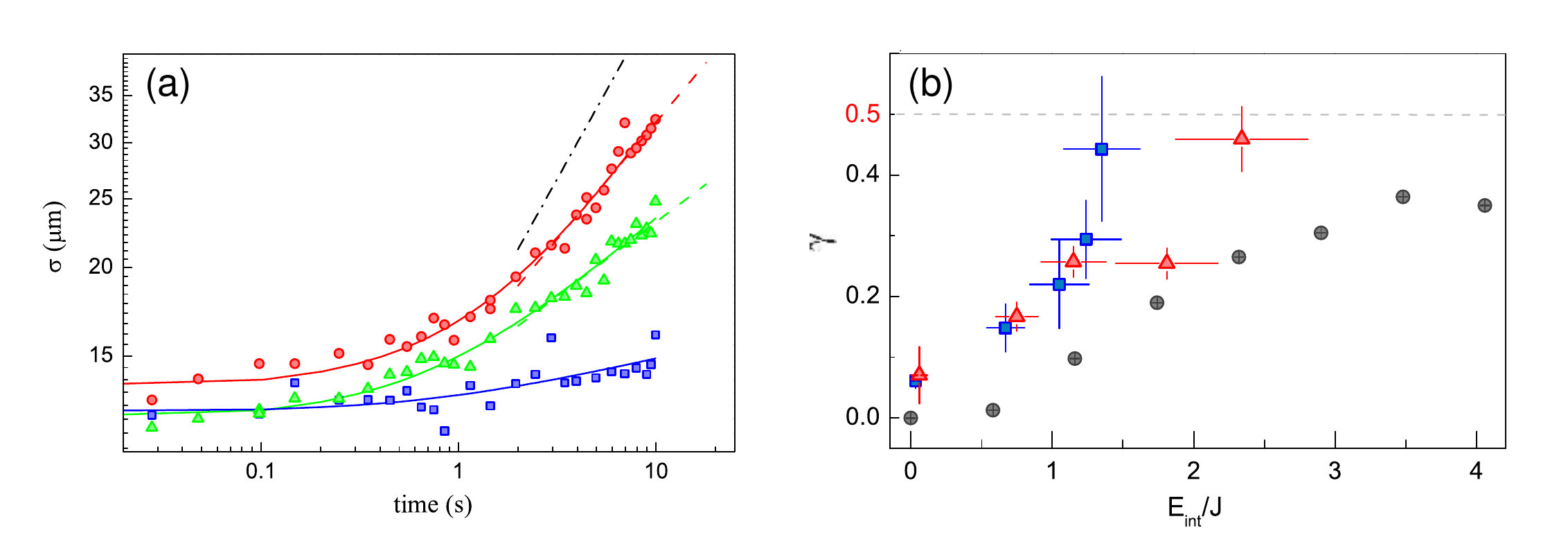}
\caption{Measurements of the spreading of an interacting BEC in a quasi-1D incommensurate lattice. Panel (a) shows the
time evolution of the width $\sigma=\sqrt{m_2}$ of the atomic cloud: squares correspond to the case of non-interacting BEC, triangles and circles -- to the cases of interacting BEC with different initial interaction energies. The continuous lines are the fit $\sigma(t) \propto t^\gamma$. The dash-dotted line shows the expected behavior
for normal diffusion ($\gamma=0.5$). Panel (b) shows the spreading exponent $\gamma$ fitted from the $\sigma(t)$ dependences (see panel (a)) versus the initial interaction energy $E_{\rm int}$ in the experiment (triangles and squares) and simulations of 1D quasi-periodic DNLS (circles) \cite{Larcher2009}. The vertical bars are the fitting error to the data, while the horizontal bars indicate the statistical error. Adopted from \cite{Lucioni2011}.} 
\label{fig:Lucioni}
\end{center}
\end{figure}

Still the experiments do demonstrate the delocalizing effect of nonlinearity {\it per se}. Particularly, \fref{fig:Lucioni} illustrates the results of an expansion experiment of a  BEC cloud of $^{39}K$ atoms with controllable nonlinearity along 1D bichromatic optical lattice. For non-interacting BEC no transport is expected, due to Anderson localization of all single-particle eigenstates. In the experiment (\fref{fig:Lucioni}a, squares), only an extremely slow expansion was observed, presumably, due to parasitic noise. Addinional  repulsive atom-atom interactions lead to 
an expansion of the BEC. The width of atomic cloud was found to grow subdiffusively according to
$\sigma(t) \propto t^\gamma$ with $\gamma \approx 0.2 - 0.4$ (\fref{fig:Lucioni}a, triangles and circles). The measured exponents $\gamma$ (see \fref{fig:Lucioni}b) show values that are larger than those obtained from theory and numerical simulations of the 1D quasi-periodic DNLS equation (note that $\gamma=\alpha_m/2$) \cite{Larcher2009, Larcher2012}.  A possible explanation of this disagreement, which we have mentioned in the \sref{NAAnum}, is partial selftrapping in minigaps between minibands and the
subsequent release process. 

Qualitatively similar conclusions can be drawn from experiments with light propagation in disordered \cite{Pertsch2004} and quasi-periodic \cite{Lahini2009} photonic lattices, where an expansion of a single waveguide excitation was studied. Other results \cite{Schwartz07,Lahini08} were mainly focused on the self-focusing effect of nonlinearity, which can halt the spreading of initial excitation as well as Anderson localization does.


\section{Nonlinear Anderson localization}
\label{NAL}
\subsection{Localization or wave propagation?}
As we have seen, nonlinearity couples Anderson modes, induces chaotic dynamics and subdiffusive wave propagation, at least for sufficiently high energy densities. The remarkable success of the wave packet spreading theory (effective noise theory and  nonlinear diffusion) are in sharp contrast with deep controversies about the fundamentals of 
the dynamics at weak nonlinearity and energy. For example, it is unknown, whether there exists a lower bound on the nonlinearity strength, beyond which wave packets do obey Anderson localization, and therefore do not spread at all \cite{Pikovsky2008}; are some time scales diverging in this limit \cite{Fishman2008,Fishman2009,Wang2009}? Another basic question comes from realizing that even with spreading wave packets one inevitably approaches the low density and, hence, the linear limit as the wave packet expands. It remains debated, whether the observed spreading will continue infinitely or slow down and even stop to restore localization, once wave packet densities become substantially small \cite{Mulansky2010,Johansson2010,Aubry2011}. 

It is possible that the theory of nonlinear Anderson localization will be probabilistic in terms of the measure of localized regular trajectories in phase space (periodic orbits and tori). Indeed, the mere assumption that a wave packet belongs to a chaotic trajectory at time zero leads to the conclusion that chaos remains forever: Arnold's conjecture, unproved but widely accepted, states the uniqueness of the chaotic region in phase space \cite{Arnold1964}. If initial conditions belong to a chaotic trajectory, then it will be unbounded in phase space, and, characterized by mixing, by visiting all parts of the chaotic domain. That corresponds to unlimited spreading of a wave packet. The spreading is absent only if the initial conditions belong to a periodic orbit or torus, and this regular trajectory is bounded in the phase space. 

Despite the clarity of the question, the studies are still giving contradicting answers and no general agreement has been achieved. The progress in rigorous mathematical analysis has been quite limited up to now due to the difficulty of the task. The downsides of any direct computational experiment are unavoidable finite size, time, energy, and precision limitations will make the most advanced numerical results not entirely convincing. The wave packet spreading theory predicts unlimited divergence of the packet width, and numerical search for a slowing down of
the subdiffusive exponent  below the theoretically predicted asymptotic value was not successful (although
some decrease of the subdiffusion exponent was claimed to be observed \cite{Mulansky2010}, though its value stays within the bounds complying with the wave packet spreading theory, and much larger time scales need be covered in computations to make the  interpretation clear). Still, it cannot be excluded that observations might change the picture at even larger integration times. Finite time measurements of Lyapunov exponents either for wave packets \cite{Johansson2010} 
or uniform energy distributions \cite{Pikovsky2011,Basko2012} have also to be interpreted carefully. Indeed, the chaotic evolutionary dynamics is expected to exhibit prolonged sticking about regular-like trajectories \cite{Aubry2011}, and so it does intermittently with strongly delocalizing behavior observed in between \cite{Ivanchenko2011}. See also the already discussed long time Lyapunov exponent measurements in 
\cite{csigsf13} which support the effective noise and nonlinear diffusion theory for spreading wave packets.

At the risk of being subjective, we think that the overall picture is already well-shaped. At weak nonlinearity
 below a certain threshold localized tori seem to persist with a non-zero probability and disappear with asymptotic probability $1$ above it \cite{Johansson2010,Aubry2011,Ivanchenko2011}. We highlight the arguments behind these statements in more detail below.  

We also stress again that at sufficiently large energy/nonlinearity  wave packets get selftrapped due to the finite width of 
the spectrum of the linear lattice wave equations. This is a nonperturbative effect of nonlinearity on the dynamics of
nonlinear lattice waves, is present even in the absence of disorder \cite{Flach2008}, and is not discussed here.

\subsection{Localization}
Up to now, the persistence of tori has been proved for a special class of weakly nonlinear infinite systems \cite{Froehlich1986} and for finite tori dimensionality \cite{Bourgain2008} only. It also remains unclear whether the continuation is uniform and what the precise dependence of existence thresholds on the energy (nonlinearity) and disorder are.

A substantial advance in the problem is due to Johansson, Kopidakis and Aubry who conjectured that these theorems have an extension to generic infinite systems and estimated the probability of existence of infinite-dimensional tori \cite{Johansson2010,Aubry2011}. 

They considered the disordered DNLS chain model
\begin{equation}
i \dot{\psi}_l=(\epsilon_l+\beta \left|\psi_l\right|^2)\psi_l-J(\psi_{l-1}+\psi_{l+1})
\label{eq:C3_eq1}
\end{equation}
with $\epsilon_l\in[-W/2,W/2]$ and conserved norm $S=\sum_l\left|\psi_l\right|^2$. They rewrote the equations in the basis of linear Anderson modes $\{A_{\nu,l}\}$ with respective eigenvalues $E_\nu$ that obey
\begin{equation}
E_\nu A_{\nu,l}=\epsilon_l A_{\nu,l}-J(A_{\nu,l-1}+A_{\nu,l+1})
\label{eq:C3_eq2}
\end{equation}
as $\psi_l(t)=\sum_\nu A_{\nu,l}\phi_\nu(t)$ to obtain
\begin{equation}
i \dot{\phi}_\nu=E_\nu\phi_\nu+\beta\sum_{\nu^\prime,\mu,\mu^\prime} I_{\nu,\nu^\prime,\mu,\mu^\prime} \phi_{\nu^\prime}^\ast \phi_{\mu}\phi_{\mu^\prime},
\label{eq:C3_eq3}
\end{equation}
where mode interaction weight coefficients are defined as $I_{\nu,\nu^\prime,\mu,\mu^\prime}=\sum_l A_{\nu,l} A_{\nu^\prime,l} A_{\mu,l} A_{\mu^\prime,l}$. Note, that the overlap coefficients for the modes that do not belong to the same localization volume are exponentially small due to their Anderson localization. 

The authors estimated the energy current between a couple of modes $\nu, \nu^\prime$ assuming that the unperturbed solutions $\phi_{\nu,\nu^\prime,\mu,\mu^\prime} \propto \e^{-iE_{\nu,\nu^\prime,\mu,\mu^\prime}t}$ are good approximations, at least, for weak nonlinearity (small energy) and over some finite time scale. For localization to hold they requested that the energy current time average is much less than the energy of the seed mode. This led them to the Chirikov-type criterion of the absence of resonances:
\begin{equation}
\left|E_\mu-E_{\mu^\prime}\pm(E_\nu-E_{\nu^\prime})\right| > \eta\left|\beta I_{\nu,\nu^\prime,\mu,\mu^\prime}\right| \cdot \left|\phi_\mu(0)\right| \cdot \left|\phi_{\mu^\prime}(0)\right|,
\label{eq:C3_eq4}
\end{equation}
where $\eta$ is a parameter of order $1$. Assuming that the eigenvalues are independent random numbers with smooth probability law and maximum density $\cP_0$, the authors bounded the probability to find at least one resonance $\cP_{\rm R}$ by
\begin{equation}
\cP_{\rm R} < \cP_0 \eta \beta A S,
\label{eq:C3_eq5}
\end{equation}
where $A$ is a finite positive value proportional to the squared localization length. Correspondingly, the probability to have no resonance is
\begin{equation}
\label{eq:C3_eq6}
\cP_{\rm N} > 1-\cP_0 \eta \beta A S,
\end{equation}
or, in the limit of a distributed wave packet with low norm/energy, $\cP_{\rm N} \approx \e^{-\cP_0 \eta \beta A S}$, which agrees well with numerical data of \cite{Krim10}.

Numerical calculations of recurrence rates and Lyapunov exponents gave supportive evidence for the increase in the fraction of regular-like localized trajectories as the wave packet energy decreases \cite{Johansson2010,Aubry2011}.

Led by qualitative arguments and numerics the authors conjectured that for sufficiently small packet energy there is always a finite measure of localized regular trajectories and a finite probability of Anderson localization which survives nonlinearity. Moreover, they proposed that even chaotic trajectories may show asymptotic slowing down of spreading, as the volume of regular motion, presumably, expands with the decay of energy density and the spreading trajectories become trapped by close vicinity tori for increasingly long times. While a systematic verification of this hypothesis is needed, there exist supportive evidence of intermittent spreading in the strong disorder limit \cite{Ivanchenko2011}. This effect, though, does not lead to slowing down of subdiffusion on average, in contrast to expectations of \cite{Aubry2011}. 

Further development of these exciting results and ideas is much awaited, in particular, improving the non-resonance criterion, working towards the existence proof of infinite dimensional tori and their non-zero measure, pushing up precision and time scales of detecting chaotic motion or recurrent quasi-regular dynamics.

\subsection{Delocalization}
Recent attempts to estimate the KAM tori probability, rigorously speaking, have not yet produced sufficient conditions for tori persistence, nor their measure in the phase space. An alternative approach, much less complicated, suggested by Ivanchenko \textit{et al.} \cite{Ivanchenko2011} is to derive sufficient conditions for wave spreading, and for
localized periodic orbits and tori to break up.

\subsubsection{Strong disorder limit}
Consider first the paradigmatic Fr\"ohlich-Spencer-Wayne (FSW) type classical chain, for which the linear eigenmodes of a disordered system are compact single-site excitations with random frequencies, and where infinite dimensional KAM tori exist \cite{Froehlich1986}.
This model can be taken as the strong disorder limit of generic classical Klein-Gordon and semi-classical discrete nonlinear Schr\"odinger arrays. 

The FSW Hamiltonian reads
\begin{equation}
\cH_{\rm FSW}=\sum_{{l}}\left[\frac{p_{{l}}^2}{2}+\frac{\epsilon_{{l}}u_{{l}}^2}{2}
+\sum_{{m=l\pm1}}\frac{(u_{{m}}-u_{{l}})^\gamma}{2\gamma} \right],
\label{eq:C3_eq1a}
\end{equation}
where $u_{{l}}$ is the displacement of the $l$-th particle from its original
position, $p_{{l}}$ its momentum, and the random uncorrelated $\epsilon_{{l}}\in[1/2,3/2]$ are uniformly distributed. Unless explicitly specified, a chain with quartic anharmonicity $\gamma=4$ is considered. Without the loss of generality one also assumes $\epsilon_0=1$.

As argued above, Anderson localization can be observed under two simultaneous conditions: existence of a localized regular trajectory and the initial conditions belonging to it. Therefore, the probability for such a trajectory to exist at a given energy, with realizations of disorder varied, will give an upper bound for localization probability. Generically, one should consider the wave packets of different initial width $L$ and energy profile and, hence, derive the existence probability of $L$-site localized periodic and quasi-periodic solutions to (\ref{eq:C3_eq1a}). 

Start with periodic orbits, which are single-site localized solutions, and derive conditions of their destruction (when regular trajectories delocalize the 
corresponding wave packets are assumed to spread). Construct an exact time-periodic orbit of (\ref{eq:C3_eq1a}) localized at $l=0$ by perturbation theory $u_l(t)=\sum_{k=0}^\infty u_l^{(k)}(t)$ in the small-amplitude limit, taking $u_0^{(0)}(t)=A_l\cos t$, $u_{l\neq 0}^{(0)}(t)=0$ as the zero-order approximation (note that $\epsilon_0=1$). In the first order one finds
\begin{equation}
u_{\pm 1}(t)=A_{\pm 1}\cos{t}, \ A_{\pm 1}=\frac{3 A_{0}^3}{4(\epsilon_{\pm 1}-\epsilon_0)},
\end{equation}
and in higher orders
\begin{equation}
u_{\pm l}(t)=A_{\pm l}\cos{t}, \ A_{\pm l}=\frac{3 A_{\pm (l-1)}^3}{4(\epsilon_{\pm l}-\epsilon_0)}.
\label{eq:C3_eq2a}
\end{equation}
The necessary condition for convergence of the perturbative solution (\ref{eq:C3_eq2}) is the the decay of amplitudes: 
\begin{equation}
\left|\frac{A_{\pm l}}{A_{\pm (l-1)}}\right|=\frac{3 A_{\pm (l-1)}^2}{4\left|\epsilon_{\pm l}-\epsilon_0\right|} < \frac{1}{\kappa}, \ \kappa>1.
\label{eq:C3_eq3a}
\end{equation}

The probability for condition (\ref{eq:C3_eq3}) to hold at the distance $\pm l$ from the central site $\cP^{(\pm l)}$ is determined by respective random onsite potentials $\epsilon_{\pm l}$ and the amplitude of the preceding oscillator: 
\begin{equation}
\cP^{(\pm l)}=1-\frac{3}{2}\kappa A_{\pm (l-1)}^2 \ge 1-\frac{3}{2} \kappa^{3-2l} A_0^2.
\label{eq:C3_eq4a}
\end{equation} 
The probability to obtain a localized time-periodic solution in the infinite chain $\cP=\prod_{l=1}^\infty \left[\cP^{(l)}\right]^2$ is bounded from above
by the probability to have decreasing amplitudes at least in the first neighbors
\begin{equation}
\cP \le \cP^{(1)}\cP^{(-1)}=\left(1-3\kappa\cE\right)^2 \equiv\cP^{+},
\label{eq:C3_eq5a}
\end{equation}
where $\cE=A_0^2/2$ is the energy on the central site.

It follows from (\ref{eq:C3_eq5a}) that above the threshold $\cE>1/3$ it is not possible to construct a {\it single-site localized} ($\left|A_l/A_0\right|\ll 1, \ \forall l\neq 0$) time-periodic orbit. Below it, the lower bound of spreading probability always remains non-zero and scales linearly with the total wave packet energy $\cE$, in agreement with \cite{Johansson2010,Aubry2011}.

Consider now $L$-site localized solutions to (\ref{eq:C3_eq1a}), which, in the zero energy limit would correspond to $L$-dimensional tori. Developing the perturbation theory in the same manner as for single-site excitations one immediately observes that the existence probability is maximized for sparse packets, when the most excited sites are separated by intervals of weakly excited ones.  In leading order the problem of a sparse excitation with $L$ sites with the energy $\cE/L$ per site, separated by at least two non-excited sites, yields $L$ independent single-site problems. With (\ref{eq:C3_eq5a}), the upper 
bound for the localization probability of {\it $L$-site localization} reads
\begin{equation}
\cP_L=\left(1-\frac{3\kappa \cE}{L}\right)^{2L}.
\label{eq:C3_eq7}
\end{equation}
$\cP_L$ is a monotonously increasing function of $L$ with the {\it infinite-size} packet asymptotic localization probability 
\begin{equation}
\cP_\infty=\e^{-6\kappa\cE}.
\label{eq:C3_eq8}
\end{equation}
Equations (\ref{eq:C3_eq7}) and (\ref{eq:C3_eq8}) constitute the central results of the nonlinear Anderson localization theory. 

Projecting these results on the original problem of Anderson localization of an initial $L$-size excitation one calculates the ratio of the volume $v_{\rm l}$ of all points in the $2L$-dimensional phase space, which yield localization, to the full available volume $v_{\rm l}+v_{\rm s}$ where $v_{\rm s}$ is the volume of all points which yield spreading: $\cP_L=v_{\rm l}/(v_{\rm l}+v_{\rm s})$ and averages over disorder realizations.
Clearly, the existence of localized trajectories gives an upper bound for this probability. 
Therefore, for a wave packet of size $L$ no regular Anderson localized states are expected if the energy density $h\equiv\cE/L > 1/3$.
But even for $h < 1/3$ there is always a finite probability to observe spreading trajectories. 

Remarkably, at a given fixed total energy $\cE$, the probability for Anderson localization has a finite limit for the infinite packet size $L$ (\ref{eq:C3_eq8}). 
Thus there remains always a nonzero probability to spread, \textit{i.e.} $v_{\rm s}/(v_{\rm l}+v_{\rm s}) \neq 0$ in this limit.
 
It is worth noting that the derivative $\partial\cP_L/\partial L \propto L^{-2}$ is vanishing as a power law, and not as an exponential for large $L$. Therefore, at variance to the case of Anderson localization, there is no new length scale emerging. In particular, already the first moment $\langle L \rangle$ obtained with such a probability distribution function diverges.
\begin{figure}[h]
\begin{center}
\includegraphics[width=0.7\columnwidth,keepaspectratio,clip]{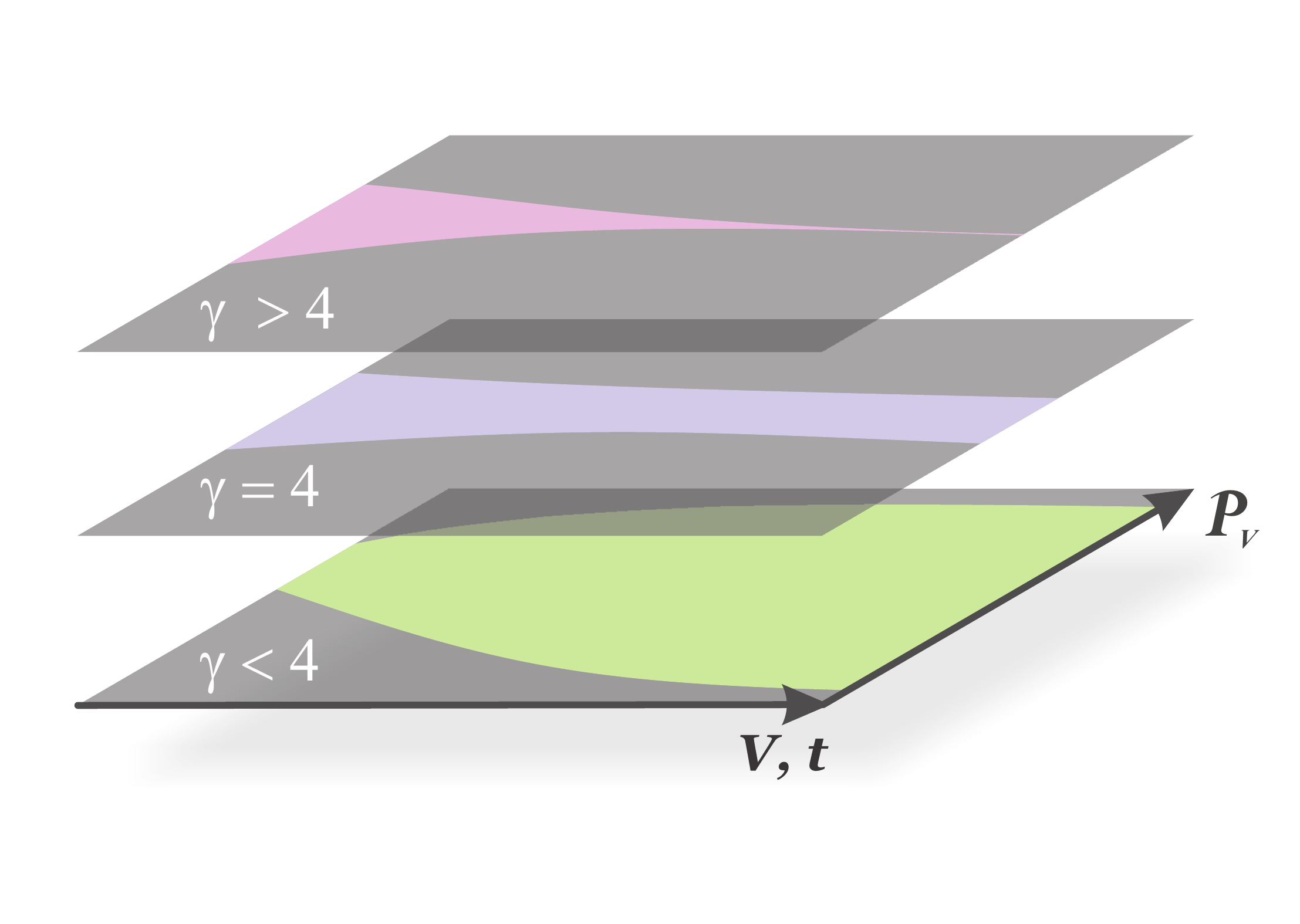}
\caption{Schematic dependence of the probability $\cP_V$ for wave packets to stay localized (dark area) together with the complementary
light area of spreading wave packets versus the wave packet volume $V$ (either initial or attained at some time $t$) for three different orders of nonlinearity
$\gamma<4$, $\gamma=4$ and $\gamma > 4$. Adapted from \cite{Ivanchenko2011}.}
\label{fig:C3_fig3}
\end{center}
\end{figure}

These results can be easily generalized for arbitrary degree of nonlinearity $\gamma$ and lattice dimensionality $\bf D$.
Here one should consider localized solutions of equally excited sites that occupy volume $V$ of characteristic size $L$. Like before, sparse solutions maximize their existence probability $\cP_V$. As the volume scales as $V \propto L^{\bf D}$ the probability to find a localized solution is given by the following product:
\begin{equation}
\cP_V=\left(1-\frac{\kappa_\gamma \cE^{\gamma/2-1}}{V^{\gamma/2-1}}\right)^{2 V {\bf D}}.
\label{eq:C3_eq9}
\end{equation}
Remarkably, for any $\gamma$ no regular localized wave packets can be obtained if the energy density $h$ exceeds a $\gamma$-dependent threshold. For smaller $h$ there is always a finite probability to launch a spreading wave packet. One further finds that the fraction of localized wave packets $v_{\rm l}/(v_{\rm l}+v_{\rm s})$ tends to zero in the limit $V \rightarrow \infty$ at fixed $\cE$ for $\gamma < 4$, and tends to unity for $\gamma > 4$ (but this is only the upper bound and the actual probability may still remain less than $1$), as shown schematically in \fref{fig:C3_fig3}.

It follows that the evolution of wave packets of different volume is sensitively controlled by the order of nonlinearity (\fref{fig:C3_fig3}). For small-order nonlinearities $\gamma<4$ the upper bound for localization probability $v_{\rm l}/(v_{\rm l}+v_{\rm s})$ converges to $0$ exponentially fast. In the case of quartic nonlinearity ($\gamma=4$) its asymptotic value lies in the interval $(0,1)$, becoming exponentially close to $0$ with the increase of the total energy $\cE$. 
Conversely, for high-order nonlinearities $\gamma>4$ the upper bound for localization probability tends to unity for $\gamma > 4$ (note, that the actual probability may still prove to be less than $1$). 

\subsubsection{General case}
Let us turn to the case of a generic nonlinear lattice with exponentially localized modes and multi-mode interactions within a localization volume. As a representative model consider the 1D Klein-Gordon lattice (\ref{eq:1DdKG}).

Define $A_{\nu,l}$ to be the spatial components of the $\nu$-th eigenvector of the linear counterpart of (\ref{eq:1DdKG}) and $\omega_\nu^2$ to be the corresponding eigenvalue. Canonical transform $u_l=\sum_\nu Q_\nu A_{\nu,l}$, $p_l=\sum_\nu P_\nu A_{\nu,l}$ gets (\ref{eq:1DdKG}) into the reciprocal (linear mode) space. The equations of motion read
\begin{equation}
\ddot{Q}_\nu=-\omega_\nu^2Q_\nu-\sum_{\nu_1,\nu_2,\nu_3}I_{\nu,\nu_1,\nu_2,\nu_3}Q_{\nu_1}Q_{\nu_2}Q_{\nu_3},
\label{eq:KGeq2}
\end{equation}
where $I_{\nu,\nu_1,\nu_2,\nu_3}=\sum_l A_{\nu,l}A_{\nu_1,l}A_{\nu_2,l}A_{\nu_3,l}$ are the overlap integrals.

As above, construct a regular periodic/quasiperiodic orbit by the perturbation theory and estimate the probability of such solution to exist. Start with constructing a localized periodic orbit centered at the mode $\nu_0$, taking a single mode excitation $Q_{\nu_0}^{(0)}=A_0\cos(\bar{\omega}_{\nu_0}t)$ and $Q_{\nu\neq\nu_0}^{(0)}=0$ as a zero-order approximation. Here the nonlinear frequency shift is taken into account $\bar{\omega}_{\nu_0}=\omega_{\nu_0}+3I_{\nu_0,\nu_0,\nu_0,\nu_0}A_0^2/(4\omega_{\nu_0})$. The first order corrections in the mode $\nu\neq\nu_0$ read:
\begin{equation}
Q_\nu=A_\nu\cos\bar{\omega}_{\nu_0}t, \quad A_\nu\approx\frac{3}{4}\frac{I_{\nu,\nu_0,\nu_0,\nu_0}A^3_0}{\omega_\nu^2-\bar{\omega}_{\nu_0}^2}.
\label{eq:KGeq3}
\end{equation}
The perturbation theory breaks up when $\left|A_\mu/A_0\right|\ge1/\kappa$, where $\kappa>1$ is some constant.
The corresponding probability is $\cP^{-}_{\nu,\nu_0}=\mbox{Prob}(\left|\omega_\nu^2-\bar{\omega}_{\nu_0}^2\right| < \frac{3}{4}\kappa I_{\nu,\nu_0,\nu_0,\nu_0}A_0^2)$. 

To proceed it one needs to know the probability distribution of the eigenvalue spacing $\mathcal{W}(s)$ inside
a localization volume, where $s\equiv\omega_\nu^2-\omega_{\nu_0}^2$. 
Due to level repulsion it is reasonable to assume a linear dependence $\mathcal{W}(s)\propto s$ similar to Gaussian Orthogonal Ensemble matrices \cite{Evers2008}. Under this conjecture the probability for localization is bounded from above by 
\begin{equation}
\cP^{+}_{\nu_0} \approx 1-C \cE^2,
\label{eq:KGeq4}
\end{equation}
where $C$ is some constant. It follows, that for small  finite energies there is a non-zero probability for the localized periodic orbit to be destroyed, which guarantees spreading from the corresponding type of initial wave packet. 

However, the mere existence of a localized periodic trajectory is not enough for the set of regular tori (localized in the mode space) to exist around this orbit. Indeed, its instability would mean the break up of the tori structure and spreading for all wave packets, whose initial conditions belong to the neighborhood of the periodic orbit, even if the latter is localized. 

Let us linearize (\ref{eq:KGeq2}) about an exact periodic solution $\bar{Q}_\mu(t)$ centered at  $\nu_0$  taking $Q_\mu(t)=\bar{Q}_\mu(t)+\xi_\mu(t)$:
\begin{equation}
\ddot{\xi}_\nu=-\omega_\nu^2\xi_\nu-3\sum_{\nu_1,\nu_2,\nu_3}I_{\nu,\nu_1,\nu_2,\nu_3}Q_{\nu_1}Q_{\nu_2}\xi_{\nu_3}.
\label{eq:KGeq5}
\end{equation}

The level repulsion conjecture is effective for pairs of eigenvalues and coexists with a finite probability of exact triplet resonances $\omega_\mu+\omega_\nu=2\omega_{\nu_0}$. Keeping the resonant terms for such triplets one gets the generalized Mathieu equation 
\begin{eqnarray}
\ddot{\xi}_\nu=-(\omega_\nu^2+\frac{3}{2}I_{\nu,\nu_0,\nu_0,\nu})\xi_\nu-\frac{3}{2}I_{\nu,\nu_0,\nu_0,\mu}A_{\nu_0}^2\cos(2\bar{\omega}_{\nu_0}t)\xi_{\mu}, \nonumber \\
\ddot{\xi}_\mu=-(\omega_\mu^2+\frac{3}{2}I_{\mu,\nu_0,\nu_0,\mu})\xi_\mu-\frac{3}{2}I_{\mu,\nu_0,\nu_0,\nu}A_{\nu_0}^2\cos(2\bar{\omega}_{\nu_0}t)\xi_{\nu}.
\label{eq:KGeq6}
\end{eqnarray}

Standard stability analysis yields the  probability of instability $\cP^{\rm u}_{\mu,\nu_0,\nu} \propto \left|I_{\nu,\nu_0,\nu_0,\mu}\right| \cE/V$. Taking into account all triplet interactions in the localization volume we obtain the probability for the localized orbit to be linearly stable:
\begin{equation}
\cP^{\rm s}_{\nu_0} \approx 1-\hat{C}\langle\left|I_{\nu,\nu_0,\nu_0,\mu}\right|\rangle \cE,
\label{eq:KGeq8}
\end{equation}
where $\hat{C}$ is some constant. It follows, that the probability for the single-mode localized periodic orbit to be unstable drops only linearly with the energy. So does the probability of spreading, same as in the FSW system.

Let us discuss multiple mode (hence, multiple frequency) excitations. It is easy to show that if at least two modes interact effectively, \textit{i.e.} belong to the same localization volume, triplet resonances become possible and break up the existence of the localized multi-mode solution with the probability linear in $\cE$. That contrasts the case of single-mode excitations, for which the energy dependence it quadratic (\ref{eq:KGeq4}). Therefore, sparse solutions consisting of locally single mode excitations minimize the probability of the break up, as before. The lower bound for this probability (and, hence, the upper bound for the existence) is given by the product of the one for single-mode localizations. Assuming $L$ effectively excited modes in the sparse wave packet, $\cE/L$ being the energy of each, one gets
\begin{equation}
\cP^{+}_L=\left[1-C\left(\frac{\cE}{L}\right)^2\right]^L.
\label{eq:KGeq9}
\end{equation}
For any finite wave packet width there exists a finite threshold in the energy density above which the upper bound of a localized regular solution probability becomes $0$, and below which is less than $1$. For infinitely sparse wave packets with the energy density below the threshold it is asymptotically close to 1, $\cP^{+}_\infty=1$. 

However, the upper bound for stability has a different scaling:
\begin{equation}
\cP^{\rm s}_L=\left[1-\hat{C}\frac{\cE}{L}\right]^L,
\label{eq:KGeq10}
\end{equation}
giving $\cP^{\rm s}_\infty=\exp(-\hat{C}\cE)<1$ independent on how sparse the solution becomes.

We conclude that the probability of a stable single-mode periodic orbit has an upper bound linearly decreasing with the increase of the energy. Above some threshold in energy it becomes zero. It also serves as an upper bound for the existence of localized infinite-dimensional tori around the orbits, and  hence for the no-spreading probability.

For multiple-site solutions the probability of existence decreases linearly which leads to the same conclusions for spreading/non-spreading probabilities. The existence criterion can be relaxed if sparse solutions of many single-mode centers of excitations are considered and the energy density is lowered. Still, the probability of instability decays only linearly with energy which again gives a non-zero ``sparsening'' limit dependent on the total energy, not its density, the final answer for localization probability being the same as for the FSW case.

\subsubsection{Numerical results}
The analytical results can be verified by direct numerical experiments. The most straightforward way is to simulate the evolution of wave packets, tracking their distribution characteristics (second moment $m_2$, participation number $P$, see the previous section). The work \cite{Ivanchenko2011} analyzed dynamics of single-mode initial excitations in dependence on energy $\cE$ in FSW (\ref{eq:C3_eq1a}) and KG (\ref{eq:1DdKG}) 1D lattices.  Numerical  integration was performed by a symplectic SABA-type scheme \cite{Skokos2009} up to $t_{\rm end}=10^9$. 
\begin{figure}[h]
\begin{center}
\includegraphics[width=0.6\columnwidth,keepaspectratio,clip]{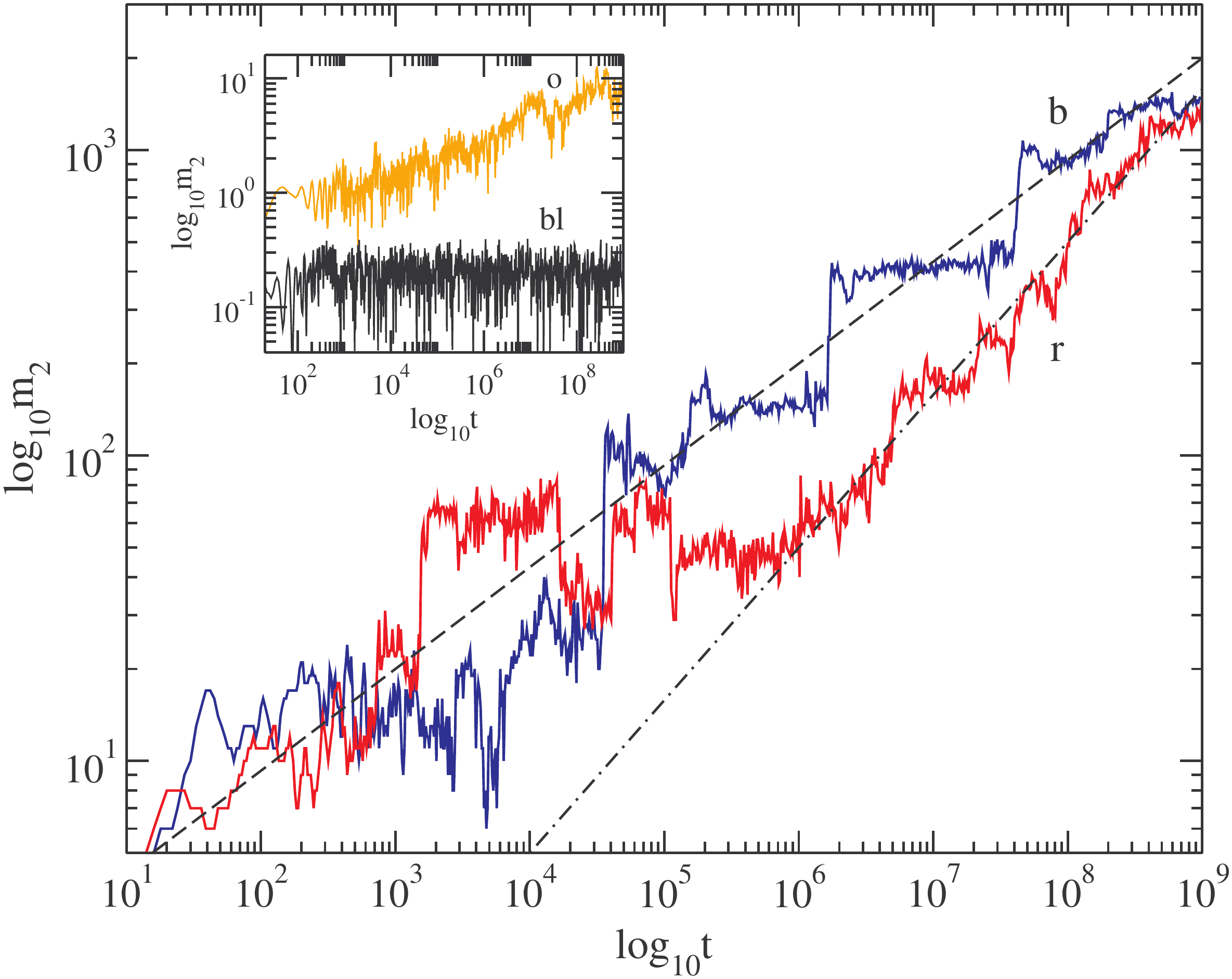}
\caption{Numerical results for FSW chain (\ref{eq:C3_eq1a}) with initial parameters $\gamma=4$ and $\cE=1$: for two different disorder realizations the second moment $m_2$ of a single-site excitation is shown \textit{vs.} time. Dashed and dash-dotted lines with the slopes $1/3$ and $1/2$, respectively, guide an eye. \textit{Inset:} for two different disorder realizations the second moment $m_2$ of a spreading and a non-spreading single-site excitation is shown \textit{vs.} time (initial energy is $\cE=0.05$). Adapted from \cite{Ivanchenko2011}.}
\label{fig:C3_fig1}
\end{center}
\end{figure}

It was found that the spreading dynamics of the FSW model depends strongly on the disorder realization. For sufficiently large energies the divergence of $m_2$ varies between $m_2 \propto t^{1/3}$ and $m_2 \propto t^{1/2}$ (\fref{fig:C3_fig1}, $\cE=1$). Although the spreading theory predicts the strong chaos regime here, a large variance makes it impossible to specify a characteristic exponent of the power-law divergence. For small enough energies some realizations manifest Anderson localization (see inset in \fref{fig:C3_fig1}, $\cE=0.05$).

\begin{figure}[h]
\begin{center}
\includegraphics[width=0.6\columnwidth,keepaspectratio,clip]{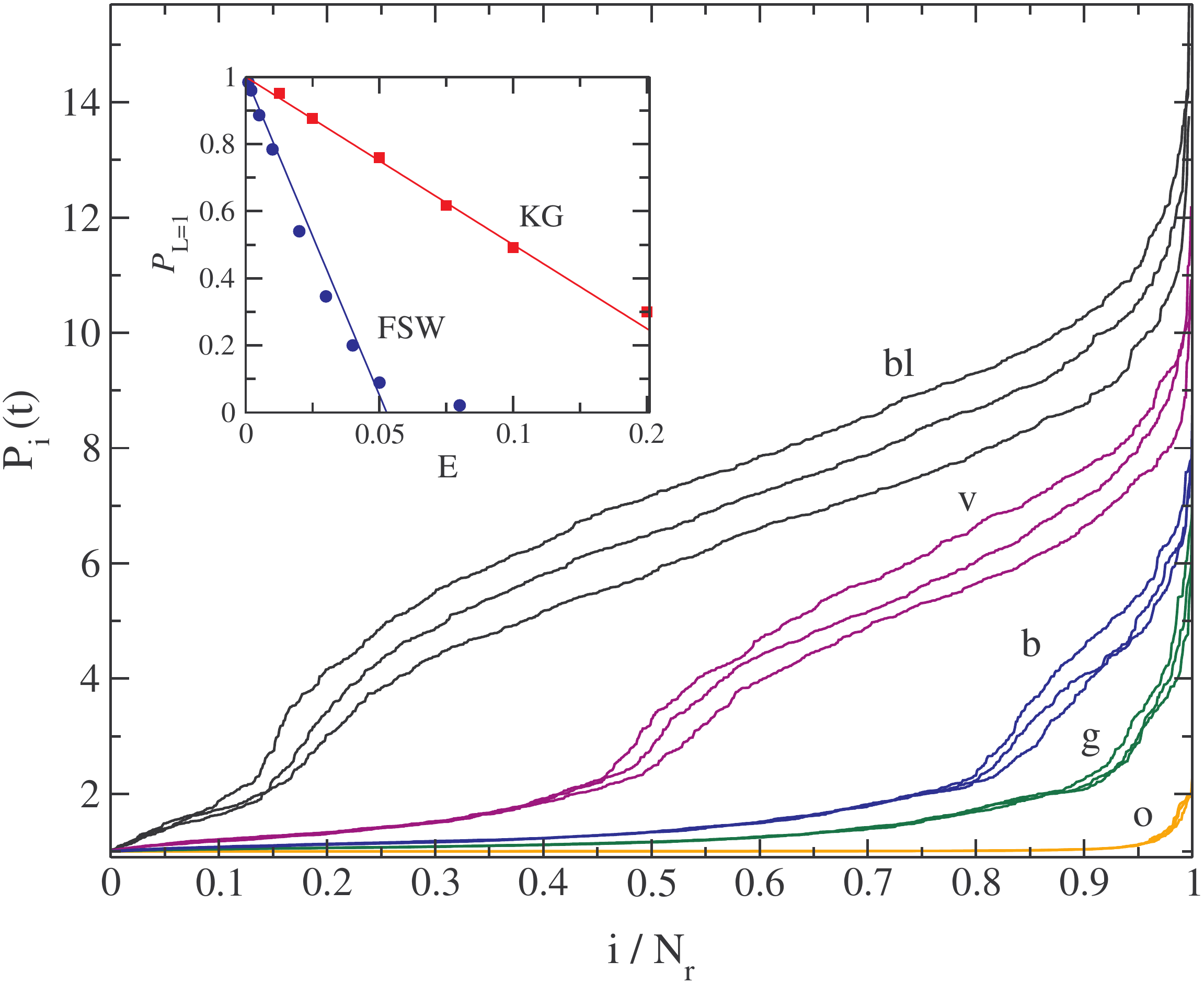}
\caption{Numerical results for FSW chain (\ref{eq:C3_eq1a}) with initial parameters $\gamma=4$ and $\cE=0.08, 0.05, 0.03, 0.02, 0.002$ (black (bl), violet (v), blue (b), green (g), orange (o)). For $N_{\rm r}=10^3$ different disorder realizations the sorted participation numbers $P_i$ with $1 \leq i \leq N_{\rm r}$ are shown at $t=10^8, 5\cdot 10^8, 10^9$ (bottom-top). \textit{Inset:} single-site localized fractions $\cP_{L=1}$, numerics by symbols, linear fits for FSW and for KG by solid lines. Adapted from \cite{Ivanchenko2011}.}
\label{fig:C3_fig2}
\end{center}
\end{figure}

Sorting realizations in increasing order of their participation numbers at $t_{\rm end}$ breaks them into spreading and non-spreading groups (\fref{fig:C3_fig2}). As predicted, the increase of energy increases the spreading fraction. Define the localized single-site (single-mode for the KG chain) trajectory as the one whose participation number at the end of integration $P(t_{\rm end})<1+\varepsilon=1.2$. Taking into account that $P \approx 1+2 (\cE_{\nu_0-1}+\cE_{\nu_0+1})/\cE_{\nu_0}$ for strongly localized trajectories, (\ref{eq:C3_eq4a}) yields for the non-spreading fraction
\begin{equation}
\cP \approx 1-6\sqrt{\frac{2}{\varepsilon}}\cE.
\label{eq:C3_eq9a}
\end{equation}

The numerically obtained dependence of the non-spreading fraction on the energy shows a linear decay and agrees well with the analytical estimate (\ref{eq:C3_eq9a}) (see inset in \fref{fig:C3_fig2}). The KG system manifests a linear decay as well, following theoretical predictions. 

\subsection{Lyapunov exponents}
Direct simulations of spreading wave packets give an obvious point for criticism: they are inevitably constrained by numerical integration time limits. A way to go about it, taken in several studies, is to calculate Lyapunov exponents for both the initially localized wave packets and uniform initial density distributions \cite{Aubry2011,Pikovsky2011,Basko2012,Basko2011}. Indeed, as we pointed out before, chaos would guarantee that the wave packet diverges infinitely. Therefore, if one obtains positive Lyapunov exponents on {\it finite} integration times, as it was clearly shown in \cite{csigsf13}, is enough to claim spreading. 

The papers \cite{Aubry2011,Pikovsky2011,Basko2012,Basko2011} convincingly demonstrate the existence of regular and chaotic trajectories at, respectively, lower and higher energies, and even coexistence at the same parameters in principle. The still open question is the scaling of the ratio between these types of behavior with energy (nonlinearity), as one faces a very complicated problem of distinguishing between very small positive and zero Lyapunov exponents.  

Extrapolation of such numerics to infinite size systems is based on the conjecture of the locality of chaos \cite{Pikovsky2011,Basko2011}, stating that for weak nonlinearity (small energy) chaos emerges locally. It relies on the assumption that the probability to observe a regular behavior in a long array is to a good approximation the product of respective probabilities for its segments: (i) if at least one segment of the array displays chaos then the whole will do as well, and (ii) if both segments produce regular oscillations so the whole array will do. Moreover, it led to propose a mechanism for the onset of chaos via mode triplets with second order resonance and the scaling of chaos probability with energy density as $h^2$ \cite{Basko2012,Basko2011}.

The main approach in numerics was to find array size independent functions and check whether the data collapse approximately on the same curve. In \cite{Pikovsky2011} the starting point was the hypothetical scaling relation between the probabilities of regular behavior $P(n,W,L)=P(n,W,L_0)^{L/L_0}$  in DNLS arrays of lengths $L$ and $L_0$ with disorder strength $W$ and  norm density $n$. An $L$-independent quantity is then $R(n,W)=P(n,W,L)^{1/L}$. Having found that the scaling holds with  a satisfactorily good accuracy, Pikovsky and Fishman fixed the chain length $L_0=16$ and analyzed the norm density scaling of the function $Q(n,W)=P_0/(1-P_0)$ to find the power law $Q \propto n^{-2.25}$, which for small $Q$ simply means the probability of chaos. Finally, they estimated the length of an array at which the probability of chaos and spreading would vanish for typical numerical values $nL \approx 1$, $1<W<10$ to obtain giant $L\sim 3\cdot 10^4 \ldots 2\cdot 10^5$. They conjectured it was the main reason why slowing down had not been computationally observable with current computing facilities. 

In turn, Basko considered the probability of chaotic dynamics in the FSW model with the energy density $h$, led by the scaling hypothesis to $P(h,L)=1-\e^{-w(h)L}$, where the size-independent function to be recovered from numerics would read $w(h)=\ln[1/(1-P(h,L))]/L$ \cite{Basko2012}. He found a nice collapse of $w(h)$ data for different sizes $L$, and, specifically, fitted $w\propto h^2$. 

The marked difference with the scaling for spreading wave packets that depends linearly on the energy \cite{Ivanchenko2011} made him propose that wave packet spreading does not necessarily implies chaos.

Very recently, Mulansky attempted to determine the scaling of the ratio of chaotic trajectories with the (finite) energy density and compare it for different order of nonlinearity in the FSW model \cite{Mulansky2014}. Reporting some difference he went further to relate the observed dependence to the wave packet spreading probability, in general, supporting conclusions of \cite{Basko2012}. However, the presented results should be treated as inconclusive since the proposed power-law scaling has been validated on the range of variables less than a single decade of magnitude. Moreover, the quality of power law fits is judged by a naked eye inspection of the proximity of scaled curves, lacking a rigorous assessment. After all, the scaling of chaos probability in disordered lattices still presents an open problem.

\section{Conclusion}

\subsection{Nonlinear Anderson localization}

It appears to be quite well established that nonlinear Anderson localization has pronounced probabilistic features. For moderate energies there exist finite probabilities of wave propagation or Anderson localization, which occur depending on whether an initial state belongs to cthe haotic or regular part of the phase space (shaped by disorder realization and parametrized by energy). This observation appears very convincing now, following from the direct measurements of wake packet profiles and Lyapunov exponents. On the other hand, there has been no clear evidence of a systematic slowing down of spreading and clear signatures of a dynamical transition from subdiffusive wave packet spreading to localization.

An intriguing question that may impact the near future research is whether all spreading wave packets are characterized by a positive Lyapunov exponent. Currently, there are multiple arguments that resonances in the wave packet occur with probability that scales linearly with small energy density $h$, correspondingly destroying localization \cite{Krim10,Aubry2011,Ivanchenko2011}. On the other hand there is a numerical observation that the measure for chaotic trajectories decays with the energy density as $h^2$ \cite{Basko2012} that is faster compared to the linear scaling above. One trivial explanation could be that the measure of trajectories with zero Lyapunov exponents  has been overestimated due to insufficient numerical integration times. Indeed, the temporal characteristics of the wave packet spreading in the FSW model show intermittent divergence, that is a fast expansion followed by prolonged stages with an almost constant second moment and participation number (\fref{fig:C3_fig1}), whose time duration may easily well exceed the whole integration time reserved for Lyapunov exponents calculations \cite{Basko2012}.

The non-trivial possibility is that finite time Lyapunov exponent are positive, as they are measured during rapid expansion stages, while the statistics of these rare events drives the asymptotic Lyapunov exponent to be zero, leading thus to chaos with zero Lyapunov exponent. The third possible answer would be the existence of a non-zero measure of regular delocalized trajectories, \textit{i.e.} spreading by regular trajectories in disordered arrays, which if proved to be true would certainly alter our current understanding of Anderson localization in one dimension.  

\subsection{Subdiffusive spreading}

If a linear wave equation generates localization with upper bounds on the localization length (degree of localization), then the corresponding nonlinear
wave equation shows destruction of this localization in a broad range of control parameters, and a subdiffusive spreading of initially localized wave packets. This observation holds
for a broad range of wave equations, e.g. with uncorrelated random potentials (Anderson localization), quasiperiodic potentials (Aubry-Andre localization),
dc fields (Wannier-Stark localization), kicked systems (dynamical localization in momentum space).  What is the cause for the observed subdiffusion?
Firstly it is
the nonintegrability of the systems, which leads to generic intrinsic deterministic chaos in the dynamics of the nonlinear system. 
Second, wave localization is inherently based on keeping the phases of participating waves coherent. Chaos is destroying phase coherence, and therefore
destroying localization. Wave packets can spread, but the densities will drop as spreading goes on. Therefore the effective nonlinearity and strength of chaos
decreases, and spreading is slowing down, becoming subdiffusive. The subdiffusive exponents are controlled by very few parameters and therefore rather universal.
Typically we only need to know the dimensionality of the system, and the power of nonlinearity (Anderson, Aubry-Andre, and dynamical localization). 
For Wannier-Stark localization the dc field strength is also becoming a control parameter, probably because the wave packet not only expands in space, but
also in the frequency (energy) domain.

\subsection{Model classes}

The more models are accumulated for the above studies, the more qualitative differences are becoming visible. For instance, models can be classified
according to the number of integrals of motion (KG - one, DNLS - two). Other models differ in the connectivity in normal mode space - while 
cubic DNLS and KG equations have connectivity $K=4$ (four modes are coupled), FSW models
have connectivity $K=2$. Again the strong disorder limit of $K=4$ models yields $K=2$ in leading order, which is one of the cases where 
analytical methods are applied. Time might be ripe to perform comparative studies.


\ack
We thank B. L. Altshuler, D. Basko, J. Bodyfelt, D. Krimer, M. Mulansky, A. Pikovsky, Ch. Skokos for many useful discussions. M.I. and T.L. acknowledge partial support of the Russian Science Foundation No. 14-12-00811 (supported preparation of Sections 1 and 2) and Russian Foundation for Basic Research No. 13-02-97028 (supported Section 3).

\clearpage
\tableofcontents
\clearpage
\section*{References}
\bibliographystyle{unsrt}
\bibliography{literatura}
\end{document}